\documentclass[final,5p,twocolumn]{elsarticle}

\usepackage[utf8]{inputenc}

\usepackage{lineno}
\usepackage[caption=false,font=footnotesize]{subfig}

\usepackage{url}

\begin{document}

\begin{frontmatter}

\title{Characterization of Three High Efficiency and Blue Sensitive Silicon Photomultipliers}

\author[gt]{Adam Nepomuk Otte\corref{mycorrespondingauthor}}
\ead{otte@gatech.edu}
\author[gt]{Distefano Garcia}
\author[gt]{Thanh Nguyen}
\author[gt]{Dhruv Purushotham}
\address[gt]{School of Physics \& Center for Relativistic Astrophysics, Georgia Institute of Technology
837 State Street NW, Atlanta, GA 30332-0430, U.S.A.
}


\cortext[mycorrespondingauthor]{Corresponding author}


\begin{abstract}
We report about the optical and electrical characterization of three high efficiency and blue sensitive Silicon photomultipliers from FBK, Hamamatsu, and SensL. Key features of the tested
devices when operated at 90\% breakdown probability are peak photon detection efficiencies between 40\% and 55\%, temperature dependencies of gain and PDE that are less than 1\%/$^{\circ}$C,
dark rates of $\sim$50\,kHz/mm$^{2}$ at room temperature, afterpulsing of about 2\%, and direct optical crosstalk between 6\% and 20\%. The characteristics of all three devices impressively
demonstrate how the Silicon-photomultiplier technology has improved over the past ten years. It is further demonstrated how the voltage and temperature characteristics of a number of
quantities can be parameterized on the basis of physical models. The models provide a deeper understanding of the device characteristics over a wide bias and temperature range. They also
serve as examples how producers could provide the characteristics of their SiPMs to users. A standardized parameterization of SiPMs would enable users to find the optimal SiPM for their
application and the operating point of SiPMs without having to perform measurements thus significantly reducing design and development cycles.
\end{abstract}


\end{frontmatter}


\section{Introduction}
Silicon photomultipliers (SiPMs) have attracted significant attention over the past few years. They are becoming increasingly popular in scientific and industrial
applications, which require fast, highly-efficient, single-photon-resolving photon detectors. Some prominent applications are in the fields of high-energy physics, astroparticle physics, and
medical imaging (s.\ \emph{e.g.}  \citep{2015NIMPA.787...85O,2012JPhCS.404a2018L,2016JInst..11C1078S,2016NIMPA.830...30B}). Reasons for the popularity of SiPMs are their high photon-detection efficiencies, mechanical and electrical robustness, low mass, low power, low bias voltages.

Another reason for the increasing popularity of SiPMs is that in recent years, they have been subject to many improvements. In particular, recent developments have successfully
addressed nuisances such as high optical crosstalk, high afterpulsing, and high dark rates, but they have also improved the photon detection efficiency, which previously limited the
usefulness of SiPMs in several applications.

We are interested in SiPMs because we aim to use them in Cherenkov telescopes to detect gamma rays from astrophysical sources. Cherenkov telescopes image the Cherenkov light emitted from
relativistic particle showers that are initiated by cosmic rays and gamma rays in the atmosphere \cite{2012EPJH...37..459L}. An in-depth understanding of photon detectors down to the level
of device physics is key in the pursuit of minimizing the systematic uncertainties present in Cherenkov telescope data.

In this paper we present an in-depth and comparative study of three recent, blue-sensitive SiPMs from FBK, SensL, and Hamamatsu, which demonstrate impressive performance improvements
compared to devices from only a few years ago, \emph{e.g.}\  \cite{2006NIMPA.563..368D}. Beside the three tested devices many more devices exist from other vendors, which could not be tested due to a lack of time and resources. 
Along with our results we give a detailed description of our test setups and discuss the measurement procedures and resulting systematic uncertainties. We, furthermore, parameterize the
overvoltage and temperature dependencies of most parameters. Where possible we use a physics-motivated model for the parameterization, which allows us to gain further insight into the device
physics of SiPMs. We hope that the parameterizations we use will help to further standardize the measurement and parameterization of SiPM characteristics.  

\begin{figure*}[!tb]
  \centering
  \subfloat[FBK NUV-HD]{\includegraphics[width=0.32\textwidth]{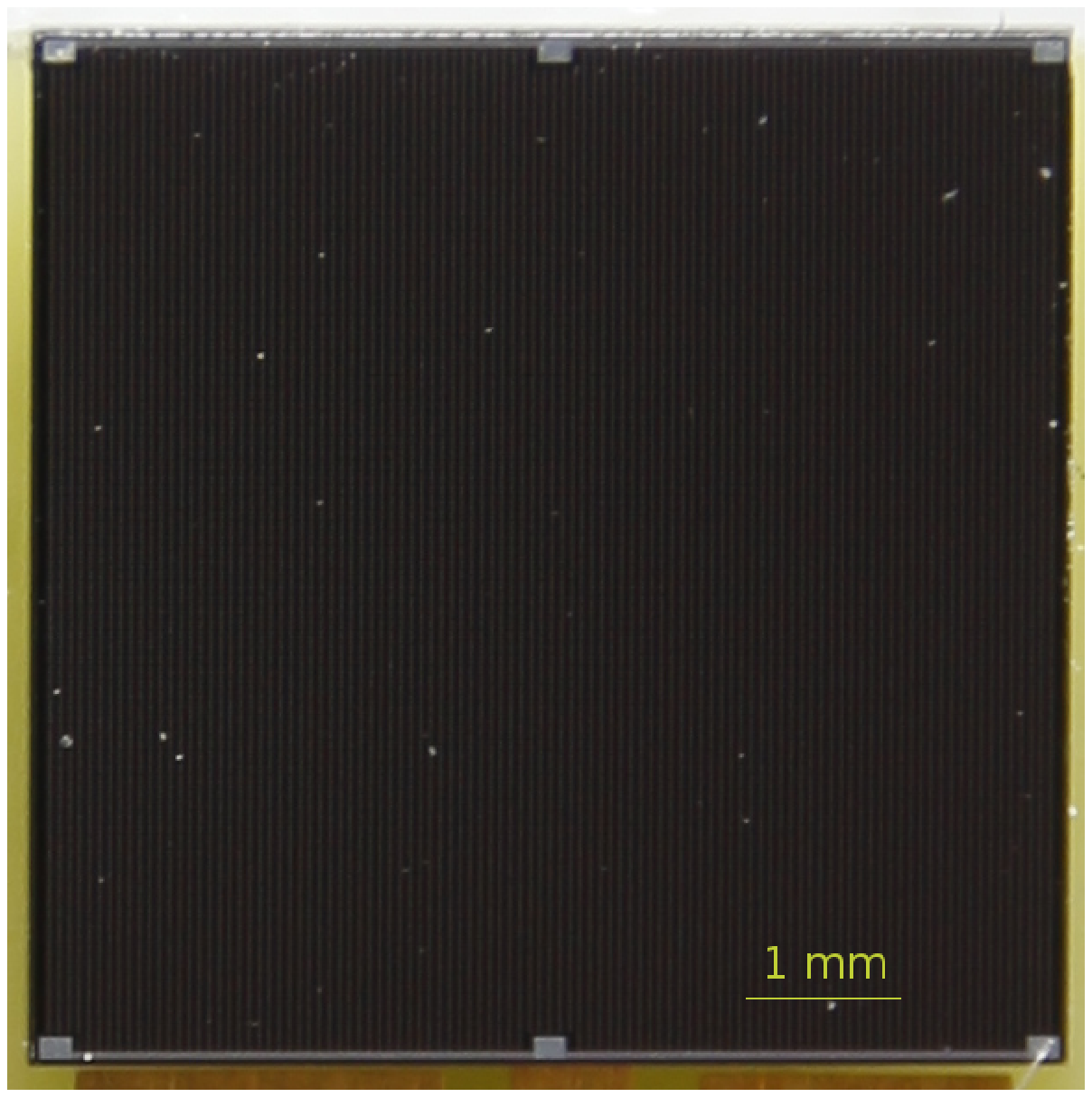}
  \label{FBKPicF}}
  \subfloat[Hamamatsu S13360-3050CS]{\includegraphics[width=0.32\textwidth]{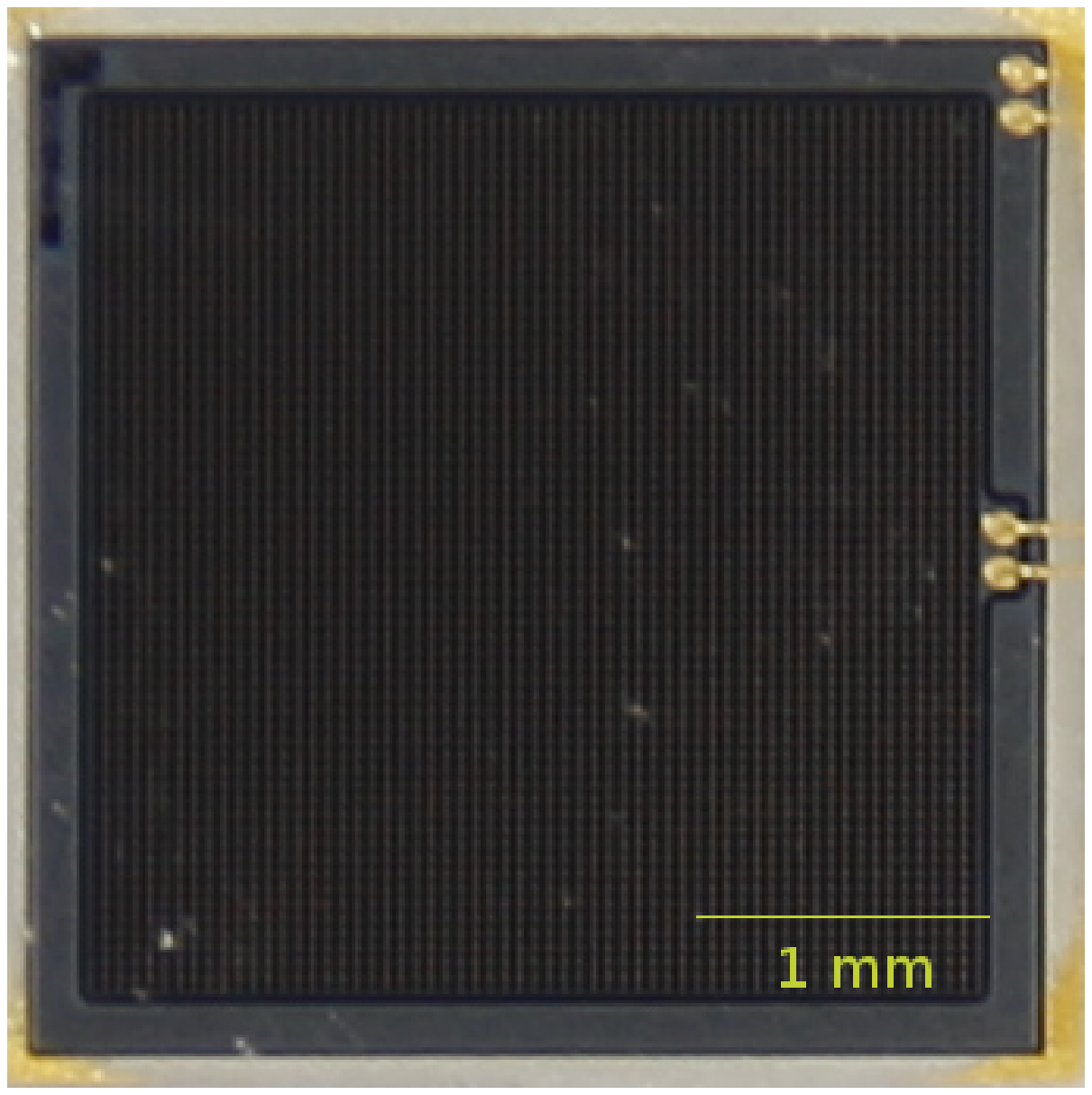}
 \label{HamamatsuPicF}}
  \subfloat[SensL J-series 30035]{\includegraphics[width=0.32\textwidth]{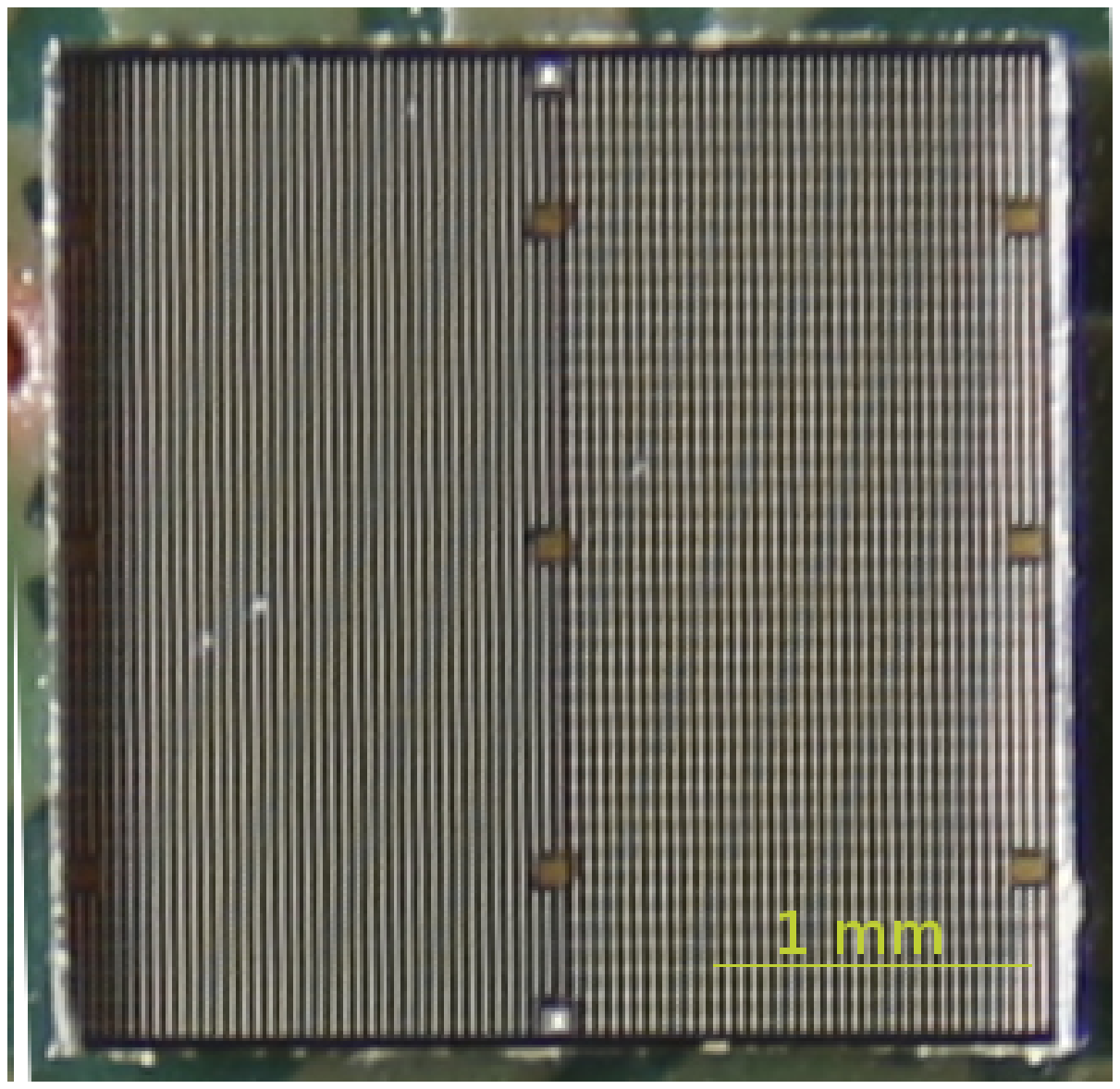}
  \label{SensLPicF}}
\caption{Full scale pictures of the three tested SiPMs. \label{fullpics}} 
\end{figure*}

\begin{figure*}[!tb]
  \centering
  \subfloat[FBK NUV-HD]{\includegraphics[width=0.32\textwidth]{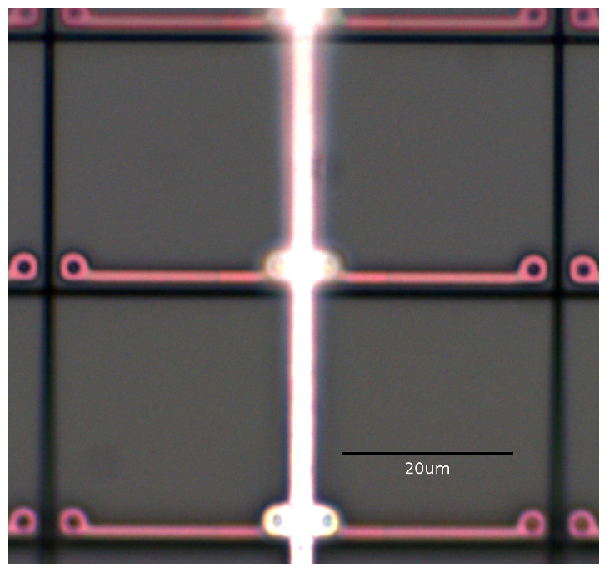}
  \label{FBKPic}}
  \subfloat[Hamamatsu S13360-3050CS]{\includegraphics[width=0.32\textwidth]{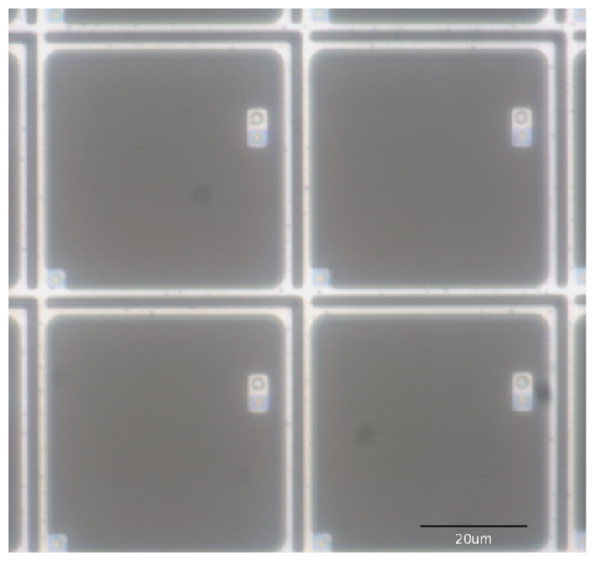}
 \label{HamamatsuPic}}
  \subfloat[SensL J-series 30035]{\includegraphics[width=0.32\textwidth]{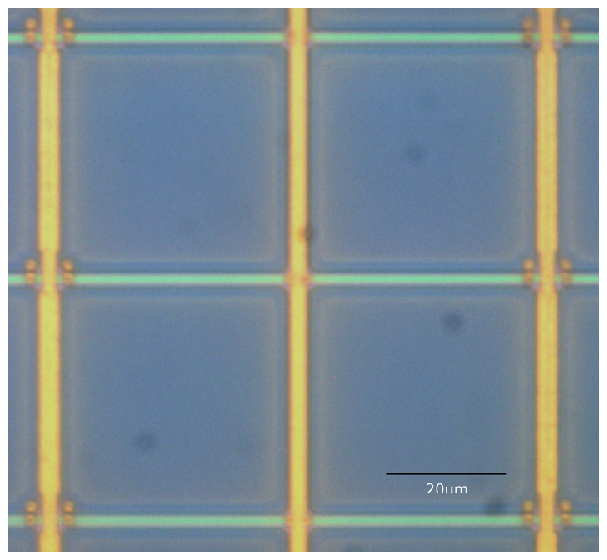}
  \label{SensLPic}}
\caption{Close-up pictures of the cells of the three tested SiPMs. The scale indicated by the black line in the images represents 20\,$\mu$m.} 
\end{figure*}

\section{Device descriptions}
SiPMs are semiconductor-based photon detectors that consist of a matrix of elementary cells, which are avalanche photodiodes operating in Geiger mode. In the conventional SiPM, which is
the type of SiPMs tested here, each cell is connected to a series resistor that limits the current flowing during the breakdown and thus ensures that the avalanche current is eventually quenched.
Furthermore, all cells are connected to one common output. For a review of the history of SiPMs and their basic functionality, the reader is referred to \cite{2009JInst...4.4004R} and
references therein.

The three tested devices are 
\begin{itemize}
\item a NUV-HD SiPM from FBK, 
\item a S13360-3050CS MPPC from Hamamatsu, 
\item and a MicroFJ-SMTPA-30035-E46 SiPM from SensL. 
\end{itemize}
A picture of each SiPM is shown in Fig.\  \ref{fullpics}.
All three devices are based on a \emph{p}-on-\emph{n} structure, which means that the avalanche structure  consists of a \emph{p}-implant in an \emph{n}-doped substrate. In this configuration the electric field
directs electrons produced by blue photons just below the surface into the high-field region, which is also why the sensitivity of all three devices peaks at wavelengths in the blue
or near UV. 

\subsection{FBK NUV-HD}

The FBK device is fabricated with NUV-HD technology \cite{2016ITED...63.1111P}.
The device investigated in this study has a custom geometry, which fits the requirements for the Cherenkov Telescope Array
(CTA) \cite{2013APh....43....3A} project. Unlike the other two devices, the NUV-HD does not have an epoxy, silicone resin, or similar protective coating. The dimensions of the FBK SiPM are
$(6.8\times6.8)$\,mm$^2$ with a micro-cell pitch of $30\,\mu$m. One SiPM has a total of 40,394 cells. The chip came glued onto a PCB carrier and is wire bonded. Fig.\  \ref{FBKPic} shows a
picture of four cells taken under a microscope. Clearly visible are the quench resistors (red) and the metal line that connects the output of all cells.

\subsection{Hamamatsu LCT5}

The SiPM from Hamamatsu is a S13360-3050CS MPPC \cite{S13360}. It is fabricated using their latest technology, which is also called LCT5 because it is the fifth iteration of a low-cross-talk development.
The dimensions of the tested device are $(3\times3)$\,mm$^2$ with a cell pitch of $50\,\mu$m (s.\ Fig.\  \ref{HamamatsuPic}) and a total of 3,600 cells. The device is mounted onto
a ceramic chip carrier and coated with UV-transparent silicon resin. Electrical contacts between the chip and the pins of the carrier are made with wire bonds. Hamamatsu produces the same
type of SiPM also with through-silicon-via (TSV) technology, which allows several chips to be packed into large matrices with minimal dead space.

\subsection{SensL J-Series}

The device from SensL is a pre-production J-Series SiPM \cite{JSeries}. The dimensions of the active area are $(3.07\times3.07)$\,mm$^2$ and the cell pitch is about $41\,\mu$m resulting in a total of 5,676
cells. The SiPM is embedded in a 4-side tileable, chip scale package with TSV that is reflow soldered onto a PCB. The SiPM came surface mounted on an evaluation board (MicroF-SMTPA). A
unique feature of SensL SiPMs is the presence of fast and slow readout terminals. The fast terminal capacitively couples directly to the cells, whereas the slow output is the conventional
readout via the quench resistor. We used the signal from the slow terminal for our measurements.

\section{Photon detection efficiency}

The photon detection efficiency (PDE) quantifies the absolute efficiency of a photon detector to absorb a photon and produce a measurable signal at its output. The PDE of SiPMs is determined
by several factors of which the three most important are the geometrical efficiency, the quantum efficiency, and the probability to produce a Geiger breakdown, hereafter breakdown probability. The breakdown probability is also referred to as triggering probability. 

We measure the PDE as a function of wavelength in three steps. In the first step, the PDE is measured at four wavelengths. In the second step, the relative spectral response is measured
between 200\,nm and 1000\,nm. 
In the last step, the spectral response is scaled to match the four PDE points and thus arrive at the PDE for all wavelengths between 200\,nm and 1000\,nm. 
In the following we walk in detail through each of these steps.
All PDE and spectral response measurements are carried out at room temperature (23$^{\circ}$C-25$^{\circ}$C).

\subsection{Concept of measuring the PDE} 

The PDE at four different wavelengths is measured with the SiPM being biased above breakdown and illuminated with fast light flashes of known intensity, and from the response of
the SiPM the PDE is calculated. 
For the measurement we use the same procedure that is described in \cite{2006NIMPA.567..360O}. 

A pulsed LED flashes fast light pulses into an integrating sphere with two exit ports, which acts as an optical splitter. The measurement of the splitting ratio is detailed in section \ref{calibSection} 
A calibrated PiN diode is mounted to one exit port, and the SiPM under test is mounted to the other port. The response of both sensors is recorded for each flash. 

After 10,000 flashes, the average number of photons at the position of the SiPM is calculated from the average PiN-diode signal, the quantum efficiency of the PiN Diode, and the splitting
ratio of the integrating sphere. The PDE of the SiPM then follows from the ratio of the average number of photons detected by the SiPM and the calculated average number of photons at the
SiPM position.

The average number of photons and dark counts detected by the SiPM $\overline{N}_{\mbox{\footnotesize Ph+DC}}$ in each flash is calculated under the assumption that the number of photons and dark counts in each flash follows a Poisson
distribution. By counting the flashes $N_0$ for which the SiPM did not detect a photon, the average number of detected photons and dark counts is
\begin{equation}
\overline{N}_{\mbox{\small Ph+DC}} = -\ln\left( \frac{N_0}{N_{\mbox{\small total}}} \right)\,,
\end{equation}
where $N_{\mbox{\small total}}$ is the number flashes. The contribution from dark counts is determined by triggering the read out $N_{\mbox{\small total}}$
times without flashing the LED. As in the previous case, the number of times the SiPM did not record a signal ($N_0^{\mbox{\tiny DC}}$) is counted. The dark-count-subtracted
average number of photons detected by the SiPM is then
\begin{equation}
\overline{N}_{\mbox{\small Ph}} = \ln\left( \frac{ N_0^{\mbox{\tiny DC}} }{N_0} \right)\,.
\end{equation}
The described procedure is commonly used to calculate the mean number of photons detected by SiPMs because it is immune to afterpulsing and optical crosstalk.

\subsection{PDE measurement setup} 

\begin{figure}[!tb]
  \centering
  \includegraphics[width=\columnwidth]{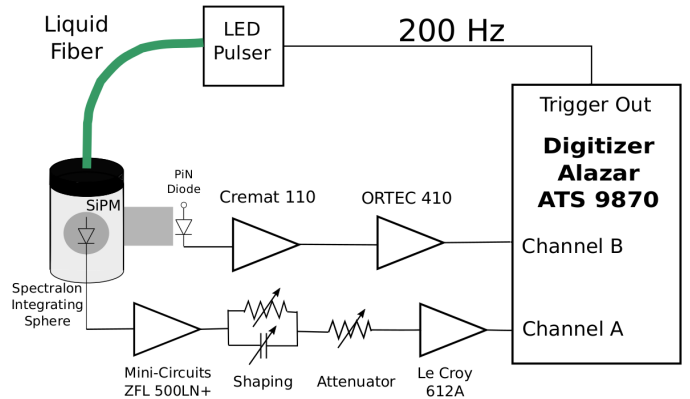}
\caption{Sketch of the PDE setup.}
  \label{PDESetup}
\end{figure}

\begin{figure*}[!tb]
  \centering
  \includegraphics[width=\textwidth]{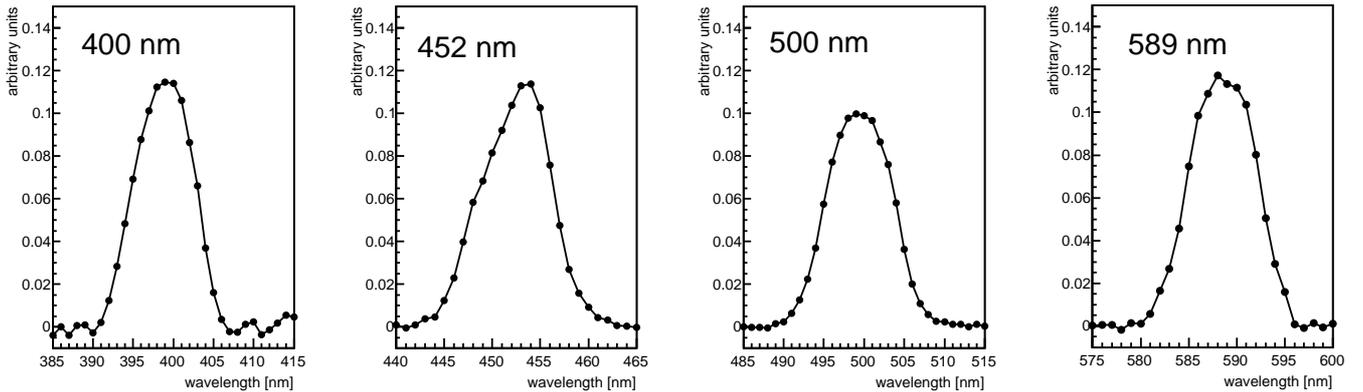}
\caption{Spectra of the four LEDs after the light has passed through a narrow bandpass filter. The LEDs are  operated in pulsed mode like in the PDE measurement. 
}
  \label{LEDspectra}
\end{figure*}

The setup of our PDE measurement is sketched in Fig.\  \ref{PDESetup}. An LED pulses 20\,ns-long flashes of light at 200\,Hz into a
UV-transparent liquid fiber that guides
the light into a hollow cylinder made out of spectralon.\footnote{The same integrating sphere that was also used in \cite{2006NIMPA.567..360O}.} The entry port and the two exit ports of the
integrating sphere are all oriented perpendicular to each other. Attached to each exit port is an aluminum cylinder with the inside of the cylinder covered with black felt. Each cylinder is
closed with a black plastic cap that has a hole in its center. A calibrated PiN diode is mounted to the cap with the larger hole ($\sim10\,$mm diameter), and the SiPM is mounted to the cap with the smaller hole ($\sim1\,$mm diameter). 

Each SiPM is held in place with an adapter that is custom designed and 3D-printed for each device. The adapter ensures that only the active area of the SiPM is illuminated by the light that
exits the integrating sphere through the end-cap of the aluminum cylinder. The diameter of the light beam is about 1\,mm. Four different LEDs fitted with narrow bandpass optical
filters are used in the PDE measurement. The spectra of the four LEDs after the filter are shown in Fig.\  \ref{LEDspectra}. The full width at half maximum (FWHM) of each spectrum is
$\sim10$\,nm.

The PiN diode used in this study is a Hamamatsu S3590-08. The noise of the PiN-diode is minimized by reverse biasing the diode at 70\,V thus decreasing the internal capacitance of the diode. The diode signal is first amplified with a Cremat
110 charge-sensitive preamplifier and then further amplified and shaped with an ORTEC Model 410 linear amplifier. The best signal-to-noise ratio is achieved with $2\,\mu$s differentiating and
integrating
shaping time constants. The noise performance of the PiN-diode signal chain is limited by the capacitance of the diode and the intrinsic noise of the preamplifier and is about 300
equivalent
noise charge (ENC). After amplification the signal is recorded with an Alazar ATS 9870 8\,bit, 1\,GS/s digitizer. 

The SiPM signal is amplified with a Mini-Circuits 500-NL amplifier and then shaped with a simple variable parallel RC circuit that differentiates the signal (C) and provides pole-zero
cancellation (R). After shaping, the typical full width of the SiPM signal is less than 10\,ns. The signal is further amplified with a LeCroy Model 612A amplifier before being digitized
with
the ATS 9870 digitizer. A switchable attenuator before the LeCroy amplifier is used to adjust the single photoelectron amplitude at the input of the digitizer to $\sim30$\,mV.

\begin{figure}[!tb]
  \centering
  \includegraphics[width=\columnwidth]{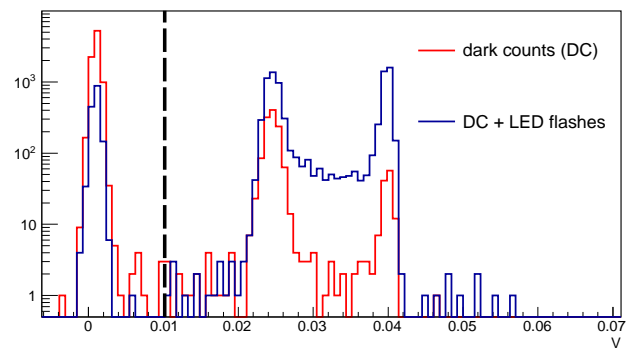}
\caption{Pulse-height distributions of Hamamatsu SiPM signals recorded in a PDE measurement. See text for details on the signal extraction. A total of 10,000 flashes contribute to each
distribution. The blue distribution is from signals recorded when the SiPM is flashed with the LED. The red distribution is from signals recorded when the LED is not flashing. Events to the
left side of the dashed vertical line can be identified as those in which the SiPM did not generate a signal.}
  \label{SiPMspectra}
\end{figure}

The LED signal is extracted from the recorded trace of the SiPM by sliding a window of three samples (3\,ns) through the trace starting before the LED signal is expected in the trace and
stopping 250\,ns later. At each position the sum of the three samples is calculated, and at the end of the scan, the maximum sum is filled into a histogram. To extract the dark count
rate, the procedure is repeated by starting 300\,ns before the LED signal and sliding the three-sample window for another 250\,ns through the trace stopping before the LED signal is expected
in the trace. The maximum of the sliding window is again filled into a histogram. Fig.\  \ref{SiPMspectra} shows the two resulting histograms for a typical measurement. Entries to the left
of the dashed vertical line correspond to events during which the SiPM did not generate a signal within the 250\,ns. The integral of these events are $N_0^{\mbox{\tiny DC}}$ (red
histogram) and $N_0$ (blue histogram), respectively. 

Note the good separation between the noise peak on the left and the first peak on the right side of the vertical line, which is necessary to keep the systematic uncertainties on the measured
mean number of detected photons low. 
In all measurements the number of events in the minimum, where the dashed vertical line is placed, is 1\% or less than the number of events in the maximum of the peak to the left.   In that
way the systematic uncertainty in the reconstructed mean number of photons is kept below 1\%.

The PiN diode signal is extracted by fitting a template pulse shape to the trace
and recording the amplitude of the fitted pulse. The template pulse shape is
obtained by averaging over 1000 pulses. The
average number of photons at the PiN-diode position is calculated from the
PiN-diode signals by taking the full LED spectrum and the wavelength-dependent quantum efficiency (QE) of the PiN diode into account.

\subsection{Calibration of the PDE setup}
\label{calibSection}
Before a PDE value can be calculated, the PiN diode, the integrating sphere, and the PiN diode signal chain need to be calibrated. The Hamamatsu S3590-08 PiN diode has been calibrated by
Hamamatsu, with a systematic uncertainty of 2-3\% between 250\,nm and 800\,nm and up to 5\% outside of that range \cite{PCHamamatsu}.

For the measurement of the splitting ratio of the integrating sphere, S3590-08 PiN diodes are placed at the end cap of each aluminum cylinder. An LED connected to a constant current source
then shines into the entrance port of the integrating sphere. After one hour the LED has stabilized such that its intensity does not vary by more than 0.1\% over the course of one
calibration measurement.

The currents of both PiN diodes are simultaneously recorded with two Keithley 6847 picoammeters. The photo current measured at the SiPM position (where the intensity is lowest) is at least
1000 times the PiN-diode dark current. In a series of measurements the PiN diodes are swapped. 

The splitting ratio is first calculated by using the currents that were measured with the same diode at the two exit ports. The ratio is then calculated a second time by using the
currents that were measured with the two diodes simultaneously. In the final calculation, the currents are corrected for the small differences in the quantum efficiencies of the two
PiN-diodes. All measurements of the splitting ratio agree within 2\%. The ratio was, furthermore, measured with all four LEDs used in the PDE measurements and found to vary within 1\%.

\begin{figure}[!tb]
  \centering
  \includegraphics[width=\columnwidth]{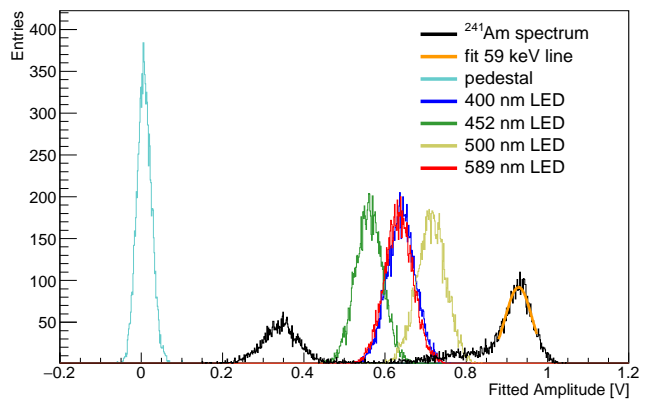}
\caption{Pulse-height distributions recorded with the calibrated PiN diode attached to the integrating sphere. Shown are distributions for all four LEDs, the $^{241}$Am source, and the
pedestal. The fit of the 59\,keV bin with a Gaussian function is also shown.}
  \label{PiNspectra}
\end{figure}

The PiN-diode signal chain is calibrated in photoelectrons by attaching a $^{241}$Am source to the diode and recording the signals of 59.54\,keV gamma rays. Using a Fano factor of 3.62
eV/eh-pair it can be shown that the gamma rays produce on average 16448 eh-pairs in the diode \cite{Gunnink}. A typical $^{241}$Am spectrum recorded with our setup is shown in Fig.\ 
\ref{PiNspectra} together with pulse-height distributions for each of the four LEDs. 

The linearity of the PiN-diode signal chain is better than 3\% down to signal amplitudes that are $\sim10$\% of an average 59\,keV signal. 

We estimate that the relative systematic uncertainty of our PDE measurements is 5\%.  The relative systematic uncertainty is dominated by systematic uncertainties of the PiN diode's QE
(3\%), uncertainties in the ratio of the
spectralon cylinder (1\%), and the signal extraction of the SiPM (1\%) and PiN diode (3\%).

\subsection{PDE measurements}\label{PDE}

\begin{figure}[!tb]
  \subfloat[FBK NUV-HD]{\includegraphics[width=\columnwidth]{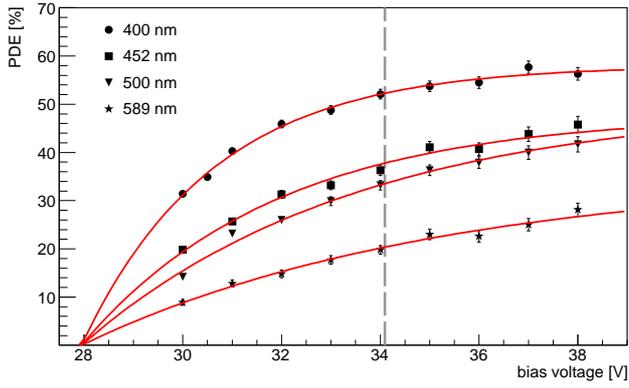}\label{FBKPDE}}\\
  \subfloat[Hamamatsu S13360-3050CS]{\includegraphics[width=\columnwidth]{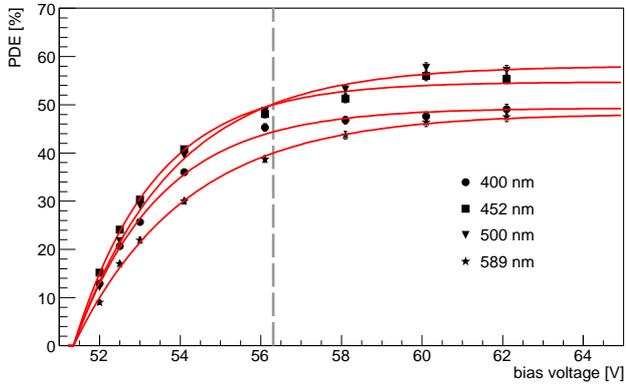}}\\
  \subfloat[SensL J-series 30035]{\includegraphics[width=\columnwidth]{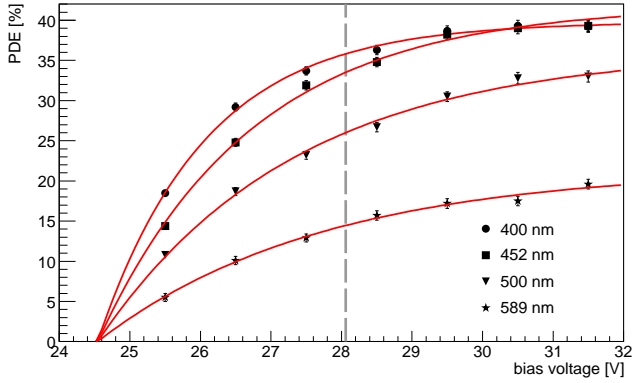}\label{SensLPDE}}
\caption{PDE measured at four different wavelengths as a function of overvoltage.\label{PDEvsBias}}
\end{figure}

The PDE of all three devices is shown as a function of bias for all four wavelengths in Fig.\  \ref{PDEvsBias}. Each of the bias-dependent PDE curves is well described by an exponential
function of the form
\begin{equation}\label{PDEFit}
PDE(U) = PDE_{\mbox{\footnotesize max}}\left[1-e^{-\left( U - U_{\mbox{\tiny BD}}\right)/a} \right]\,
\end{equation}
with fit probabilities that are in all but one case better than 60\%. The good agreement indicates that the chosen analytical function is an appropriate empirical model of the data. The
breakdown voltage $U_{\mbox{\tiny BD}}$ is determined from the best fit of the 400\,nm data and fixed in the fits of the data for the remaining wavelengths. The reasons for fixing the
breakdown voltage are twofold. Firstly, the uncertainty of the best fit breakdown voltage is smallest in the fits of the 400\,nm data, and secondly, the breakdown voltage does not depend
on photon wavelength. We note that the breakdown voltages obtained here are in agreement with the dedicated breakdown-voltage measurements presented later.

The dashed vertical lines in Fig.\  \ref{PDEvsBias} denote the bias at which each device reaches 90\% of the maximum PDE at 400\,nm as inferred from the fit of the data. For the remainder of
this paper we refer to this bias voltage as the operating point of an SiPM and mark it accordingly in all figures with a downward pointing arrow. Note that the bias where the PDE reaches
90\% of its maximum depends on wavelength as will be discussed next. 

The term in the square brackets in Equation \ref{PDEFit} has to be interpreted as the breakdown probability, because the breakdown probability is the only contribution to the PDE that depends on
bias,
so long as the active volume of a cell is fully depleted (which can be safely
assumed for the tested devices). After rewriting the exponent in units of relative overvoltage
\begin{equation}
 U_{\mbox{\tiny rel}} =  \frac{U - U_{\mbox{\tiny BD}}}{U_{\mbox{\tiny BD}}}\,,
\end{equation}
which can in fact also be interpreted as the relative electric field strength above the critical electric field strength, the breakdown probability becomes
\begin{equation}\label{breakdownPro}
P_{\mbox{\footnotesize BD}}(U_{\mbox{\tiny rel}}) = 1-e^{-U_{\mbox{\tiny rel}}/\alpha}\,.
\end{equation}

It is interesting to note that one parameter, $\alpha=a/U_{\mbox{\tiny BD}}$, is sufficient to properly describe the electric field/bias dependence of the breakdown probability. The parameter $\alpha$ depends, of
course, on the
geometry of the avalanche region, where in the avalanche region a photon is absorbed, on the impact ionization factors of electrons and holes, and other factors and is thus device and wavelength specific.
A small $\alpha$ value means that the
breakdown probability rises quickly with bias as opposed to a slow rise if $\alpha$ is large. We discuss the interpretation of $\alpha$ in more detail in the following.

\begin{figure}[b]
  \centering
  \includegraphics[width=\columnwidth]{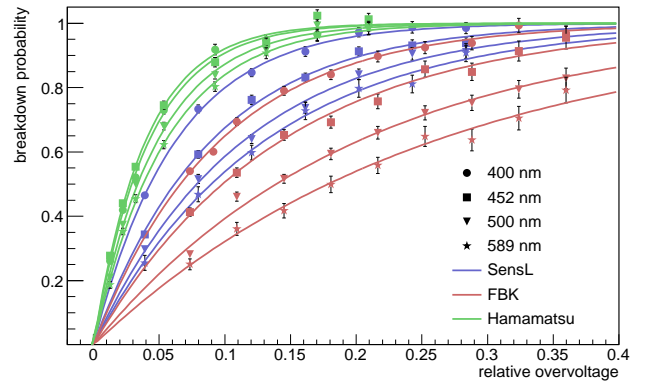}
\caption{Breakdown probability as a function of relative overvoltage above breakdown for all three SiPMs and for all four wavelengths. The corresponding $\alpha$ values are listed in Table
\ref{alphvals}.} 
  \label{BreakdownProb}
\end{figure}

Fig.\  \ref{BreakdownProb} shows the breakdown probability as a function of relative overvoltage / relative excess electric field for all three SiPMs and all four tested wavelengths. The
corresponding values for $\alpha$ are listed in Table \ref{alphvals}. All three devices have in common that $\alpha$ increases with increasing wavelength. 
This behaviour can be explained with the absorption length of photons, which increases with wavelength. For photons absorbed close to the surface of the SiPM (blue photons), it is the
photoelectron that drifts into the avalanche region in \emph{p}-on-\emph{n} devices. For photons absorbed below the avalanche region (redder photons), it is the hole that drifts upward into
the
avalanche region and initiates a breakdown. Because holes have always lower ionization factors than electrons, the breakdown probability for hole-dominated breakdowns is lower than for
electron-dominated ones. 

The ionization factors for electrons and holes grow rapidly with bias,
therefore, the breakdown probability also increases until saturation is reached.
Even though the ionization factor of holes increases faster with bias than the
one for electrons it never becomes larger than the ionization factor of
electrons. Thus the breakdown probability is always less for longer wavelengths
than for shorter wavelengths and saturation is reached at a higher bias.

The Hamamatsu SiPM has the lowest $\alpha$ of the three devices at all wavelengths, while the FBK device features the largest $\alpha$ values. These differences can be qualitatively
attributed to differences in the location of the avalanche region (how close it is to the surface), spatial extent of the avalanche region, the geometry of the avalanche region, and
variations of it when the bias is being changed.

It is evident that all three devices can be operated at a breakdown probability of 90\% or more---at least in the blue. This is a significant improvement compared to a few years ago
when most devices could only operate at a maximum overvoltage of 5\%-10\%, and, therefore, yielded much lower breakdown probabilities \cite{2006NIMPA.563..368D}.

\begin{table}
\caption{$\alpha$ Values of the Fit Results in Fig.\  \ref{BreakdownProb}.}
\centering
\begin{tabular}[!htb]{c|c|c}
Device&Wavelength&$\alpha$\\\hline\hline
FBK&400\,nm&0.095$\pm$0.001\\ 
   &452\,nm&0.142$\pm$0.003\\
   &500\,nm&0.200$\pm$0.004\\
   &589\,nm&0.258$\pm$0.007\\\hline
Hamamatsu&400\,nm&0.0420$\pm$0.0005\\
   &452\,nm&0.0395$\pm$0.0006\\
   &500\,nm&0.0485$\pm$0.0007\\
   &589\,nm&0.0546$\pm$0.0010\\\hline
SensL&400\,nm&0.062$\pm$0.001\\
   &452\,nm&0.089$\pm$0.002\\
   &500\,nm&0.113$\pm$0.002\\
   &589\,nm&0.129$\pm$0.004\\
\end{tabular}
\label{alphvals}
\end{table}

\begin{figure}[!tb]
  \centering
  \includegraphics[width=\columnwidth]{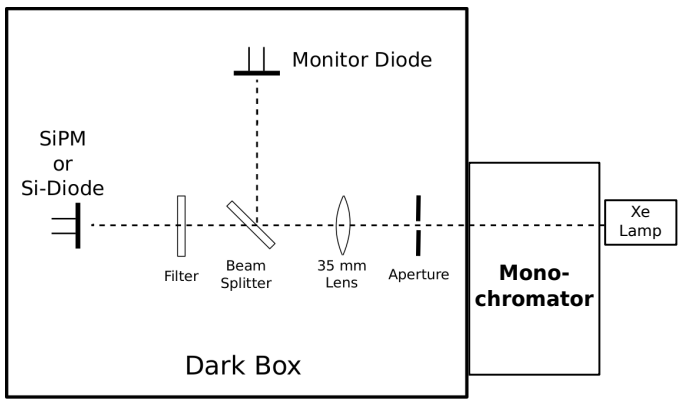}
\caption{Sketch of the spectral response setup.}
  \label{specres}
\end{figure}

\subsection{Concept of the spectral response measurement}
For the spectral response measurement, we use the setup that is sketched in Fig.\  \ref{specres}.
The SiPM is biased at the voltage that yields a 90\% breakdown probability at 400\,nm as defined in the previous section. 
The SiPM is measured first and then replaced with the reference detector instead of measuring both sensors simultaneously like in the PDE measurement. Doing the spectral response measurement
in this way eliminates optical elements that split the light between the two sensors and therefore would have to be calibrated. A main source of systematic uncertainties is thus
eliminated. 

Any variability of the light source is monitored and recorded with a permanently installed PiN diode. Further corrections that are applied in the data analysis are a) subtraction of dark
currents of all sensors and b) subtraction of stray light transmitted through the monochromator, which affects measurements mainly below 350\,nm.

The intensity of the light source is adjusted throughout a measurement by controlling the slits of the monochromator such that the SiPM current is within 50 to 75 times the dark
current of the SiPM. Keeping the current of the SiPM quasi-constant guarantees that the fraction of SiPM cells that are in recovery remains about the same, and thus the geometrical
efficiency of the
SiPM also remains constant. The current limits are such that only a small fraction of the cells of an SiPM ($<$1\%) are always in recovery and, therefore, saturation effects of the SiPM
are
avoided. The light spot at the position of the SiPM is larger than the sensor itself. Each spectral response measurement is cross-checked by increasing the current limits to be between 100
and 150 times the dark current and making sure that the residuals between the two measurements remain less than 2\%.

The spectral response measurement is a relative one and is converted into an absolute PDE measurement by fitting it to the PDE measurements presented earlier. Corrections for optical
crosstalk
and afterpulsing, therefore, do not have to be applied to the spectral response measurements.

\subsection{Setup of the spectral response measurement}

The light source in the spectral response measurement is a 300\,W UV-enhanced Xenon arc lamp (PE300BUV from Cermax). The light of the lamp is air-coupled into a Czerny-Turner single-grating
monochromator Digikröm DK\,240 1/4$\lambda$ from Spectral Products. The grating of the monochromator that is used for all measurements has 1200 grooves per millimeter and a 300\,nm blaze
wavelength. The output of the monochromator is coupled into a dark box where the light beam is further conditioned before it illuminates the monitoring diode and the SiPM or reference
sensor.

Inside the dark box the light first passes an adjustable aperture followed by a lens with a focal length of 35\,mm. The beam is then split by a polka dot beamsplitter. The reflected part of
the beam illuminates the monitoring diode---an unbiased Hamamatsu S3590-08 PiN diode. 
The size of the beam spot matches the size of the monitoring diode.

The transmitted part of the beam passes through an optical long-pass filter that is mounted onto a filter wheel, followed by an optional broadband polarizer (UBB01A from Moxtek) before the
beam
illuminates either the SiPM or the reference sensor. The beam spot is larger
than the size of the reference sensor or the SiPM.
  The reference sensor is a UV-enhanced, Si-diode from Hamamatsu (type S1227-1010BQ, calibrated by Hamamatsu).  All optical elements are UV transparent down to 200\,nm. 

A total of three long-pass filters with cut-off wavelengths at 280\,nm, 400\,nm, and 750\,nm are mounted into a computer-controlled filter wheel. The 280\,nm filter is used to quantify stray
light with wavelengths above the cut-off wavelength that gets transmitted through the monochromator and affects measurements below 270\,nm. The 400\,nm filter is used to quantify the
stray-light component that affects measurements between 270\,nm and 350\,nm. The 400\,nm filter is also used to suppress higher-order diffraction above 430\,nm. The 700\,nm filter suppresses
higher-order diffraction above 770\,nm.

The current of the monitoring diode is recorded with a Keithley 6845 picoammeter, and the currents of the reference sensor and the SiPM are measured with a Keithley 6847 picoammeter. The
readings of both instruments are transfered via serial link to a computer, which also controls the monochromator and the filter wheel.

For the spectral response measurement, the SiPM is fixed on a rotary mount that allows making spectral response measurements as a function of the angle of incidence between 0 degrees
(normal
incidence) and 90 degrees. The SiPM is biased with the internal voltage source of the Keithley 6847 picoammeter. 

In the measurement the monochromator output is changed between 200\,nm and
1000\,nm and for each wavelength the exit and entrance slits of the
monochromator are adjusted 
to keep the SiPM current within the previously discussed limit of 50-75 times the SiPM's dark current. The long-pass filters are inserted at the above mentioned wavelengths. The SiPM is then swapped out with the calibrated Si-diode, and the photocurrent of the diode is recorded at the same wavelengths and with the same monochromator
slit settings used in the SiPM measurement.

The spectral response $S$ at a given wavelength is calculated as
\begin{equation}
   S = \frac{I_{\mbox{\tiny SiPM}}}{I_{\mbox{\tiny Si-Diode}}}\cdot 
       \frac{I_{\mbox{\tiny Mon. Si-Diode}}}{I_{\mbox{\tiny Mon. SiPM}}}\cdot 
       QE_{\mbox{\tiny Si-Diode}}\,,
\end{equation}
where $I_{\mbox{\tiny SiPM}}$ and $I_{\mbox{\tiny Si-Diode}}$ are the dark and the stray-light corrected currents of the SiPM and the calibrated Si-diode, respectively. The factor in
the
middle is the ratio of the dark-current-subtracted currents of the monitoring diode that corrects for fluctuations of the Xe lamp. The last factor $QE_{\mbox{\tiny Si-Diode}}$ is the quantum
efficiency of the reference sensor.

The systematic uncertainties between 300\,nm and 800\,nm are dominated by uncertainties in the wavelength-dependent response of the calibrated Si-diode ($\sim$3\%) and variations in the SiPM
photocurrent that cause the fraction of recovering SiPM cells to vary accordingly ($\sim$1\%). Below 300\,nm the systematic uncertainties are dominated by residuals in the stray-light
correction when the PDE of the SiPM drops below 10\%. They reach 100\% when the PDE of the SiPM drops below
a couple of percent. Above 800\,nm the uncertainties are dominated by the uncertainty in the QE of the reference sensor, which is $\sim4\%$.

\begin{figure}[!tb]
  \centering
  \includegraphics[width=\columnwidth]{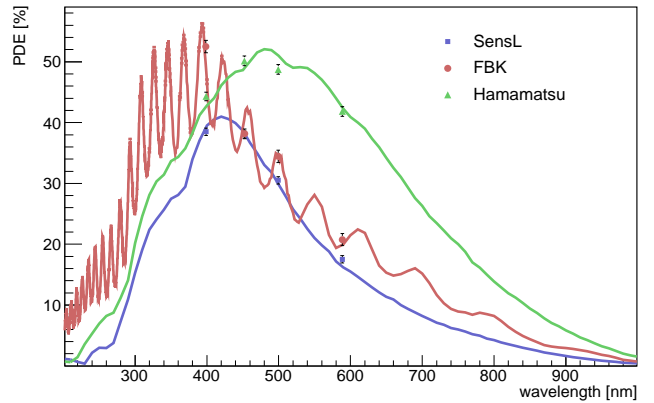}
\caption{PDE vs.\ wavelength for all three devices between 200\,nm and 1000\,nm. The bias voltage for each device results in a 90\% breakdown probability at 400\,nm, the operating point of
each SiPM.} 
  \label{PDEvsWL}
\end{figure}

\subsection{Wavelength dependent PDE}

The spectral response measurement is a relative one and converted into an absolute PDE measurement by fitting it to the previously discussed PDE measurements at four wavelengths. The fit is
done by invoking a scaling factor that minimizes $\chi^2$ between the four PDE points and the spectral response measurements. In the fit it is taken into account that the spectral response
of the SiPM varies across the spectra of the LEDs that have been used in the PDE measurements. In order to find the correct wavelength that corresponds to the measured PDE, an LED spectrum
is weighted with the spectral response of the SiPM, and the mean wavelength of the weighted spectrum is used as the wavelength of the PDE measurement. The correction, however, is small, and
the shift with respect to the mean LED wavelength is $<1\,$nm. Afterpulsing and optical crosstalk do not affect the outcome of the scaling because both result in a wavelength-independent
factor that gets marginalized in the fit.

The spectral response measurements scaled to absolute PDE are shown in Fig.\  \ref{PDEvsWL}. Also shown are the four PDE measurements for each device to which the spectral response
measurements have been scaled.

The FBK device has the highest peak PDE of the three tested SiPMs with 56\% at 395\,nm, even though it has the smallest pitch between cells. The oscillations in the PDE are due to
interference caused by the thin passivation layer and the lack of a coating on top of the device like in the other two devices. In a previous study we tested an NUV-HD device with
coating that shows a comparable PDE down to 300\,nm. Below 300\,nm the FBK device presented here has a better efficiency because it is not coated with silicon resin. The full width at
half
maximum (FWHM) of the FBK PDE extends from 280\,nm to 560\,nm. The Hamamatsu device has a peak PDE of 52\% at 455\,nm and a FWHM of the PDE response that extends from 310\,nm to 700\,nm,
which is significantly more red sensitive than the FBK SiPM. The SensL device
has a peak PDE of 41\% at 420\,nm and a FWHM of the PDE response from 310\,nm to 560\,nm, which is
similar to the response of the FBK SiPM. 

Compared to similar SiPMs from only a few years ago \cite{2006NIMPA.563..368D}, all three devices are testaments to the major improvements that have been made in increasing the PDE and shifting the response of
SiPMs to shorter
wavelengths. 

\begin{figure}[!tb]
  \subfloat[FBK NUV-HD]{\includegraphics[width=\columnwidth]{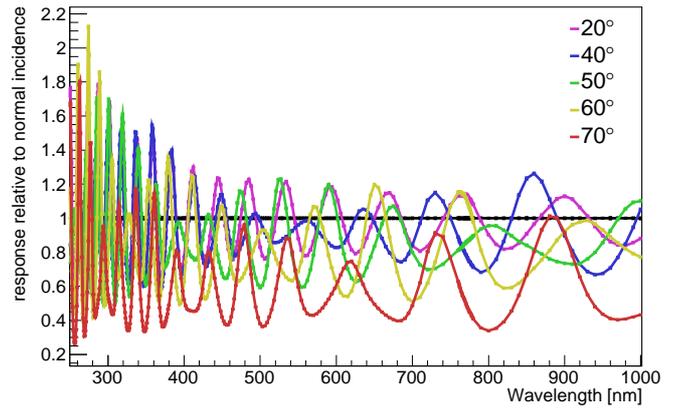}}\\
  \subfloat[Hamamatsu S13360-3050CS]{\includegraphics[width=\columnwidth]{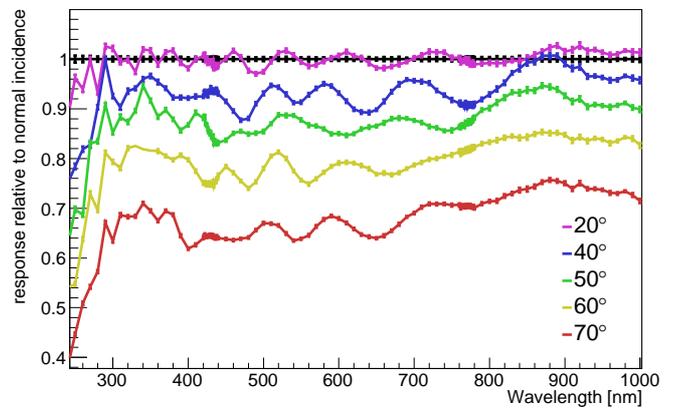}}\\
  \subfloat[SensL J-series 30035]{\includegraphics[width=\columnwidth]{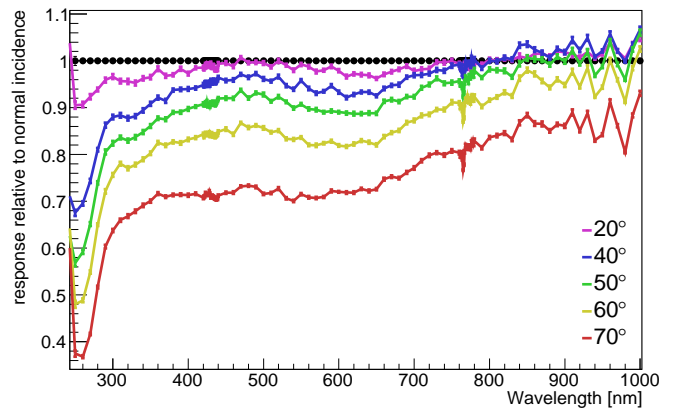}}
\caption{Response as a function of angle of incidence relative to normal incidence with light polarized perpendicular to the plane of incidence.\label{AngleOfIncidencePer}}
\end{figure}

\begin{figure}[!tb]
  \subfloat[FBK NUV-HD]{\includegraphics[width=\columnwidth]{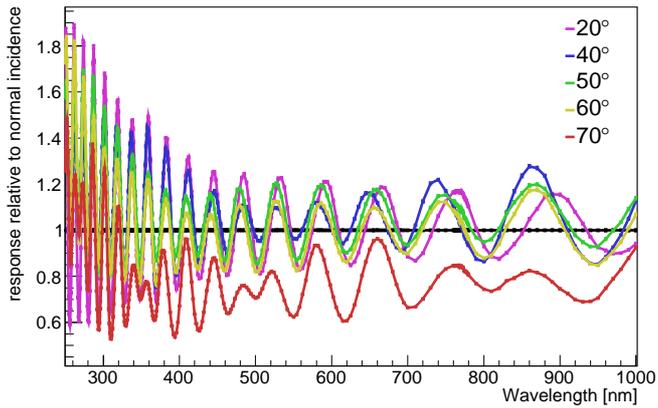}}\\
  \subfloat[Hamamatsu S13360-3050CS]{\includegraphics[width=\columnwidth]{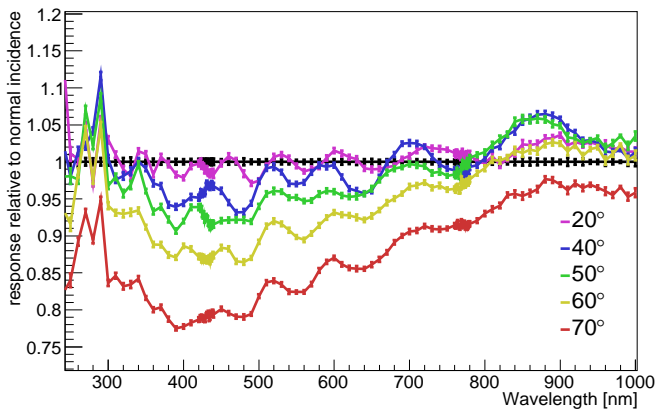}}\\
  \subfloat[SensL J-series 30035]{\includegraphics[width=\columnwidth]{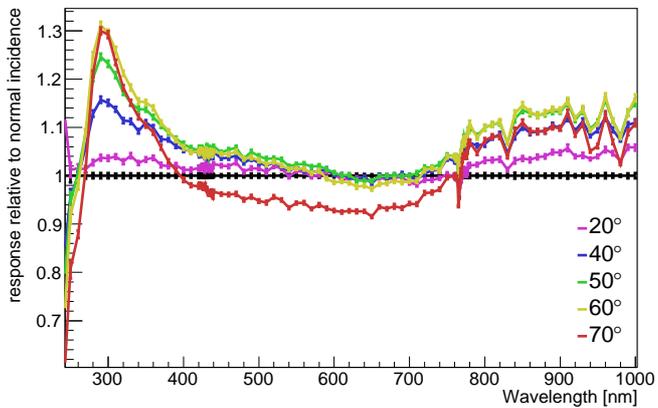}}
\caption{Response as a function of angle of incidence relative to normal incidence with light polarized parallel to the plane of incidence.\label{AngleOfIncidencePar}}
\end{figure}

\subsection{Dependence of SiPM response on angle of incidence}

The dependence of the PDE on the angle of incidence was tested for light polarized in the plane of incidence (parallel polarization) and perpendicular to the plane of incidence for angles
incidence angles of $20^{\circ}$, $40^{\circ}$, $50^{\circ}$, $60^{\circ}$, and $70^{\circ}$. For this measurement a broadband polarizer UBB01A from Moxtek was inserted after the beam
splitter. 
Fig.\  \ref{AngleOfIncidencePer} shows the response of the three SiPMs relative to normal incidence for polarization perpendicular to the plane of incidence and in Fig.\
\ref{AngleOfIncidencePar} for light polarized parallel to the plane of incidence. The measurements are corrected for the change in the projected area of the light beam onto the SiPM with
different angle of incidence. We estimate a maximum uncertainty on the angle of incidence of $2^{\circ}$, which translates into a maximum systematic uncertainty of 10\% on the measurements
done at $70^{\circ}$ and less at smaller angles.

The response to different angles of incidence depends to a large fraction on the coating of the chip and also how the chip is packaged. In order to reduce effects from stray light that
reflects off the chip carrier into the edges of the chip or light that directly enters through the edges of the chip under larger angles, the boundaries of the Hamamatsu and the SensL SiPM
were covered with thin copper tape. Unfortunately, the FBK SiPM could not be taped because the chip is not protected, thus edge effects are included in the measurement.

The response of all devices is relatively constant up to angles of $60^{\circ}$, when the response is still about 80\% and better than 90\% for perpendicular and parallel polarized light,
respectively. At larger angles the sensitivity starts to quickly drop. Note that there is a steep increase in sensitivity of the SensL device to parallel polarized light between 300\,nm and
400\,nm for larger angles of incidence. 

\begin{figure}[!tb]
  \centering
  \includegraphics[width=\columnwidth]{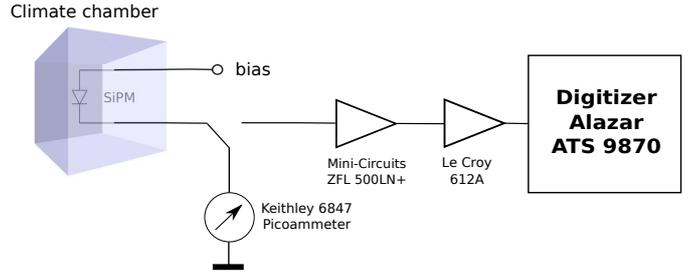}
\caption{Sketch of the basic measurement setup} 
  \label{SketchBasicMeas}
\end{figure}

\section{IV curves}

For the measurement of the electrical characteristics, the SiPMs are placed in a thermal chamber, and their performance is measured between -40$^{\circ}$C and 40$^{\circ}$C in steps of
20$^{\circ}$C. Fig.\  \ref{SketchBasicMeas} shows a sketch of the setup.

In this section the \emph{IV}-curve measurements are discussed. For each
measurement, an SiPM is connected to a Keithley 6847 picoammeter that biases the SiPM and records the
current. The measurements
are done in DC mode as opposed to a pulsed mode, which is acceptable given the small amount of power dissipated by the SiPM ($<20$\,mW when biased in the forward direction and $<1$\,nW when
biased in reverse). From the \emph{IV}-curves the average value of the quench resistor and the breakdown voltage are derived.

\subsection{Quench resistor values}

\begin{figure}[!tb]
  \centering
  \subfloat[FBK NUV-HD]{\includegraphics[width=\columnwidth]{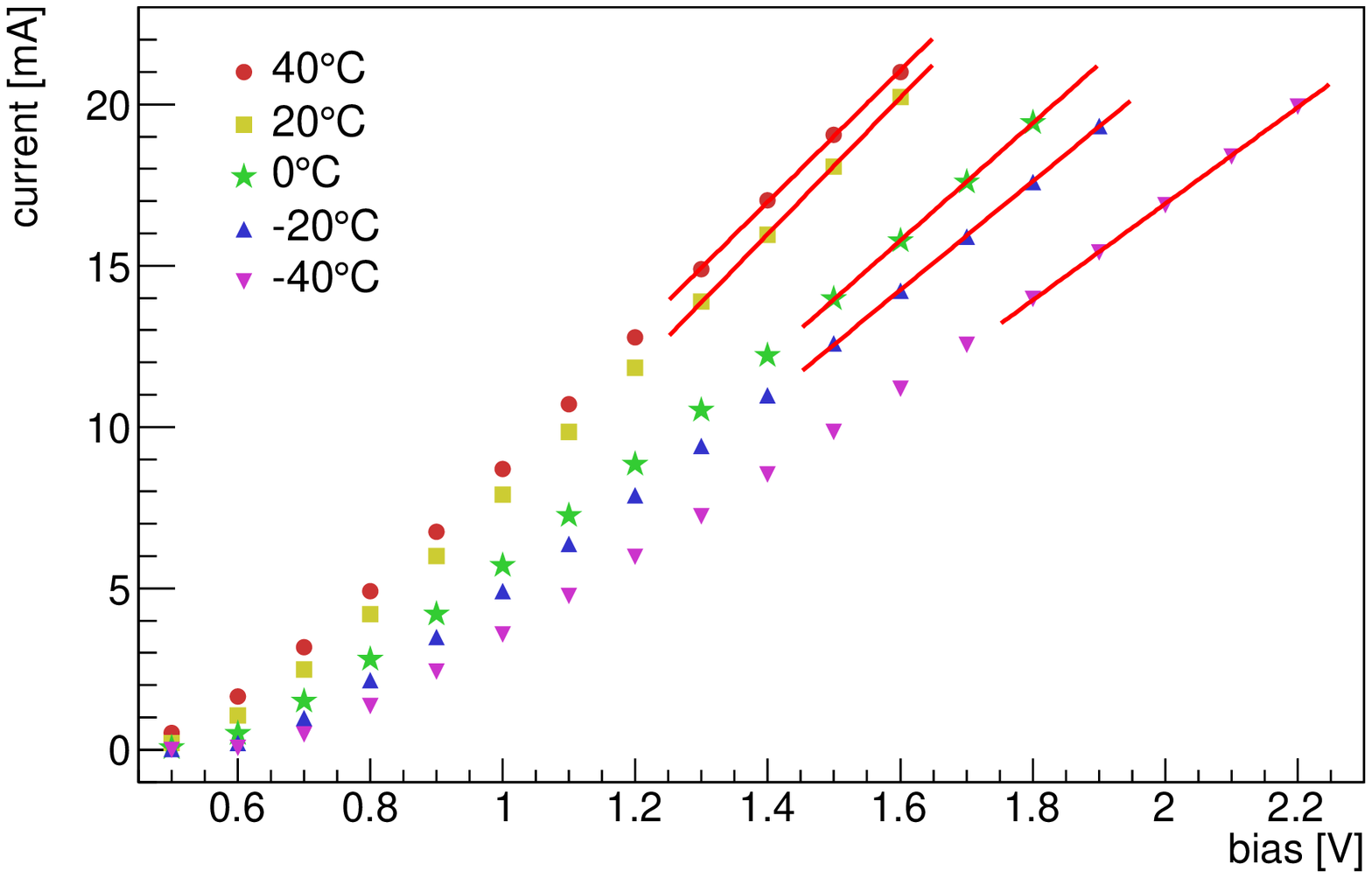}}\\
  \subfloat[Hamamatsu S13360-3050CS]{\includegraphics[width=\columnwidth]{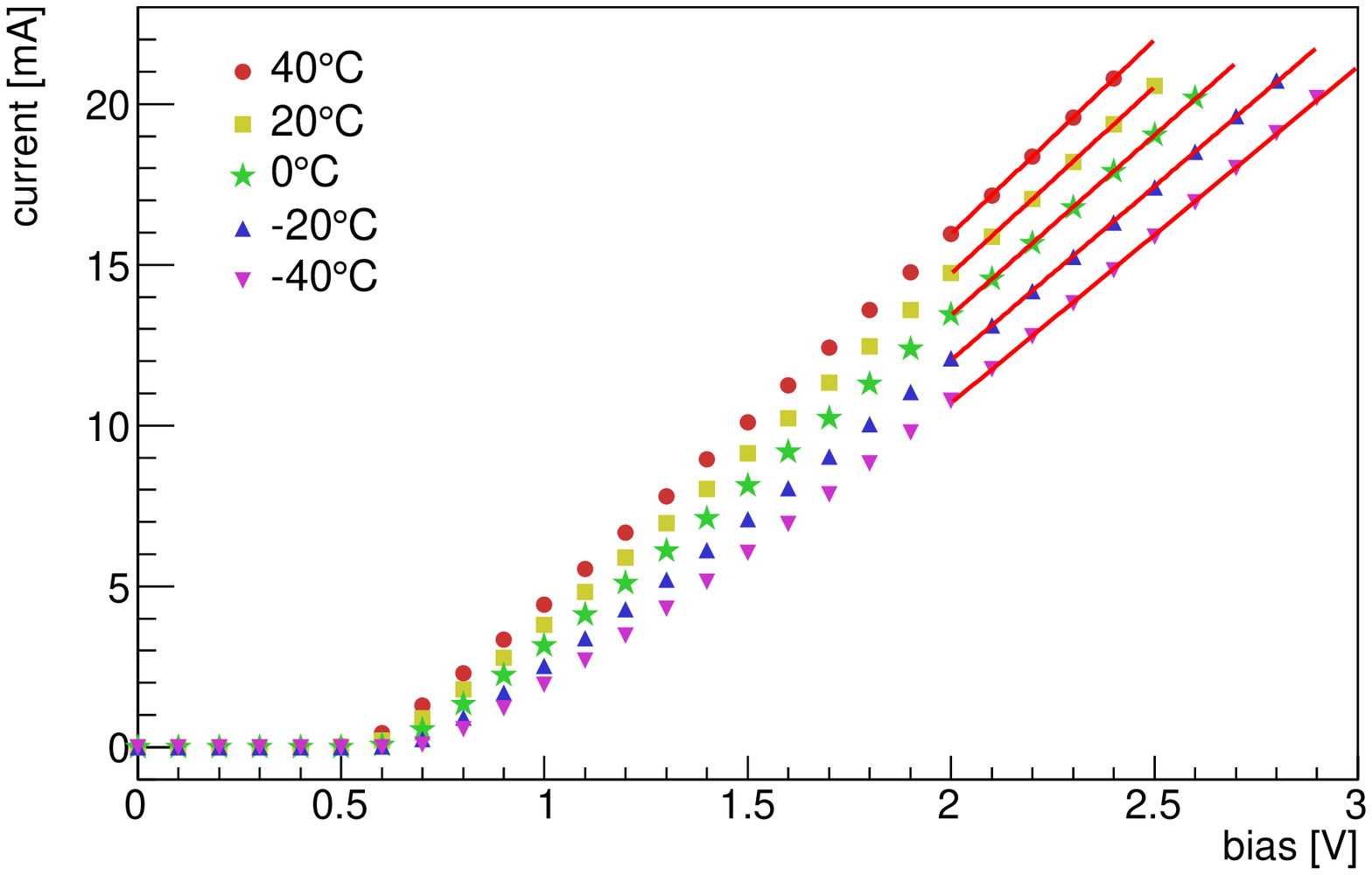}}\\
  \subfloat[SensL J-series 30035]{\includegraphics[width=\columnwidth]{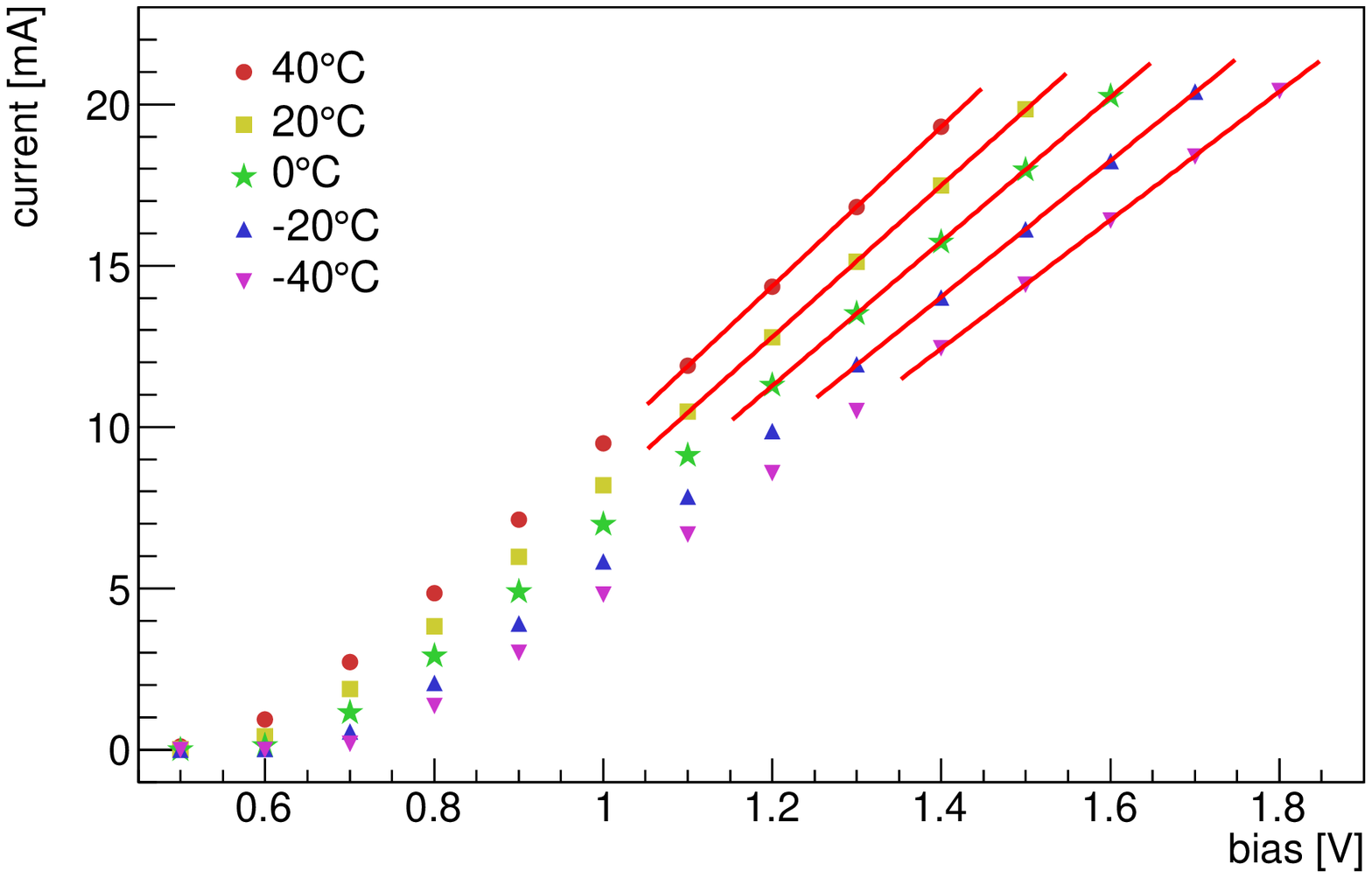}}
\caption{\emph{IV} curves of all SiPMs biased in the forward direction at five different temperatures. The solid lines are fits with linear functions, which are used to derive the average
quench
resistor value. The measured quench resistor values are shown in Fig.\  \ref{QRs}.  See text for further details.} 
  \label{IVcurvesForward}
\end{figure}

The quench resistor values are derived from the linear part of the forward biased \emph{IV} curves (see Fig.\  \ref{IVcurvesForward}), \emph{i.e.}\, in the regime where the resistance of the
\emph{pn}-junction of a cell becomes negligible, and the total resistance is dominated by that of the quench resistor. 

\begin{figure}[!tb]
  \centering
  \subfloat[FBK NUV-HD]{\includegraphics[width=\columnwidth]{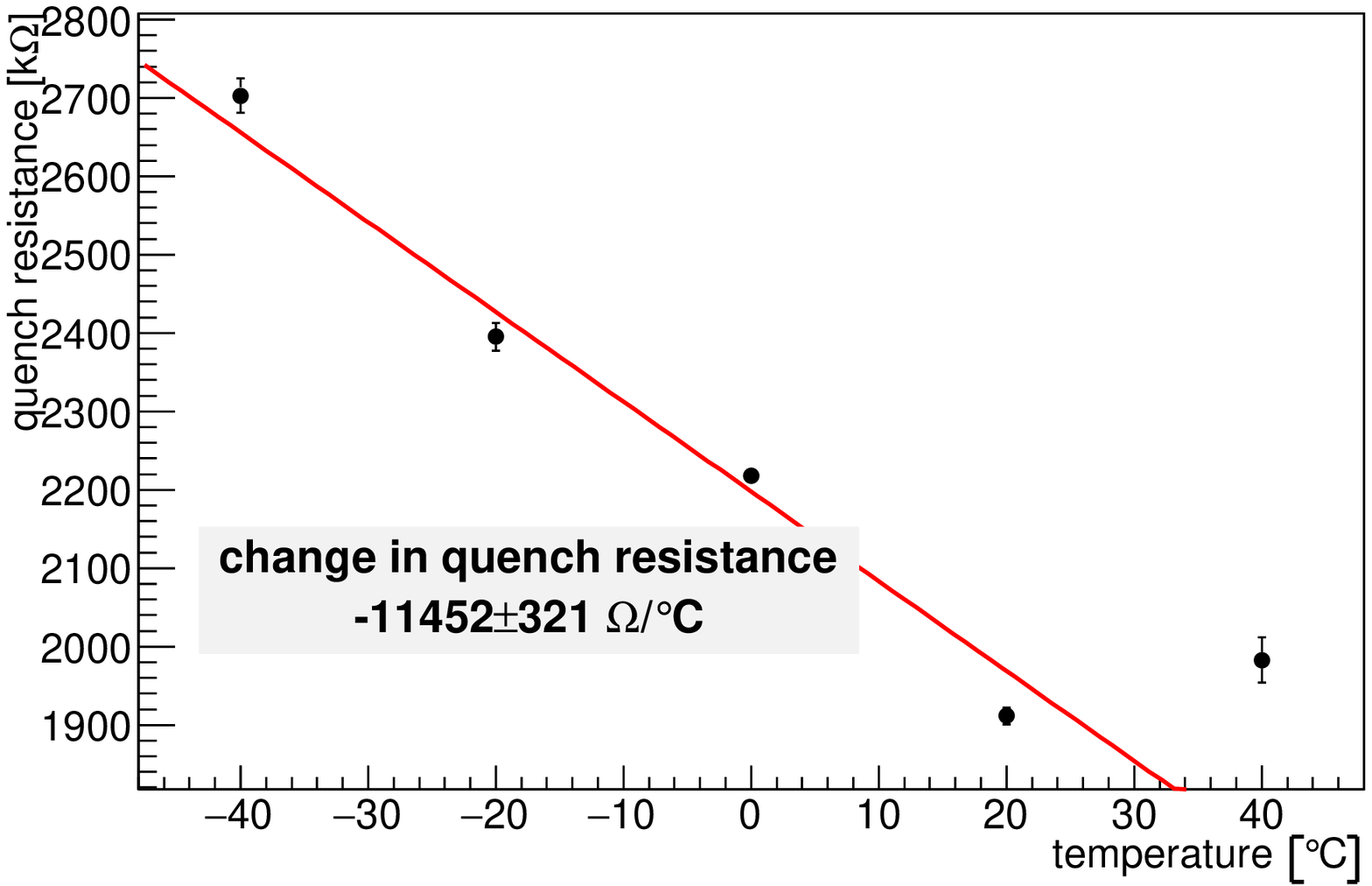}}\\
  \subfloat[Hamamatsu S13360-3050CS]{\includegraphics[width=\columnwidth]{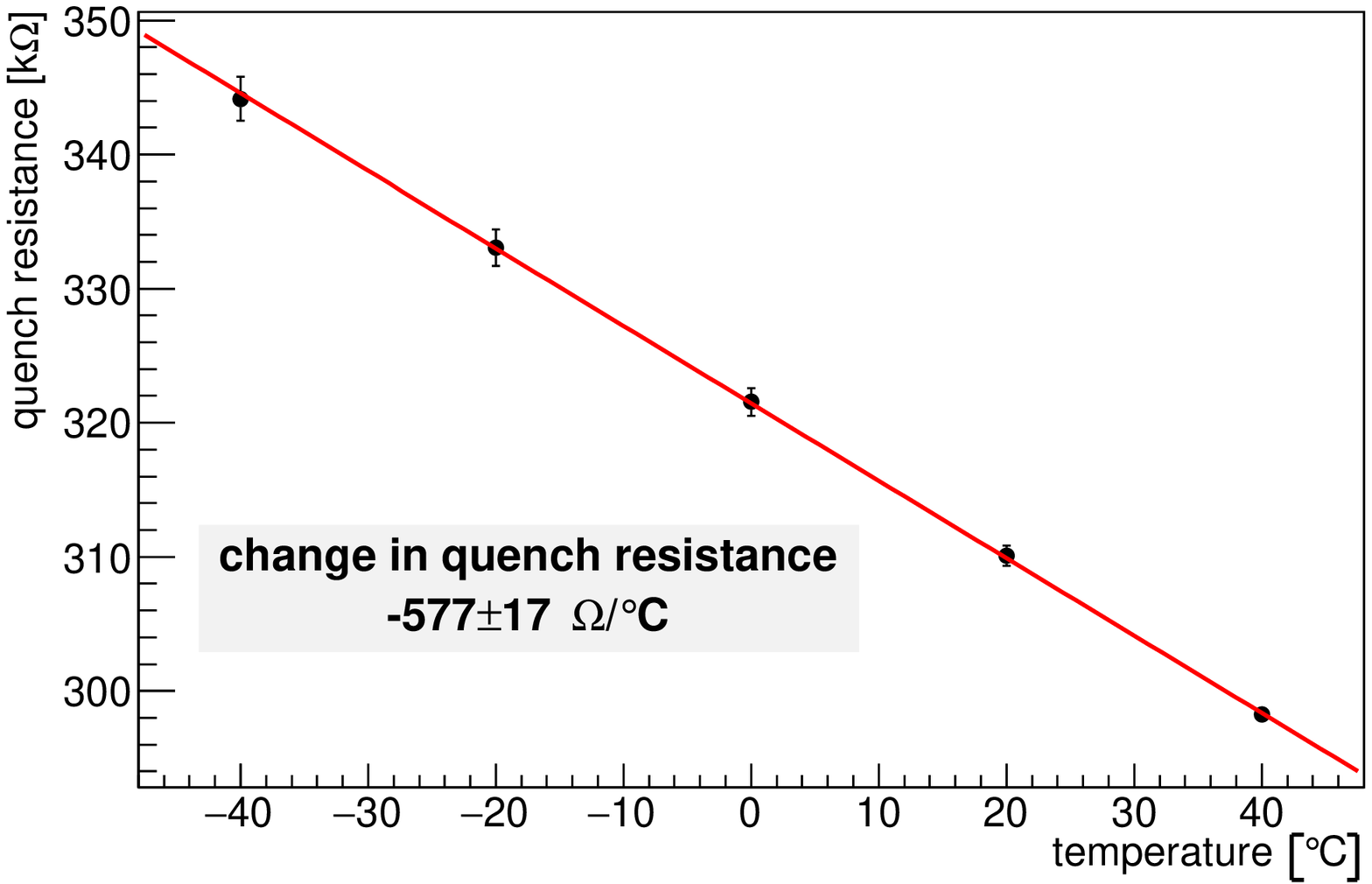}}\\
  \subfloat[SensL J-series 30035]{\includegraphics[width=\columnwidth]{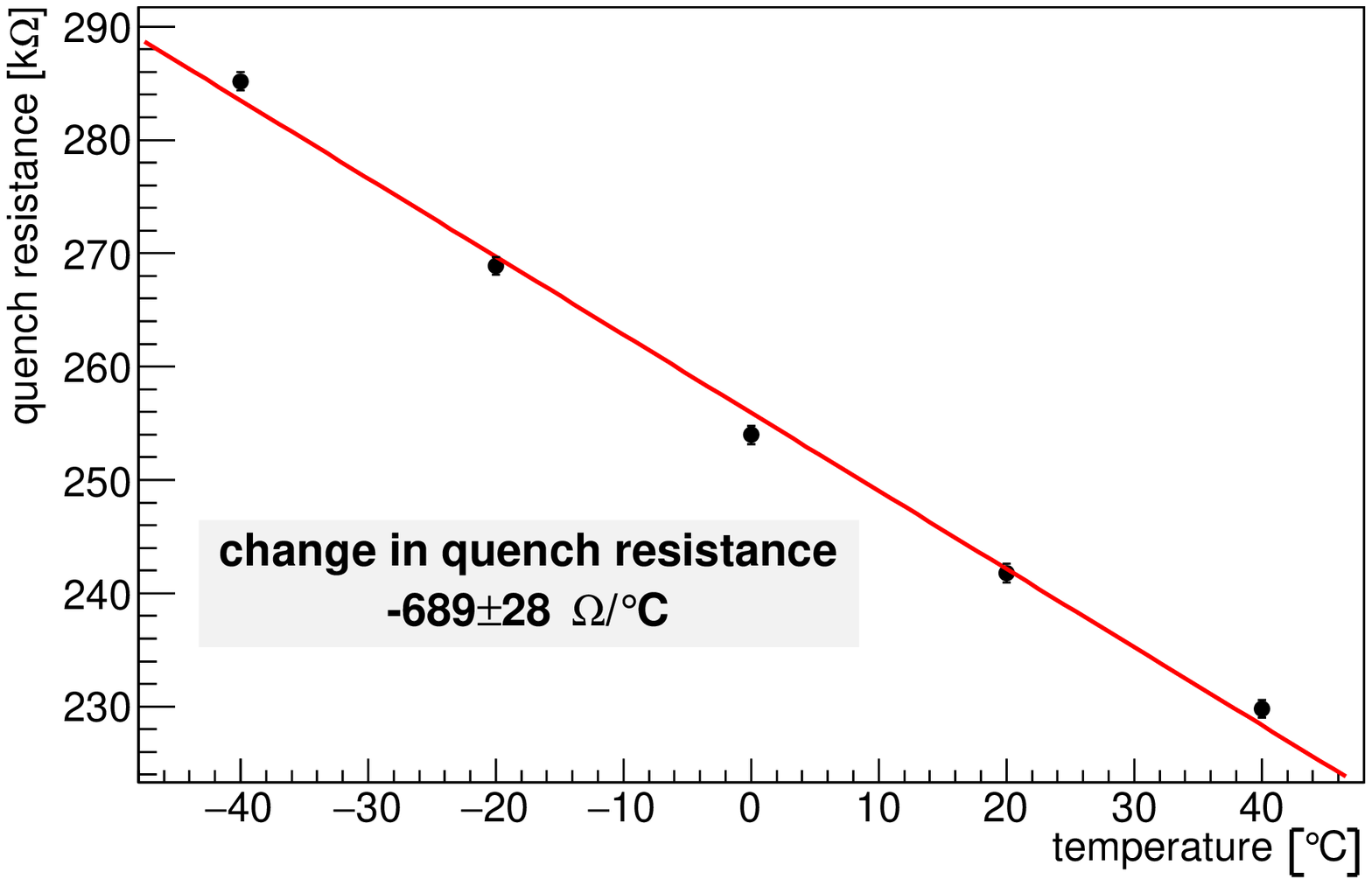}}
\caption{Average quench-resistor values for all three SiPMs at five different temperatures. The change in resistance with temperature shown in each figure is determined from a fit of the
data points with a linear function.} 
  \label{QRs}
\end{figure}

The inverse of the slope of the \emph{IV} curve yields the resistance of all quench resistors of the SiPM connected in parallel. Multiplying the total parallel resistance with the number of cells
of an SiPM thus gives the average value of a quench resistor, which is shown in Fig.\  \ref{QRs} as a function of temperature for all three SiPMs.

The figure also gives the temperature coefficients of the quench resistors, which are determined by fitting a linear function to the data points, which is a good approximation for the
Hamamatsu and SensL data. For the FBK SiPM, the quench resistor values fluctuate significantly. In particular the value at 40$^{\circ}$C is higher than one would expect by extrapolating the
quench resistor values from lower temperatures. We can not exclude that a contamination of the uncoated device during handling or residual humidity is responsible for these effects.  

The quench resistors of the Hamamatsu device have the smallest relative dependence on temperature with $2\cdot10^{-3}$, followed by $3\cdot10^{-3}$ for the SensL device, and $5\cdot10^{-3}$
for the FBK device. The temperature coefficient and the absolute value of the quench resistor determine the maximum temperature and bias at which a device can be operated before a breakdown
cannot be reliably quenched anymore. It, furthermore, determines how the recovery time of a cell changes with temperature. The temperature coefficients of all three SiPMs, however, are too
small to have any practical impact on the maximum operating temperature or cell recovery times.

\subsection{Breakdown voltages}

The second characteristic derived from the \emph{IV}-curves is the breakdown voltage. We took a close look at three different proposed methods
\cite{ChenXu,Musienko,2011ITNS...58.1233S} to extract the breakdown voltage, and we compare them with the classical method that uses gain vs.\ bias measurements. 
Based on our findings we propose yet another method that is based on
\cite{ChenXu,Musienko} and yields breakdown voltages within $\pm0.2\%$ of the
breakdown voltage derived from gain bias measurements.

It has been noted, based on empirical evidence, that the \emph{IV} curve of single SiPM cells (also called SPADs) can be described by a parabola above breakdown \cite{2007ITNS...54..236P}.
Here we give a physical explanation why a parabola is in fact expected for the \emph{IV} curve just above breakdown.

Biased just above breakdown, the current is proportional to the product of gain $G = C\cdot\Delta U = C\cdot U_{\mbox{\footnotesize BD}}\cdot U_{\mbox{\footnotesize rel}}$  and breakdown probability $1-\exp\left( -U_{\mbox{\footnotesize rel}}/\alpha\right)$, where $C$ is the effective capacitance of one SiPM cell\footnote{The cell capacitance is determined from gain vs.\ bias measurements and is discussed later.} and $\Delta U = U - U_{\mbox{\footnotesize BD}}$. 
The proportionality constant is the sum of the dark current $I_{\mbox{\footnotesize DC}}$ and the current due to external light sources $I_{\mbox{\footnotesize ext}}$ multiplied by the
optical crosstalk probability $P_{\mbox{\footnotesize OC}}$ and afterpulsing probability $P_{\mbox{\footnotesize AP}}$. The total current above breakdown is then
\begin{eqnarray}
I(U_{\mbox{\footnotesize rel}}) &= &\left[I_{\mbox{\footnotesize DC}}(U_{\mbox{\footnotesize rel}}) + I_{\mbox{\footnotesize ext}}\right]\nonumber\\
&& \cdot\left[1+P_{\mbox{\footnotesize OC}}(U_{\mbox{\footnotesize rel}})+P_{\mbox{\footnotesize AP}}(U_{\mbox{\footnotesize rel}})\right]\nonumber\\
       && \cdot C\cdot U_{\mbox{\footnotesize BD}} \cdot U_{\mbox{\footnotesize rel}}\cdot\left[1-e^{\left( -U_{\mbox{\footnotesize rel}}/\alpha\right)}\right]\label{IVmodel}.
\end{eqnarray}

The dark current changes much less with bias than the breakdown probability, and the gain and can thus be assumed constant if only a small range around the breakdown voltage is considered.
The impact of a varying dark current is further suppressed by illuminating the SiPM with an external light source that produces a current that is ten times or more than the SiPM dark current.\footnote{An
external light source that produces a current 100 times the dark current will not affect the response of the SiPM (see spectral response measurement section).} In fact, for this method to
also work
at low temperatures when the dark current becomes too low to provide a large enough primary signal, an external light source is needed.

Optical crosstalk and afterpulsing are only a few percent around the breakdown voltage and can, therefore, be neglected. With these simplifications the total current becomes
\begin{eqnarray}\label{Eqn7}
I(U_{\mbox{\footnotesize rel}})&\approx&\left[I_{\mbox{\footnotesize DC}} + I_{\mbox{\footnotesize ext}}\right]\cdot C\cdot U_{\mbox{\footnotesize BD}}\nonumber\\
    &&  \cdot U_{\mbox{\footnotesize rel}}\cdot\left[1-e^{\left( -U_{\mbox{\footnotesize rel}}/\alpha\right)}\right]\,.
\end{eqnarray}
Doing a series expansion of the exponential function to second order in $U_{\mbox{\footnotesize rel}}/\alpha$ gives
\begin{eqnarray}\label{Eqn8}
I(U_{\mbox{\footnotesize rel}})&\approx&\left[I_{\mbox{\footnotesize DC}} + I_{\mbox{\footnotesize ext}}\right] \cdot C \cdot U_{\mbox{\footnotesize BD}}\nonumber\\
 &&\cdot \left[ U_{\mbox{\footnotesize rel}}^2/\alpha+ U_{\mbox{\footnotesize rel}}^3/2\alpha^2 + \ldots\right]\,.
\end{eqnarray}
Thus in leading order the current above breakdown is indeed proportional to $\Delta U^2$ as long as $U_{\mbox{\footnotesize rel}}/\alpha<1$, which is the case for overvoltages that are less than 5\%-10\% (see Table \ref{alphvals}). 

To obtain the breakdown voltage from an \emph{IV} curve, \cite{Musienko} proposes using the voltage where $($d$I/$d$U)/I$ is maximal, whereas \cite{ChenXu} proposes using the maximum of $
\mbox{d}\ln\left(I(U)\right)/\mbox{d}U$. Both methods are equivalent because if applied to Eqn.\ \ref{Eqn7} both yield
\begin{equation}\label{Eqn10}
\frac{\mbox{d}I/\mbox{d}U}{I}=\frac{\mbox{d}\ln\left(I(U)\right)}{\mbox{d}U}=\frac{2 + f(y)}{U-U_{\mbox{\footnotesize BD}}}\,.
\end{equation}
The function $f(y)=(y+1-\exp(y))/(\exp(y)-1)$, with $y=U_{\mbox{\footnotesize rel}}/\alpha$, is about -0.2 for values of $y$ that are typical for the tested devices. 

\begin{figure}[!tb]
  \centering
  \subfloat[FBK NUV-HD]{\includegraphics[width=\columnwidth]{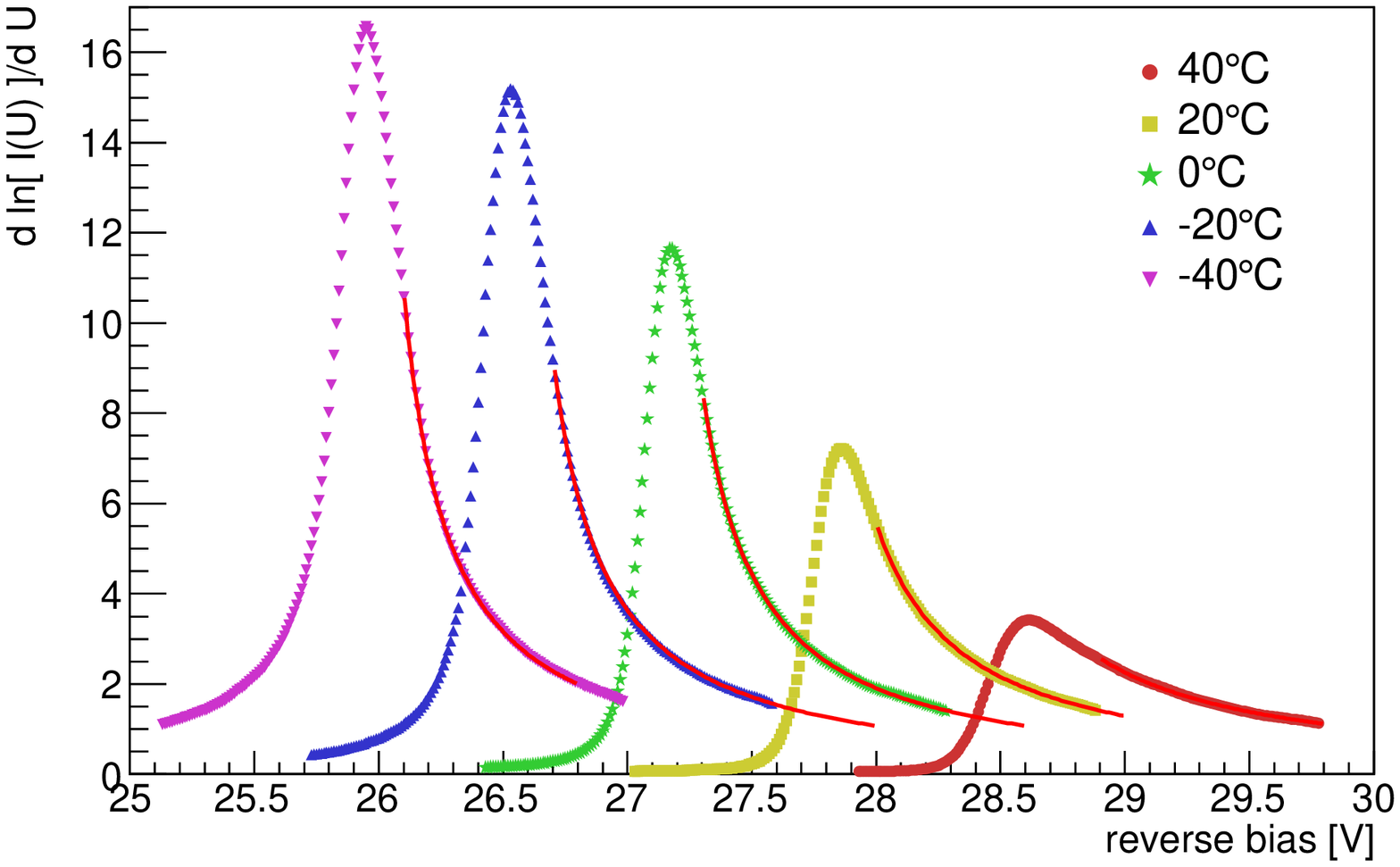}}\\
  \subfloat[Hamamatsu S13360-3050CS]{\includegraphics[width=\columnwidth]{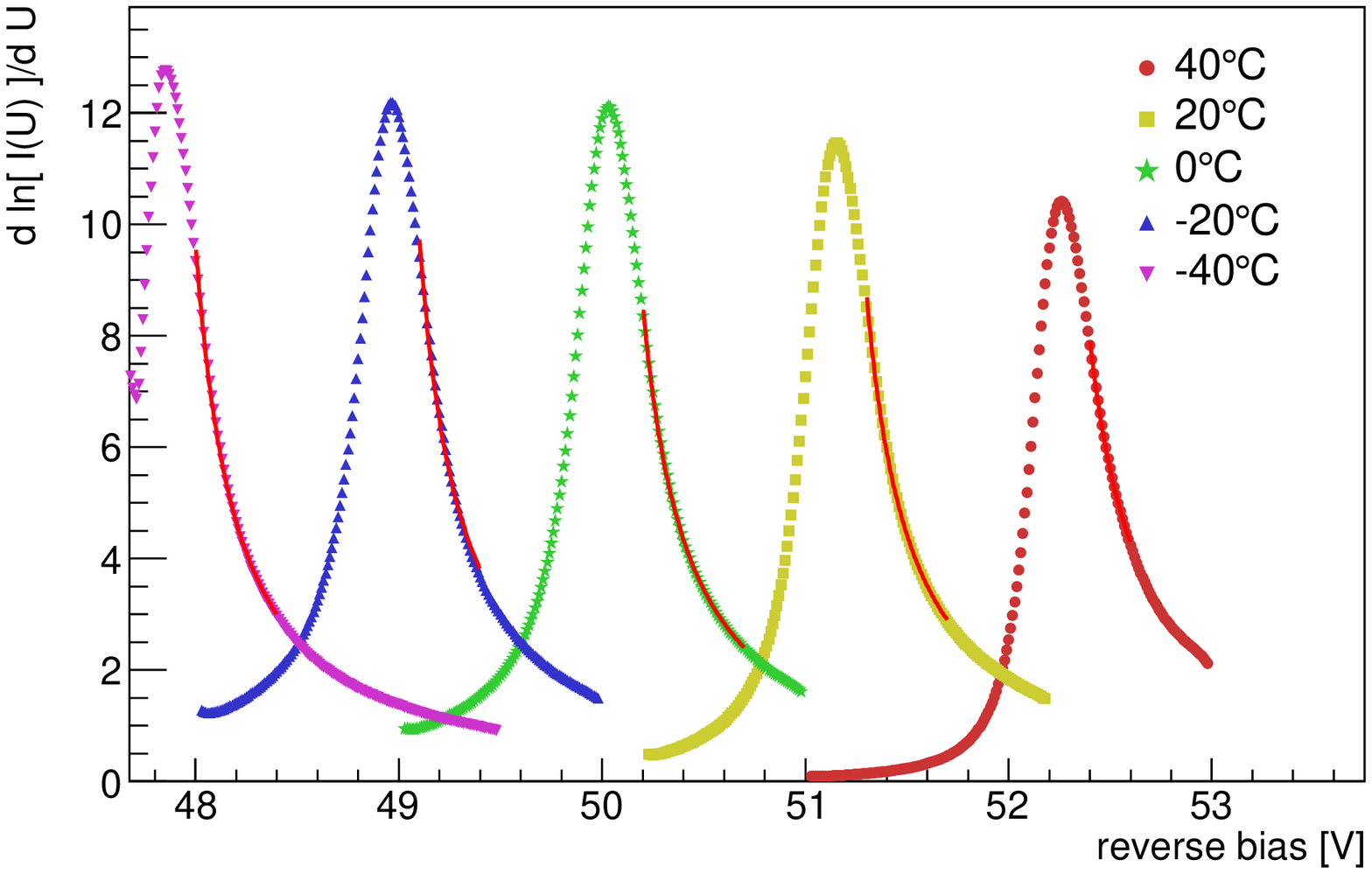}}\\
  \subfloat[SensL J-series 30035]{\includegraphics[width=\columnwidth]{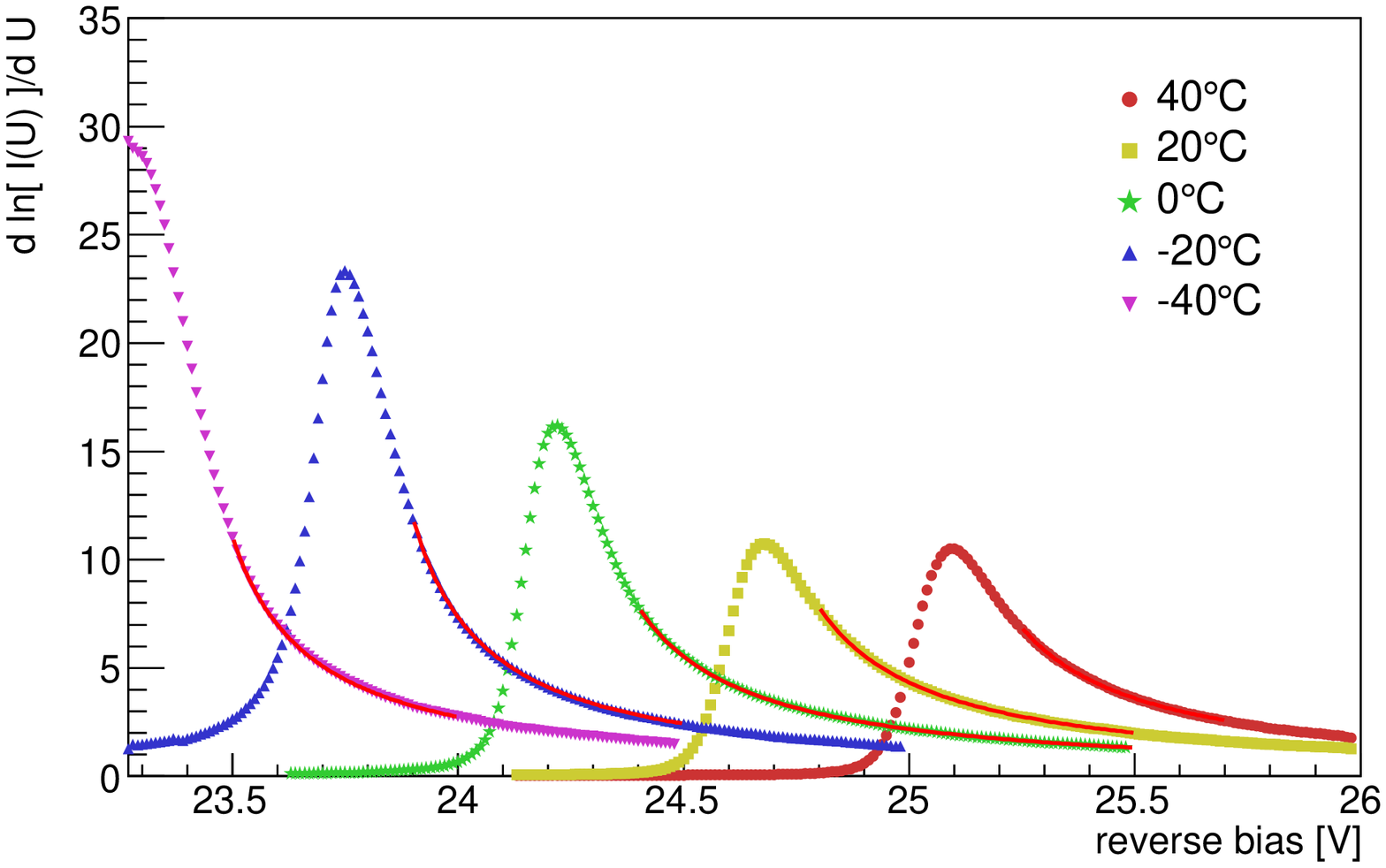}}
\caption{Derivative of the logarithm of the current around the breakdown voltage. The solid lines are fits to the curves from which the breakdown voltage is determined.} 
  \label{deriv}
\end{figure}

We verified that processing our \emph{IV} measurements in both ways does indeed yield identical results.
Fig.\ \ref{deriv} shows the outcome when they are processed according to
$\mbox{d}\ln\left(I(U)\right)/\mbox{d}U$. In all of these measurements the SiPMs were illuminated with a dimmed 400\,nm LED. 

The peak positions shown in Fig.\ \ref{deriv} are systematically above the breakdown voltage derived from gain vs.\ bias measurements by about 0.7\%, which is not acceptable in some
applications. In an effort to obtain a better estimate of the breakdown voltage, we fit each curve in Fig.\ \ref{deriv} with Eqn.\ \ref{Eqn10}. The results of the fits are shown in Fig.\ \ref{deriv} on top of the data.

\begin{figure}[!tb]
  \centering
  \subfloat[FBK NUV-HD]{\includegraphics[width=\columnwidth]{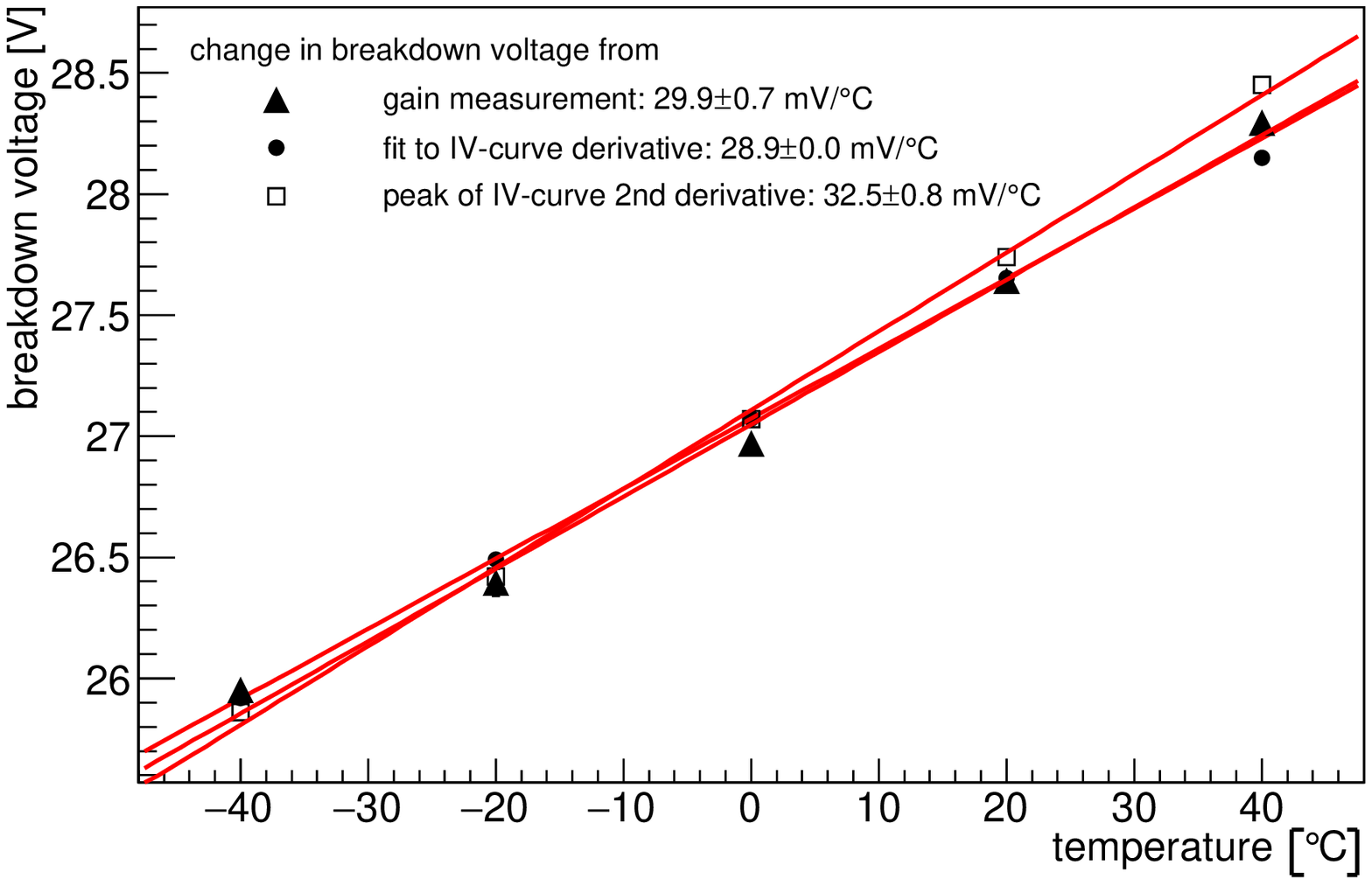}}\\
  \subfloat[Hamamatsu S13360-3050CS]{\includegraphics[width=\columnwidth]{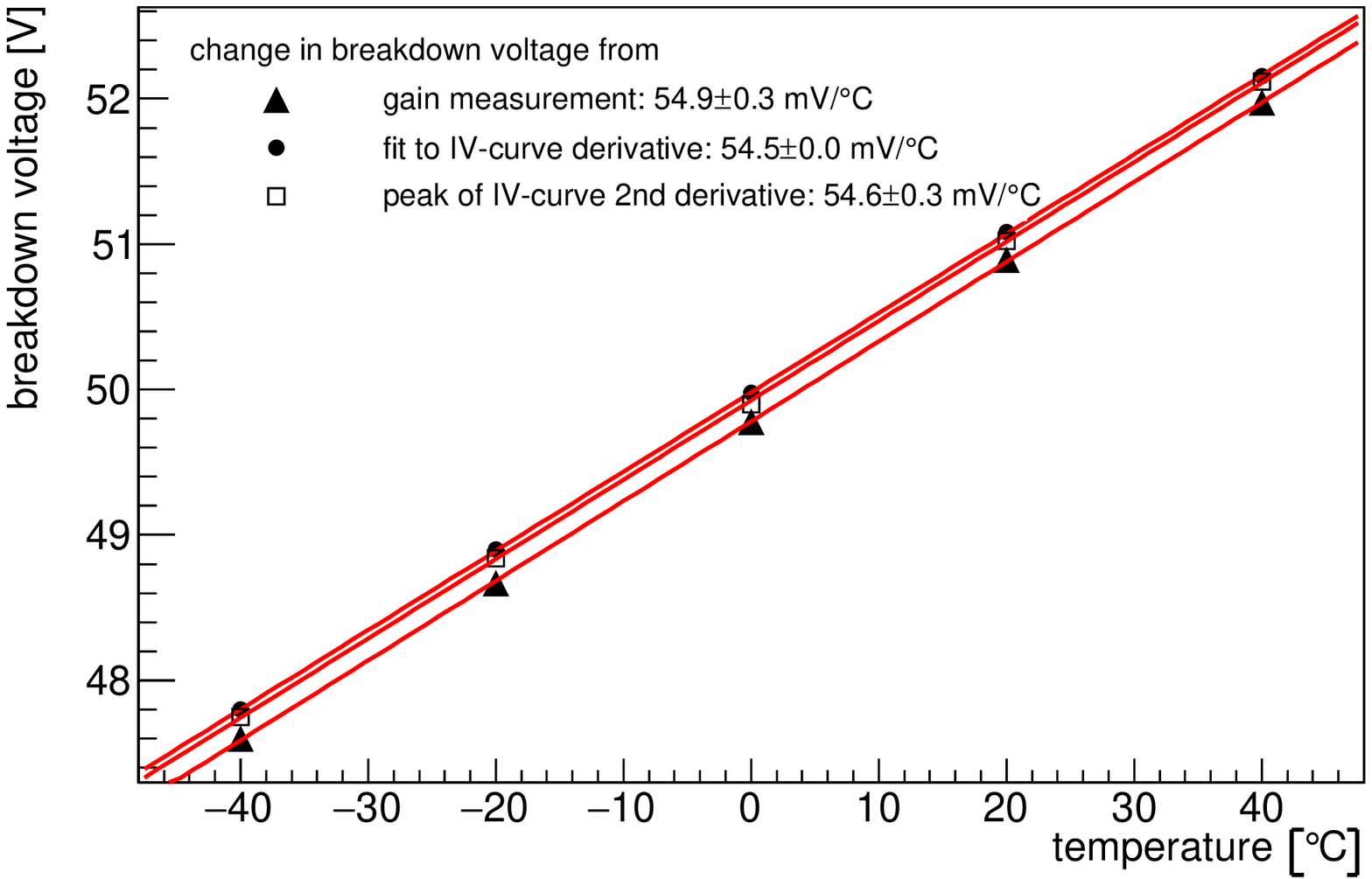}}\\
  \subfloat[SensL J-series 30035]{\includegraphics[width=\columnwidth]{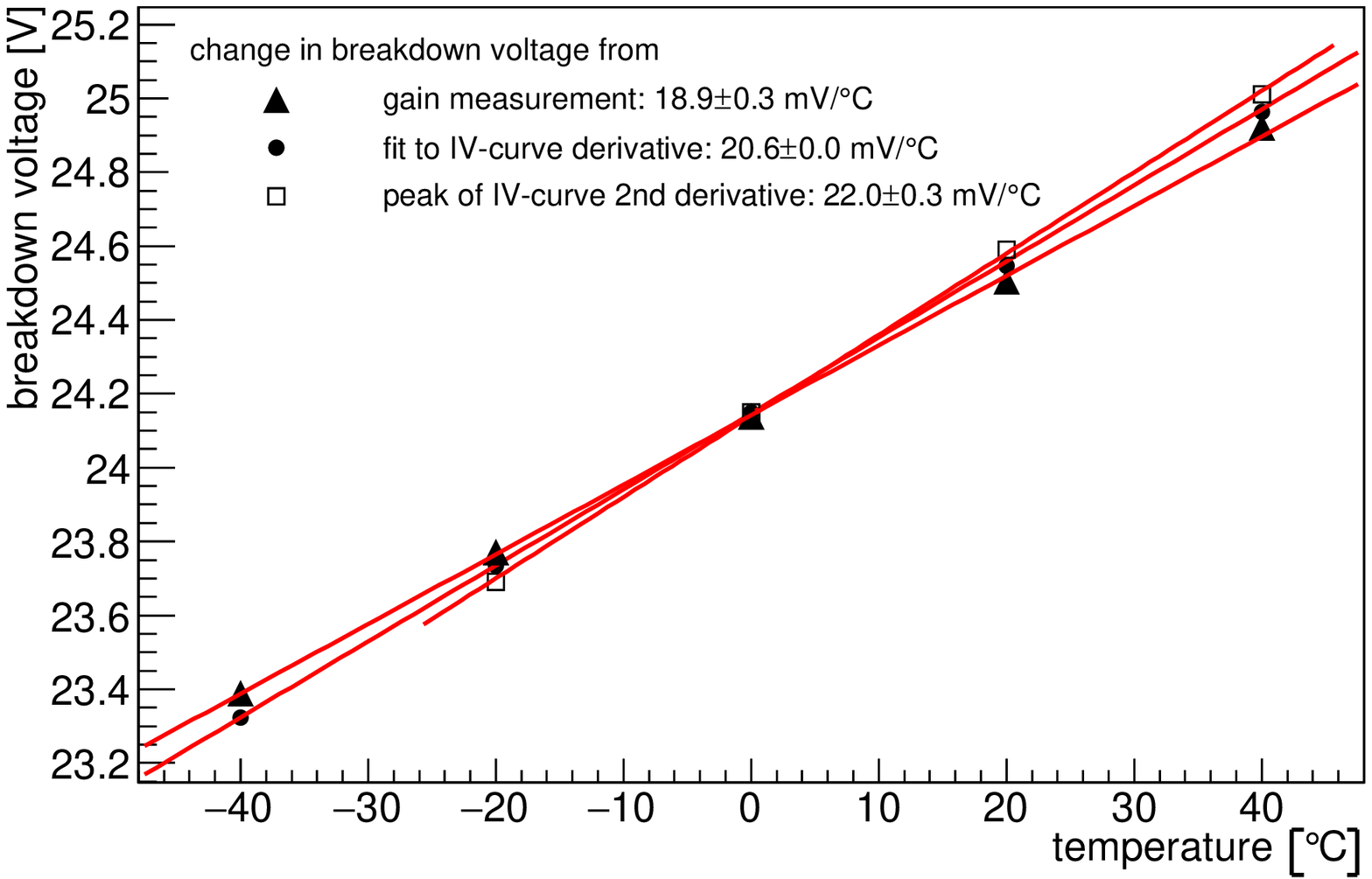}}
\caption{Breakdown voltage derived from the derivative of the IV-curves (solid dots), the second derivative of the IV-curves (empty squares), and gain measurements (triangles).} 
  \label{fBD}
\end{figure}

The breakdown voltages extracted from the fit are shown together with those from the gain measurements in Fig.\ \ref{fBD}. Differences between the fitting method and the gain method are less than
$\pm0.2$\,\%, which is significantly better than the 0.7\% offset observed in the peak-derivative method. Some of the remaining offset can be explained with systematic uncertainties in the
calibration of the signal chain that is used in the gain vs.\ bias measurements. 

An obvious outlier is the result obtained for the Hamamatsu SiPM where all breakdown voltages derived from the \emph{IV} curve have a relative offset of 0.4\% from the gain vs.\ bias
derived breakdown voltages, which is too large an offset to be explained by calibration uncertainties. The measurement of the breakdown voltage done by Hamamatsu agrees with the one from our
gain
vs.\ bias measurement.

We cannot exclude with certainty that variations of the cell capacitance with bias might be a possible cause for the discrepancy in the breakdown voltage measurements. But we note that the gain vs.\ bias curves in Figure \ref{gainbiasexample} are linear down to 1 Volt overvoltage. Thus any significant change in the cell capacitance must happen around the breakdown voltage and thus invalidate the model of the \emph{IV} curve (Equation \ref{IVmodel}) and the gain method, which both assume a constant cell capacitance. 

An additional benefit of the fit is that it also extracts values for $\alpha$. For all three devices the fit produces $\alpha$-values at room temperature that are consistent with those listed in Table \ref{alphvals}. The data seem to indicate a weak increase of $\alpha$ with temperature but the uncertainties are too large to make a more quantitative statement.

The last method we investigated to extract the breakdown voltage from the \emph{IV} curve is to use the maximum of the second derivative of the logarithm of the current
\cite{2011ITNS...58.1233S}. The estimated breakdown voltages are shown in Fig.\ \ref{fBD} as open squares and yield a similarly good estimate of the breakdown voltage as our fitting
method. For the Hamamatsu SiPM the position of the maximum of the second
derivative gives slightly better results, but it is still offset from the
breakdown voltage obtained with the gain bias method. 

The breakdown voltages in Fig.\  \ref{fBD} change proportionally with temperature for all three devices. The temperature coefficients of the breakdown voltage are given in the same figure.
The
relative change in breakdown voltage with temperature is about the same for all three devices, namely $10^{-3}$ per one degree Celsius. 


\subsubsection{IV curve simulations in the breakdown region}
We simulated \emph{IV} curves for two reasons. First we want to explain why the position of the maximum in the derivative of the logarithm of the \emph{IV} curve does not match with the breakdown voltage derived
from the gain measurement. The second reason is that we want to validate the other two methods to derive the breakdown voltage.

The model of the simulated \emph{IV} curve is based on Equation \ref{IVmodel} extended by the fraction of the dark current, which does not get amplified. The additional term allows one to simulate the
\emph{IV}
curve below the breakdown voltage. As before, contributions from optical crosstalk and afterpulsing have again been neglected. Equation \ref{IVmodel} is a model of the absolute current, whereas relevant for the derivation of the breakdown voltage is only the relative change of the current, see Eqn.\ \ref{Eqn10}. Therefore, only the relative current versus bias curve is simulated:
\begin{equation}
\hspace{-2ex}
I_{\mbox{\tiny rel}}(U_{\mbox{\tiny rel}})=
\frac{I(U_{\mbox{\tiny rel}})}{I_{\mbox{\tiny ampl}}} =
h + U_{\mbox{\tiny rel}}\cdot G\cdot\left[1-e^{\left( -U_{\mbox{\tiny rel}}/\alpha\right)}\right]\,.
\end{equation}
Where the normalization $I_{\mbox{\tiny ampl}}$ is the part of $I_{\mbox{\tiny DC}}+I_{\mbox{\tiny ext}}$ that makes it into the avalanche region and gets amplified. Note that in Eqn.\ \ref{IVmodel} and subsequent equations $I_{\mbox{\tiny DC}}+I_{\mbox{\tiny ext}}$ implicitly denote only the amplified part of the total dark and external generated current. $G$ becomes the product of the cell capacitance and the breakdown
voltage and is $6.4\cdot10^6$, $3.5\cdot10^7$, and $2.5\cdot10^7$ for the FBK, Hamamatsu, and SensL device, respectively. 
Note that we restrict ourselves to measurements done at 20$^{\circ}$C. 
The quantity $h$ is the ratio of the unamplified and amplified part of $I_{\mbox{\tiny DC}}+I_{\mbox{\tiny ext}}$. The value for $h$ is adjusted in the model until the simulated ratio of the currents at 10\% overvoltage and before breakdown matches the data and typically assumes values of 1000 or more.

Cell-to-cell variations of the breakdown voltage are included by simulating 10,000 cells each with a different breakdown voltage that is randomly picked from a normal distribution with a
mean of zero and a standard deviation that is a free parameter in the simulation. The simulated \emph{IV} curve is the sum of the currents of all 10,000 cells. 

The last parameter in the simulation is $\alpha$. A small $\alpha$ is expected if the majority of the dark current enters the multiplication region from the front, such as photoelectrons
generated by blue photons, and a large $\alpha$ is expected if the dark current is generated behind the avalanche region, \emph{e.g.}\, in the bulk. Increasing $\alpha$ in the model shifts
the position
of the maximum of the derivative of the logarithm of the \emph{IV} curve towards higher relative overvoltages and can thus be used to tune the simulations to get a match with the data. A
good agreement with measurements is achieved if $\alpha$ is 0.015, 0.05, and 0.1 for the FBK, Hamamatsu, and SensL devices, respectively. The agreement remains good if $\alpha$ is varied
within the range of values listed for each device in Table \ref{alphvals}. 

The width of the peak of the derivative of the logarithm of the \emph{IV} curve is tuned by changing the standard deviation of the cell-to-cell variations of the breakdown voltage. A value
of 0.001 reproduces the FWHM of the measurements of all three SiPMs.

We remark that we did not perform a rigorous tuning of the model parameters. Therefore, we cannot exclude that a completely different set of model parameters with different physics
implications can equally well reproduce the data. However, we are confident that  the model and its parameterization is good enough to discuss the validity of the different methods to
extract the
breakdown voltage.

The simulations confirm that the peak position of the derivative of the logarithm of the \emph{IV} curve is systematically above the breakdown voltage. We also find that fitting the
derivative reproduces the true breakdown voltage within 0.1\%. The maximum of the second derivative also lies within 0.1\% of the breakdown voltage. 

Our fitting method and the second-derivative method to extract the breakdown voltage, therefore, seem to be on solid footing. However, we emphasize that the breakdown voltages extracted from
the \emph{IV} curves of the Hamamatsu SiPM are inconsistent with the ones from the gain vs.\ bias measurements on the level of 0.4\% (200\,mV) for which we do not have an explanation.

\begin{figure}[!tb]
  \centering
  \includegraphics[width=\columnwidth]{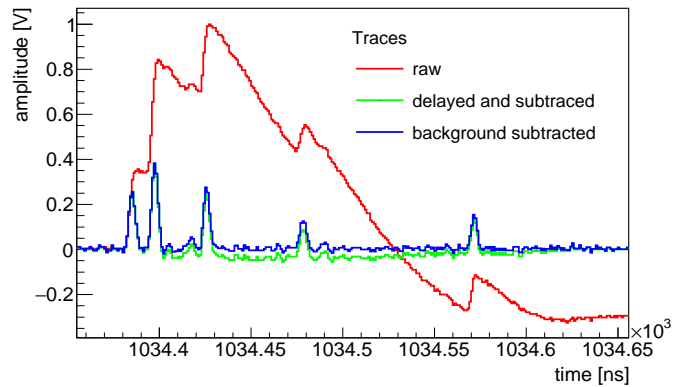}
\caption{Snapshot of an SiPM trace recorded with 1GS/s and 8\,bit resolution after amplification (red). The remaining two curves show the trace at two different stages of its processing to
reduce the signal widths. See text for details.} 
  \label{sampleTrace}
\end{figure}

\section{Signal trace analysis}

In the remainder of the paper, we discuss the analysis of SiPM signals recorded with the Alazar ATS 9870 digitizer after amplifying the signal with a Mini-Circuits ZFL 500LN+ amplifier and a
LeCroy Model 612A amplifier (see Fig.\  \ref{SketchBasicMeas}). For the absolute calibration of the gain measurement, the SiPM signals were recorded in parallel with a Tektronix TDS
3054C
oscilloscope after amplification of the SiPM signals with the Mini-Circuits ZFL 500LN+ preamplifier.

The SiPM signals need to be processed to eliminate the long tails of the individual signals. Fig.\  \ref{sampleTrace} shows an example of a recorded SiPM trace before (red) and after (blue)
processing. Long tails are a general feature of SiPMs with surface areas larger than 1\,mm$^2$ because their terminal capacitance increases with sensor area which, combined with a
50\,Ohm input impedance preamplifier, results in long tails. Long tails are also the result of cell recovery times that are less than a few hundred nanoseconds long. 

To process the signals, we follow a two-step procedure similar to the approach used in \cite{Piemonte:2012rka}. In the first step, a copy of the original trace is shifted by three
nanoseconds and subtracted from the original trace. This step results in a significant shortening of individual SiPM signals down to a full width of about 9\,ns. An example of the outcome of
this processing step is shown as the green trace in Fig.\  \ref{sampleTrace}. A small remaining undershoot is subtracted from the trace by applying a background-subtraction algorithm that is
implemented in the ROOT analysis framework \cite{root}. The final result is shown as the blue trace in the figure.

\begin{figure}[!tb]
  \centering
  \includegraphics[width=\columnwidth]{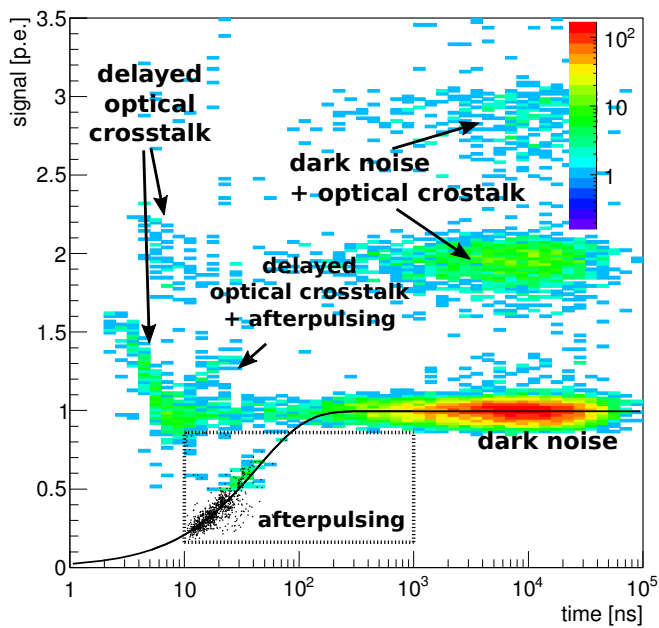}
\caption{The histogram shows the time difference between two consecutive SiPM signals on the x-axis and the amplitude of the second signal on the y-axis. Note the logarithmic scale of the
x-axis. The colors represent the number of events in each bin on a logarithmic scale. Several populations can be identified and are correspondingly labeled. } 
  \label{ampltimes}
\end{figure}

The general procedure of the signal trace analysis is to record randomly
triggered 10\,ms long signal traces until enough pulses are accumulated to reconstruct all parameters of interest
with high enough precision. The measurement of the afterpulsing is typically the bottleneck and defines how many traces need to be recorded. At low temperatures a dimmed LED is used to
increase the SiPM signal rate and thus speed up the afterpulsing measurement. Measurements of the dark rate are made with the LED turned off.

After a trace is processed, all SiPM signals with an amplitude of at least 0.5\,photoelectrons (p.e.) are identified, \emph{i.e.}\, signals with at least half the amplitude that is generated
when one cell of an SiPM breaks down. The amplitudes and times of the identified signals are then used to extract the SiPM parameters (similar to how it is described in \cite{Piemonte:2012rka}).

An illustrative example of the type of information that can be extracted from the amplitudes and times is given in Fig.\  \ref{ampltimes}. It is a two-dimensional histogram that has the time
difference between two consecutive signals on the x-axis and the amplitude of the second signal in units of p.e.\ on the y-axis. The color gives the number of events per bin.
In this figure only signal pairs have been selected in which the first signal has an amplitude of one photoelectron.

A number of different populations can easily be identified. The biggest population is made up by signals in which only one cell of the SiPM fires. That population peaks at a time difference
of $\approx10\,\mu$s, which is the expected average time difference between two uncorrelated dark count signals, \emph{i.e.}\, the inverse of the dark count rate for that specific device
and
temperature. The bands above that population are from signals where one cell fires due to an uncorrelated dark count, and one or two additional cells fire in coincidence due to direct
optical
crosstalk.

To the left of the main blob is a smaller population that is due to delayed optical crosstalk signals. The amplitudes of the delayed optical-crosstalk signals to the very left depend on the
time when the signal appears because there is significant overlap with the preceding signal, and the signal-extraction algorithm is not able to properly handle the overlap.\footnote{The width
of one signal is 9\,ns after a trace is processed.}

Also visible are afterpulsing events that generate a second signal from the same cell before it is fully recharged. The solid black line shows a fit to the afterpulsing events in the dashed
box and is used to measure the recovery time of one cell.

\subsection{Gain, Cell Capacitance and Breakdown Voltage}

\begin{figure}[!tb]
  \centering
  \includegraphics[width=\columnwidth]{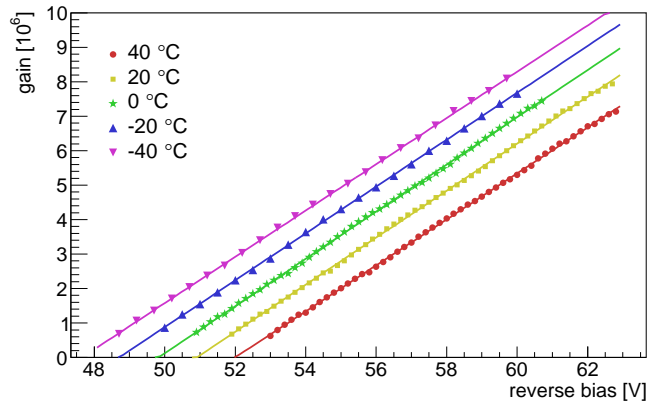}
\caption{Gain vs.\ bias for the Hamamatsu SiPM for five different temperatures.}
  \label{gainbiasexample}
\end{figure}

The first information extracted from the signal amplitudes is the signal charge in units of electrons, which is commonly referred to as the gain of an SiPM. The amplitudes of signals between
0.5 and 1.5 p.e.\ are averaged and then converted into signal charge. For this conversion, a separate calibration of the entire signal chain was performed for
each SiPM. 

In the first step of the calibration, the average single p.e.\ amplitude was read off a Tektronix TDS 3054C oscilloscope at a temperature of $-20^{\circ}$C and at two different bias voltages
after amplification of the raw signals with a Mini-Circuits ZFL-500LN+. The uncertainty in reading the amplitude off the oscilloscope is 0.2\% and dominates the uncertainty of the absolute
gain and breakdown voltage measurement. In the second step, the signal amplitudes are divided by the gain of the amplifier (30\,dB). In the third step, the calibrated amplitudes are
multiplied with the integral of the normalized raw signal shape,\footnote{The signal shape was normalized to a peak amplitude of one.} thus obtaining two absolute gain measurements. These
two absolute gain measurements and the average single-cell amplitudes that were extracted from the processed traces at the same bias and temperature  are then used to define a linear
transformation from processed signal amplitude to absolute charge.

An example of a calibrated gain measurement is shown in Fig.\  \ref{gainbiasexample}. The solid lines are linear fits to the data. A closer inspection of the data points reveals small
residuals
with respect to the fits, which can be attributed to non-linearities in the front-end amplifier of the digitizer. 

The linear dependence of the gain on bias can be explained in the small-signal model of SiPMs where the cell of an SiPM is represented by a capacitance $C_{\mbox{\footnotesize cell}}$ that
is discharged to the breakdown voltage in a breakdown. The total charge $G$ of the signal is then
\begin{equation}\label{gain}
G = C_{\mbox{\footnotesize cell}} \cdot \left(U - U_{\mbox{\footnotesize BD}}\right)\,.
\end{equation}
If $G$ is given in units of electrons, it is usually referred to as the gain of the device, which is the definition of $G$ we adopt in this paper. 


Based on Equation \ref{gain} the breakdown voltage can be measured from the gain vs.\ bias curve as the voltage where the gain is zero. The determined breakdown voltage is shown in Fig.\ 
\ref{fBD} together with those extracted from the \emph{IV}-curves.

\begin{figure}[!tb]
  \centering
\subfloat[FBK NUV-HD]{\includegraphics[width=\columnwidth]{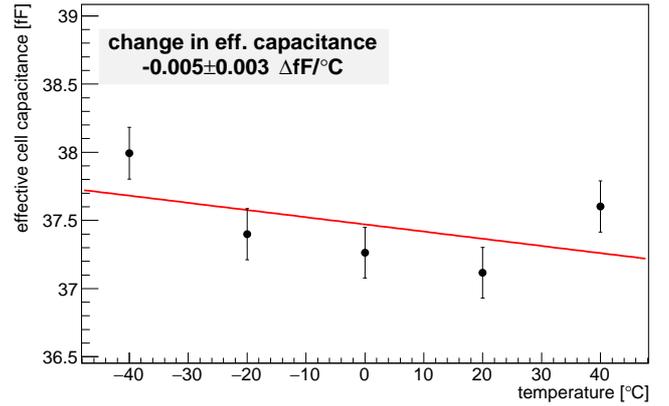}}\\
  \subfloat[Hamamatsu S13360-3050CS]{\includegraphics[width=\columnwidth]{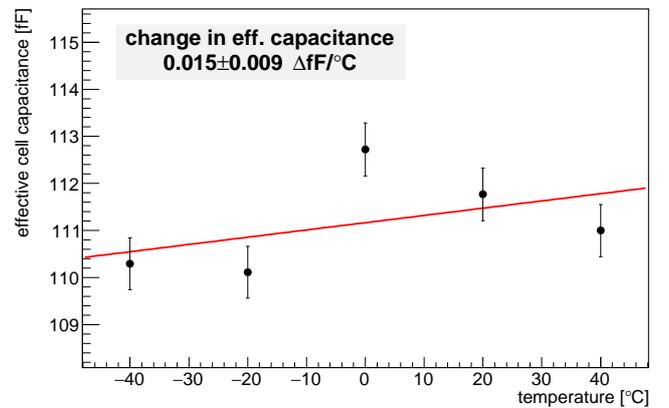}}\\
  \subfloat[SensL J-series 30035]{\includegraphics[width=\columnwidth]{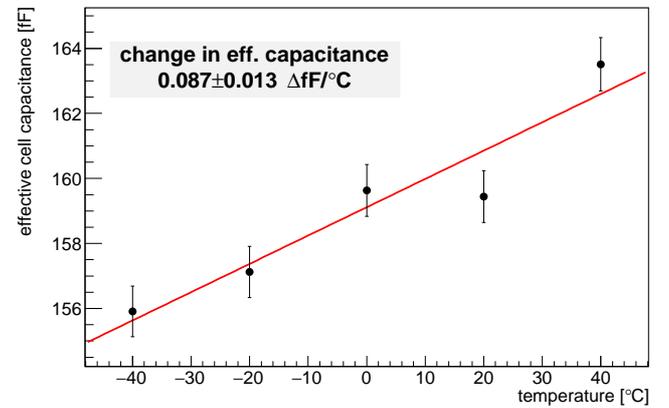}}
\caption{Cell capacitance.} 
  \label{cellcap}
\end{figure}

The cell capacitance $C_{\mbox{\small cell}}$ is given by the slope of the gain vs.\ bias measurement and is shown in Fig.\  \ref{cellcap}. For the Hamamatsu and the FBK SiPM the cell
capacitance remains constant, whereas a 5\% change is seen in the SensL SiPM between $-40^{\circ}$C and $40^{\circ}$C. The gain vs.\ bias curves are well described by linear functions, and
aside from the residuals that can be attributed to the digitizer, no further deviation from linearity is observed that would point to a dependence of the cell capacitance on bias for any of
the
tested devices.

\subsection{Dark count rates}

The dark count rates are measured by counting all signals with an amplitude larger than 0.5\,p.e.\ and dividing that number by the total duration of all analyzed traces. Included in this measurement are, therefore, thermal generated dark counts as well as delayed optical crosstalk and afterpulsing. However, the latter two contribute only minor to the total dark count rate as they are less than 2\% at 90\% breakdown probability. Two pulses have to
be
at least $\approx3$\,ns apart in order to be identified as separate signals. 

\begin{figure}[!tb]
  \centering
\subfloat[FBK NUV-HD]{\includegraphics[width=\columnwidth]{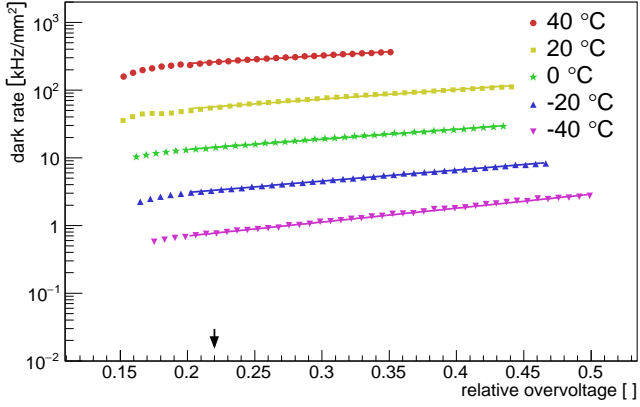}}\\
 \subfloat[Hamamatsu S13360-3050CS]{\includegraphics[width=\columnwidth]{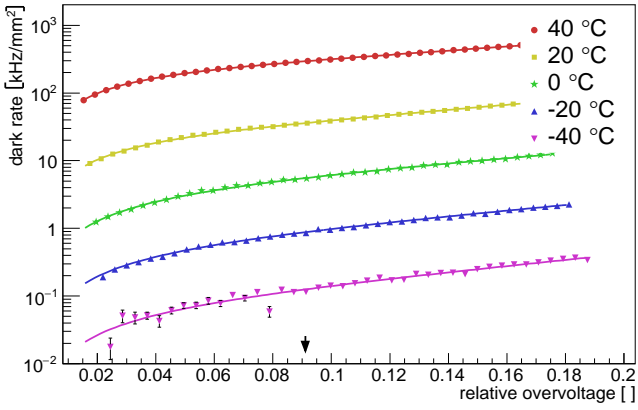}}\\
  \subfloat[SensL J-series 30035]{\includegraphics[width=\columnwidth]{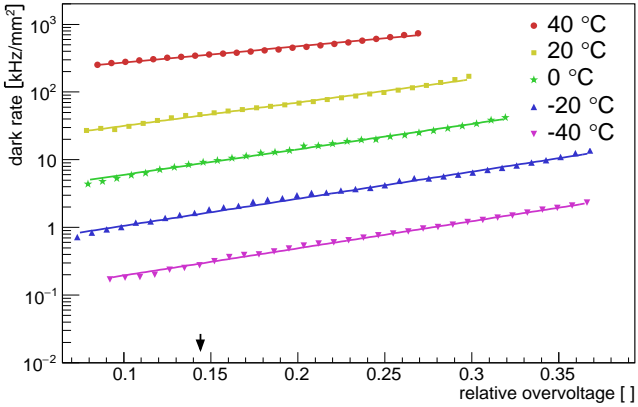}}
\caption{Dark count rates. The arrow marks the nominal operating bias of each device.} 
  \label{dcabsolute}
\end{figure}

Fig.\  \ref{dcabsolute} shows the dark count rates per one square millimeter sensor area for all temperatures and for all three devices. The solid lines are fits to the data with the
function
\begin{equation}
DC\left(U_{\mbox{\footnotesize rel}}\right ) = e^{a+b\cdot U_{\mbox{\tiny rel}}} \cdot \left[1-e^{\left(-U_{\mbox{\footnotesize rel}}/\alpha\right)}\right]\,,
\end{equation}
where the last term is the breakdown probability and is only used in the fit of the dark rate measurement of the Hamamatsu SiPM. For the SensL and FBK SiPMs the dark-rate measurements start
at an overvoltage where the breakdown probability is already close to 90\% (check the position of the arrow). The turnover in the data for the FBK device occurs because the small cell
capacitance results in signals too small to be reliably detected with our signal chain at low overvoltages.  The results of the fits are shown in Table \ref{dctable}. The $\alpha$ values
extracted for the Hamamatsu SiPM are consistent with the $\alpha$ extracted from the PDE measurements (s.\ Table \ref{alphvals}) for short photon wavelengths, which indicates that the
majority of the dark noise enters the avalanche region from the surface of the device.

\begin{table}
\caption{Best Fit Values Obtained From the Fit of the Dark Rate Measurements Shown in Fig.\  \ref{dcabsolute}.}
\centering
\begin{tabular}[!htb]{c|c|c|c|c}
Dev.&Temp.& $a$ & $b$ & $\alpha$ $\left[10^{-2}\right]$\\\hline\hline
FBK&-40$^{\circ}$C&  -1.30$\pm$0.01& 4.72$\pm$0.02 & \\
   &-20$^{\circ}$C&   0.368$\pm$0.003&3.79$\pm$0.01 & \\
   &0$^{\circ}$C&  1.91$\pm$0.01&3.41$\pm$0.01 & \\
   &20$^{\circ}$C&  3.31$\pm$0.01&3.26$\pm$0.01&\\
   &40$^{\circ}$C&  4.97$\pm$0.01&2.68$\pm$0.01&\\\hline
Ham.&-40$^{\circ}$C&  -2.9$\pm$0.1& 10.1$\pm$0.7 & 4$\pm$1\\
   &-20$^{\circ}$C&   -0.84$\pm$0.03&9.2$\pm$0.2 & 4.3$\pm$0.2\\
   &0$^{\circ}$C&  1.11$\pm$0.02&8.2$\pm$0.1 & 4.5$\pm$0.1\\
   &20$^{\circ}$C&  2.86$\pm$0.01&8.43$\pm$0.06&2.8$\pm$0.1\\
   &40$^{\circ}$C&  5.100$\pm$0.003&6.83$\pm$0.02&2.7$\pm$0.1\\\hline
SensL&-40$^{\circ}$C&  -2.56$\pm$0.01& 9.22$\pm$0.04 & \\
   &-20$^{\circ}$C&   -0.86$\pm$0.01&9.19$\pm$0.03 &\\
   &0$^{\circ}$C&  0.92$\pm$0.01&8.65$\pm$0.02 & \\
   &20$^{\circ}$C&  2.662$\pm$0.001&7.92$\pm$0.01&\\
   &40$^{\circ}$C&  5.055$\pm$0.001&6.71$\pm$0.01&\\\hline
\end{tabular}
\label{dctable}
\end{table}

\begin{figure}[!tb]
  \centering
  \subfloat[FBK NUV-HD]{\includegraphics[width=\columnwidth]{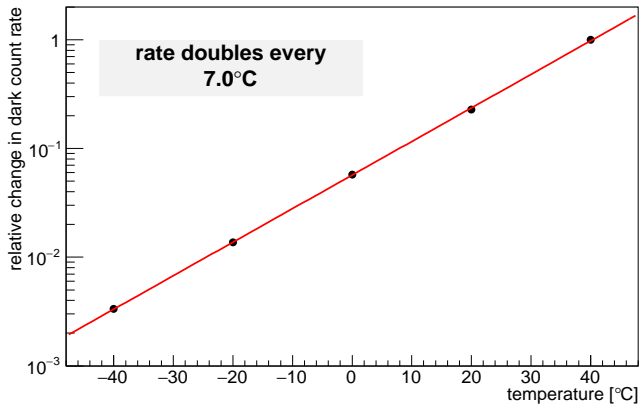}}\\
  \subfloat[Hamamatsu S13360-3050CS]{\includegraphics[width=\columnwidth]{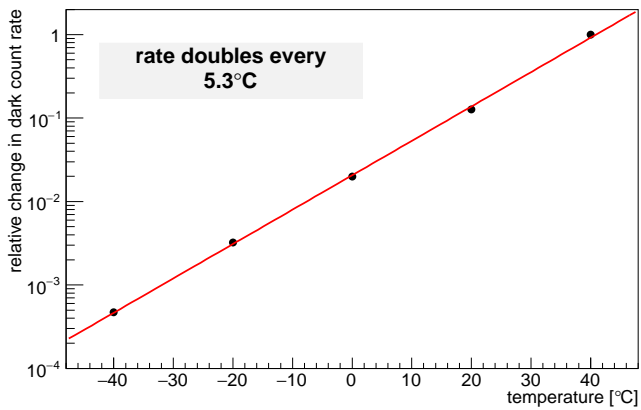}}\\
  \subfloat[SensL J-series 30035]{\includegraphics[width=\columnwidth]{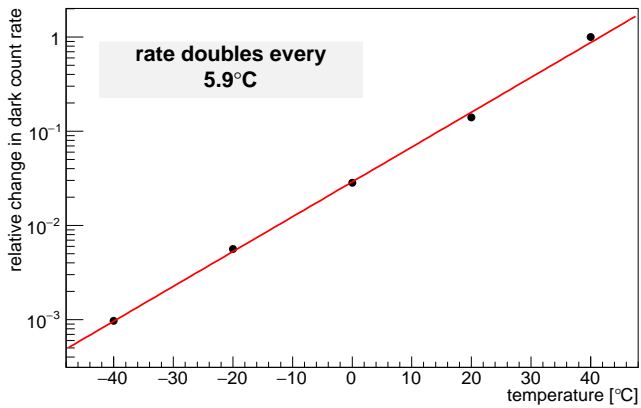}}
\caption{Relative change in dark count rates.} 
  \label{reldcchange}
\end{figure}


The rates in Fig.\  \ref{dcabsolute} are shown versus relative overvoltage. For a fixed relative overvoltage, any change in the dark rate with temperature can be attributed to changes in the
thermal generation of charge carriers. Fig.\  \ref{reldcchange} shows how the dark count rate changes with temperature for a fixed overvoltage relative to the dark count rate at
40$^{\circ}$C  and averaged over the operating voltage range at 40$^{\circ}$C.  The relative change in dark count rate with temperature for all three devices is well described by
$e^{a+b\cdot T}$. The change in temperature needed to change the dark count rate by a factor of two is stated in the inserts in the figure.  

\subsection{Optical crosstalk}

Optical crosstalk (OC) is the correlated firing of cells due to photons emitted in the breakdown of one cell. Any of these photons can initiate the breakdown of a neighboring cell. Two types
of optical crosstalk can be distinguished. Direct OC is due to crosstalk photons that get absorbed in the active volume of a neighboring cell and cause the breakdown of that cell,
which happens quasi-simultaneous to the first one.  Delayed OC is due to crosstalk photons that convert in the non-depleted bulk. In this case the generated charge carrier has to first diffuse
into the active volume of the cell \citep{Piemonte:2012rka,NepomukOtte2009105,Rosado2015153}. The diffusion
process introduces a measurable time delay between the breakdown of the first cell and the breakdown of the second cell. 

Measurements of direct OC are presented in this section and the delayed OC measurements are discussed together with afterpulsing measurements in the next section. 

\begin{figure}[!tb]
  \centering
  \includegraphics[width=\columnwidth]{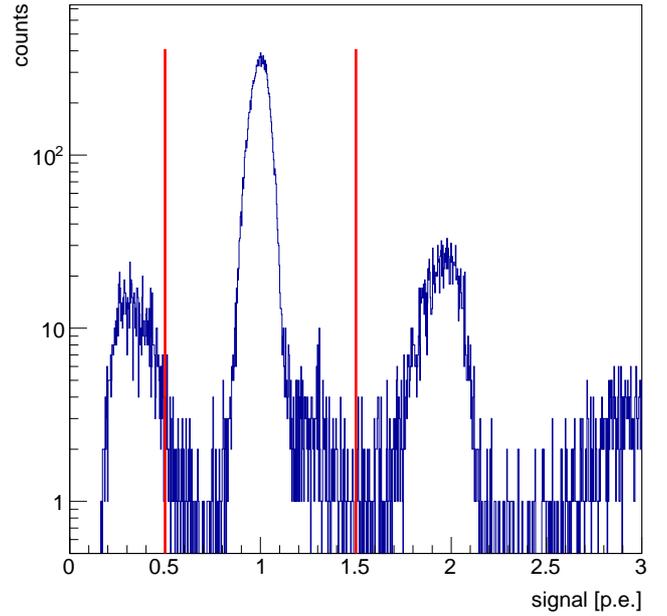}
\caption{Example of a pulse-height distribution of signals from the SensL device. The vertical line at 1.5\,p.e.\ marks the boundary between signals in which only one cell fired (left) and more than one
(right). The small peak at the left is due to afterpulsing events that can also be identified in Fig.\  \ref{ampltimes}. Only signals with an amplitude of at least 0.5\,p.e.\ are used in the
optical crosstalk analysis. } 
  \label{OCPHD}
\end{figure}

Direct OC is extracted from the pulse-height distribution of the SiPM signals. Fig.\  \ref{OCPHD} shows an example of such a distribution where events can be clearly identified that are due
to 1, 2, or 3 cells firing simultaneously. The small peak on the left is due to afterpulses, which are the same events that are also marked as afterpulses in Fig.\  \ref{ampltimes}. The
OC probability is determined by counting all events with an amplitude larger than 1.5\,p.e.\ and dividing that number by the total number of events with an amplitude larger than 0.5\,p.e.

\begin{figure}[!tb]
  \centering
  \subfloat[FBK NUV-HD]{\includegraphics[width=\columnwidth]{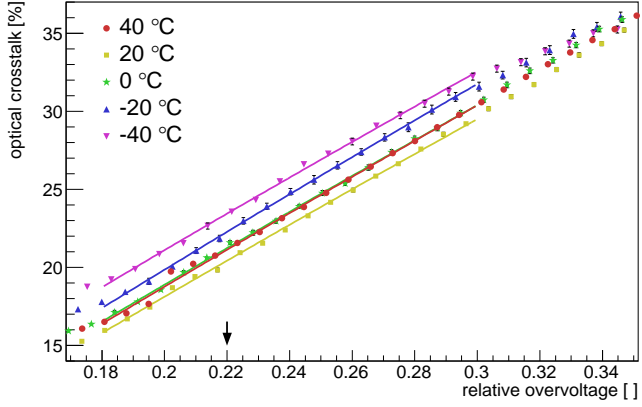}}\\
  \subfloat[Hamamatsu S13360-3050CS]{\includegraphics[width=\columnwidth]{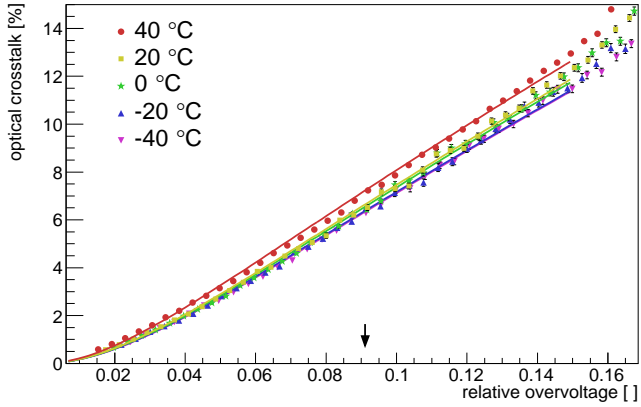}}\\
  \subfloat[SensL J-series 30035]{\includegraphics[width=\columnwidth]{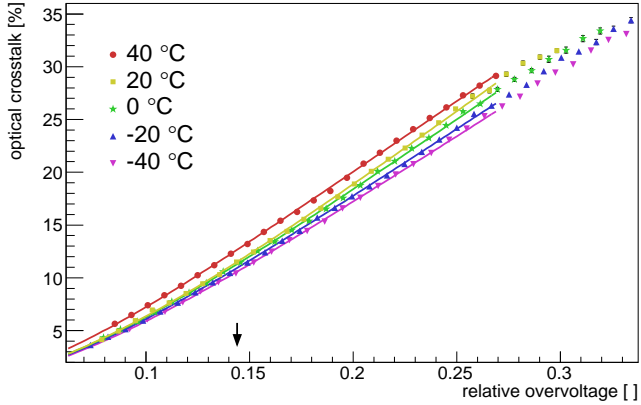}}
\caption{Direct optical crosstalk. The arrow marks the nominal operating bias of each device.} 
  \label{opticalXtalk}
\end{figure}

Fig.\  \ref{opticalXtalk} shows the direct OC for all three SiPMs as a function of relative overvoltage. At their respective operating voltages, marked by the arrow, the FBK device has the
highest OC at 23\% followed by the SensL and the Hamamatsu SiPM, which has the lowest OC (6\%). 

The OC of the Hamamatsu device does not depend on temperature, whereas the SensL OC increases with temperature; both behaviors can be explained with a constant and increasing cell
capacitance, respectively, as will be detailed later.

The OC measured for the FBK device on the other hand shows a clear offset of the curves that is about $\pm$5\,\%. Upon further investigation we came to the conclusion that the offset is a
systematic effect due to the partial overlap of the individual peaks in the pulse-height distribution of the FBK device. The same effect also explains the small offset of the OC measurement
at 40$^{\circ}$C for the SensL and the Hamamatsu device. 

We note that the FBK device is by far the largest of the three tested devices, which is why the absolute dark count rates are also highest and the probability of overlapping pulses is,
therefore, more frequent than in the other two devices. We also remark that optical crosstalk increases with the size of the device, and our measurements are not corrected for that effect.

The overvoltage dependence of the OC can be understood in the following way. The number of photons emitted in the breakdown of one cell is $f\cdot C_{\mbox{\footnotesize eff}}\cdot  \Delta
U$, where $f$ is about $3\cdot10^{-5}$ photons per electron in the avalanche \cite{NepomukOtte2009105} and $C_{\mbox{\footnotesize eff}}\cdot  \Delta U$ is the gain in units of electrons.
Each emitted photon has a probability $\gamma$ to absorb in the active volume of a neighboring cell and generate a charge carrier. The likelihood of that charge carrier to initiate a
breakdown is given by the breakdown probability $1-\exp\left(-U_{\mbox{rel}}/\alpha\right)$. Combining all factors, the OC as a function of relative overvoltage becomes
\begin{equation}\label{OCf}
OC(U_{\mbox{\footnotesize rel}}) = f\cdot  C_{\mbox{\footnotesize eff}}\cdot U_{\mbox{\footnotesize rel}}\cdot U_{\mbox{\footnotesize BD}}\cdot \gamma\cdot
\left[1-e^{\left(-U_{\mbox{\footnotesize rel}}/\alpha\right)}\right]\,.
\end{equation}
The probability $\gamma$ is thus a device-specific number that quantifies how well a given structure suppresses OC and is hereafter referred to as optical crosstalk efficiency. While our
specific
parameterization of the OC is different, it is conceptually equivalent to the one used in \cite{0031-9155-59-13-3615}.

The measured OC curves are fit with the above function, and the best fit $\gamma$ and $\alpha$ values are listed in Table \ref{OCvals}. All OC curves including the FBK curve are well
described
by the fit function. 
For the Hamamatsu
and the FBK device the small $\alpha$ value indicates that the majority of
optical crosstalk photons enter the avalanche region from the surface
whereas the large $\alpha$ value for the SensL device indicates that the
majority of the crosstalk photons enter the cell from below.
With a $\gamma$ of 0.08, the OC efficiency is lowest for the Hamamatsu device, which
has filled trenches between cells to prevent photons from crossing into a neighboring cell. For the SensL device, which does not have trenches, the OC efficiency is twice as high.  

We note that the $\gamma$ values of 0.5  and the $\alpha$ values for the FBK device are likely affected by the above mentioned systematic effects caused by the reduced separability of the
peaks in the pulse-height distribution and thus should be interpreted with caution.
 
\begin{table}
\caption{Best Fit Values Obtained From Fitting the Direct Optical Crosstalk Measurements Shown in Fig.\  \ref{opticalXtalk}. The Last Column Shows the Probability That a Photon Emitted in a
Breakdown Results in a Breakdown of a Neighboring Cell.}
\centering
\begin{tabular}[!htb]{c|c|c|c}
Device&Temp.&$\alpha$&OC efficiency $\gamma$\\\hline\hline
FBK&-40$^{\circ}$C&0.059$\pm$0.002&0.590$\pm$0.002\\
   &-20$^{\circ}$C&0.082$\pm$0.004&0.584$\pm$0.005\\
   &0$^{\circ}$C&0.085$\pm$0.002&0.551$\pm$0.003\\
   &20$^{\circ}$C&0.092$\pm$0.001&0.531$\pm$0.002\\
   &40$^{\circ}$C&0.089$\pm$0.001&0.528$\pm$0.001\\\hline
Hamamatsu&-40$^{\circ}$C&0.040$\pm$0.001&0.079$\pm$0.001\\
   &-20$^{\circ}$C&0.040$\pm$0.001&0.076$\pm$0.001\\
   &0$^{\circ}$C&0.041$\pm$0.001&0.076$\pm$0.001\\
   &20$^{\circ}$C&0.039$\pm$0.001&0.076$\pm$0.001\\
   &40$^{\circ}$C&0.034$\pm$0.001&0.078$\pm$0.001\\\hline
SensL&-40$^{\circ}$C&0.161$\pm$0.001&0.129$\pm$0.002\\
   &-20$^{\circ}$C&0.160$\pm$0.001&0.127$\pm$0.001\\
   &0$^{\circ}$C&0.162$\pm$0.001&0.130$\pm$0.002\\
   &20$^{\circ}$C&0.168$\pm$0.001&0.137$\pm$0.002\\
   &40$^{\circ}$C&0.154$\pm$0.001&0.105$\pm$0.001\\\hline
\end{tabular}
\label{OCvals}
\end{table}

\subsection{Afterpulsing and Delayed Optical Crosstalk}
\begin{figure*}[!tb]
  \centering
  \subfloat[Only next pulse]{\includegraphics[width=0.32\textwidth]{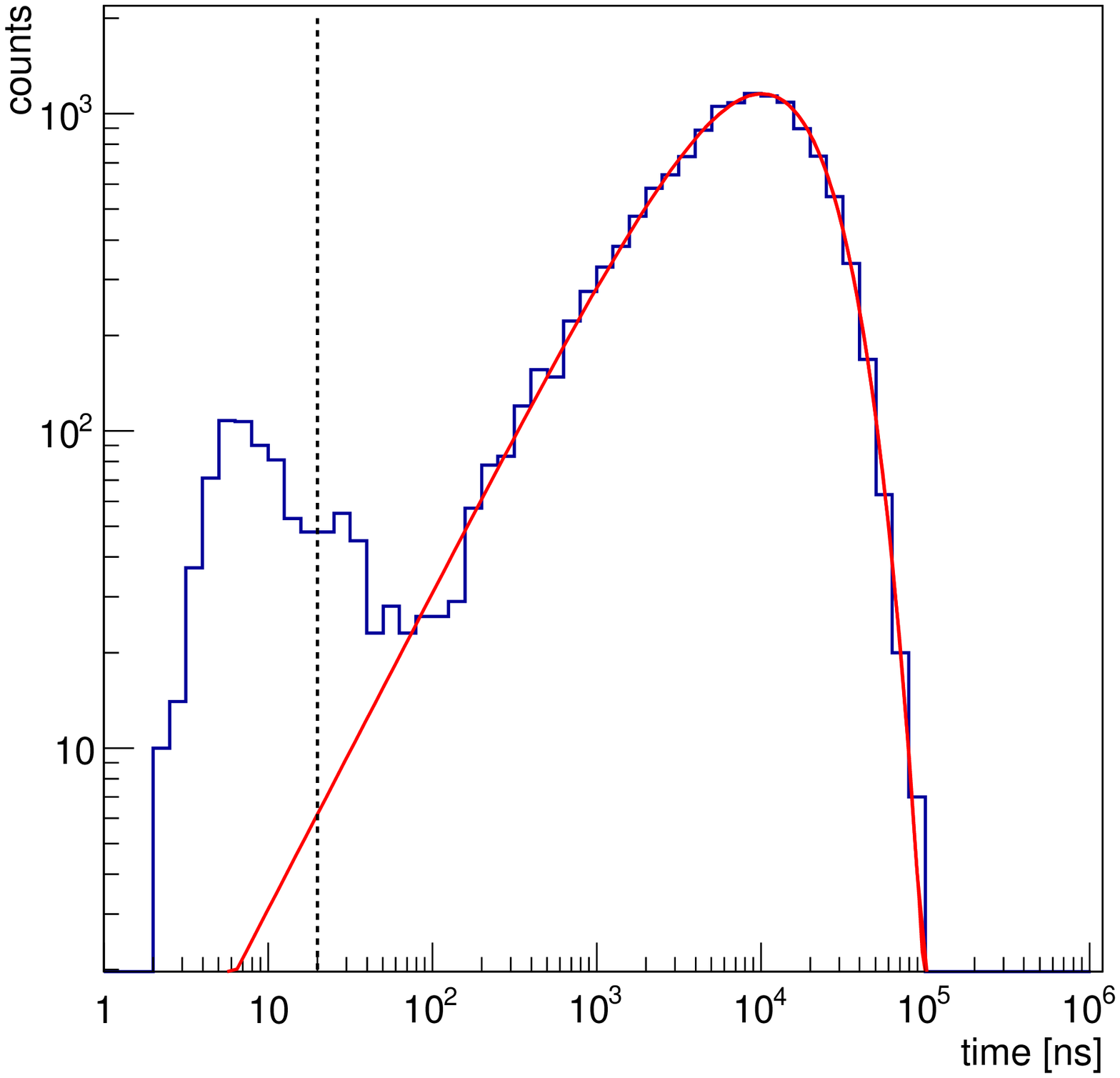}
  \label{APPHDa}}
  \subfloat[All following pulses]{\includegraphics[width=0.32\textwidth]{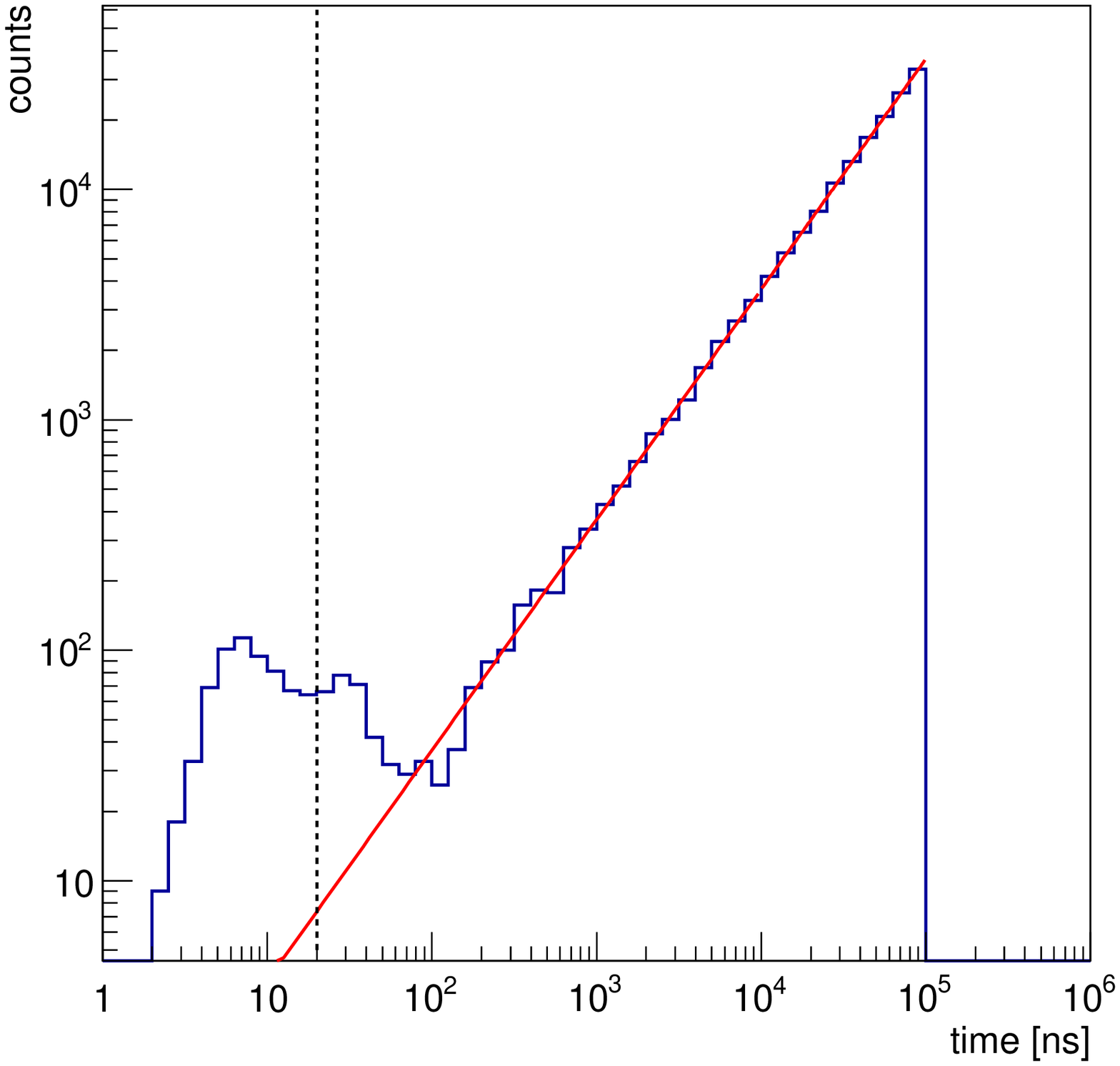}
\label{APPHDb}} 
  \subfloat[Residuals]{\includegraphics[width=0.32\textwidth]{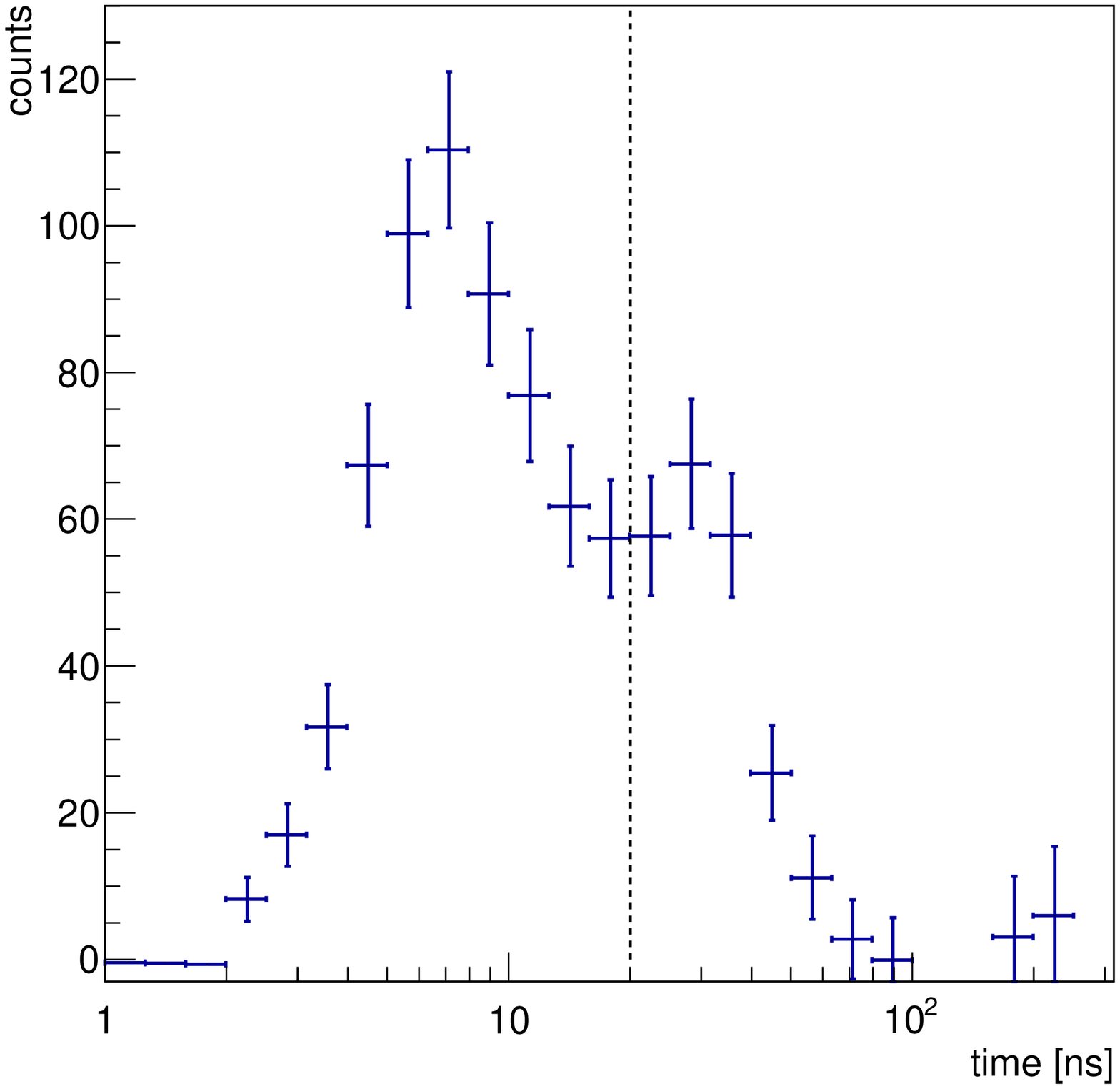}}
\caption{Example of distributions of time differences between two pulses from the SensL device. See text for details. } 
  \label{APPHD}
\end{figure*}

Afterpulsing and delayed OC both produce signals that are correlated in time with respect to a previous SiPM signal. Both effects are quantified by selecting SiPM signals with amplitudes
between 0.5 and 1.5\,p.e.\ and recording the time to the next signal. Fig.\  \ref{APPHDa} shows an example of the distribution of the time differences. The main peak is due to uncorrelated,
Poisson-distributed dark-noise counts. The position of the peak is at the average time difference between two dark counts, which is equal to the inverse of the dark-count rate. We note that
the binning of the histograms is logarithmic, and as a result of the binning, the Poisson distribution takes the form $a\cdot t  \cdot \exp(-t/\tau)$ instead of a pure exponential function.
The main peak is well fit with a Poisson distribution, and the residuals due to delayed optical crosstalk and afterpulses at small time differences are clearly visible.

For the extraction of the delayed OC and afterpulsing probabilities, however, we histogram not only the time difference between the first and the next pulse, but all following pulses up to a
time difference of 100$\,\mu$s. In this way we eliminate the need to consider cases in which an afterpulse or delayed OC signal is missed because of an earlier dark count.  Fig.\ 
\ref{APPHDb} shows the corresponding pulse-height distribution. The Poisson-distributed dark counts follow a line through the origin now. The fit of the distribution with a line was
performed between 10\,$\mu$s and 100\,$\mu$s. The figure to the right shows the residuals between the data and the fit, which are due to delayed OC and afterpulses. 

The residuals consist of two components. The left component is due to delayed OC, and the right is due to afterpulses. The two components are better visible in the amplitude vs.\ time
distribution shown in Fig.\  \ref{ampltimes}. Delayed OC produces signals with amplitudes of 1 p.e.\ or larger, whereas afterpulses have amplitudes between 0 and 1. 

For the measurement of the afterpulsing probability, we select all the events in the residual distribution that are to the right-hand side of the time when the amplitude of afterpulses
reaches 0.5\,p.e. Residuals with shorter time delays are assumed to be due to delayed OC. The vertical lines in Fig.\  \ref{APPHD} give an example of where the boundary between the two
components is placed for the SensL SiPM. The dividing time delay is 50\,ns for the FBK, 17\,ns for the Hamamatsu, and 20\,ns for the SensL device. 

The method is robust but does not provide a clean separation between the two components. A more rigorous approach would also include the amplitude information, which allows a clear
separation between the two components  (see Fig.\  \ref{ampltimes}). Such an approach would also allow extracting the trapping times of the afterpulses. We did not implement such an analysis
because our method to extract the amplitudes and times becomes increasingly inefficient if two pulses are separated by less than 10\,ns. This inefficiency introduces a considerable
systematic effect and results in an underestimation of the delayed optical crosstalk, which dominates the uncertainty in our measurements.

\begin{figure}[!tb]
  \centering
\subfloat[FBK NUV-HD]{\includegraphics[width=\columnwidth]{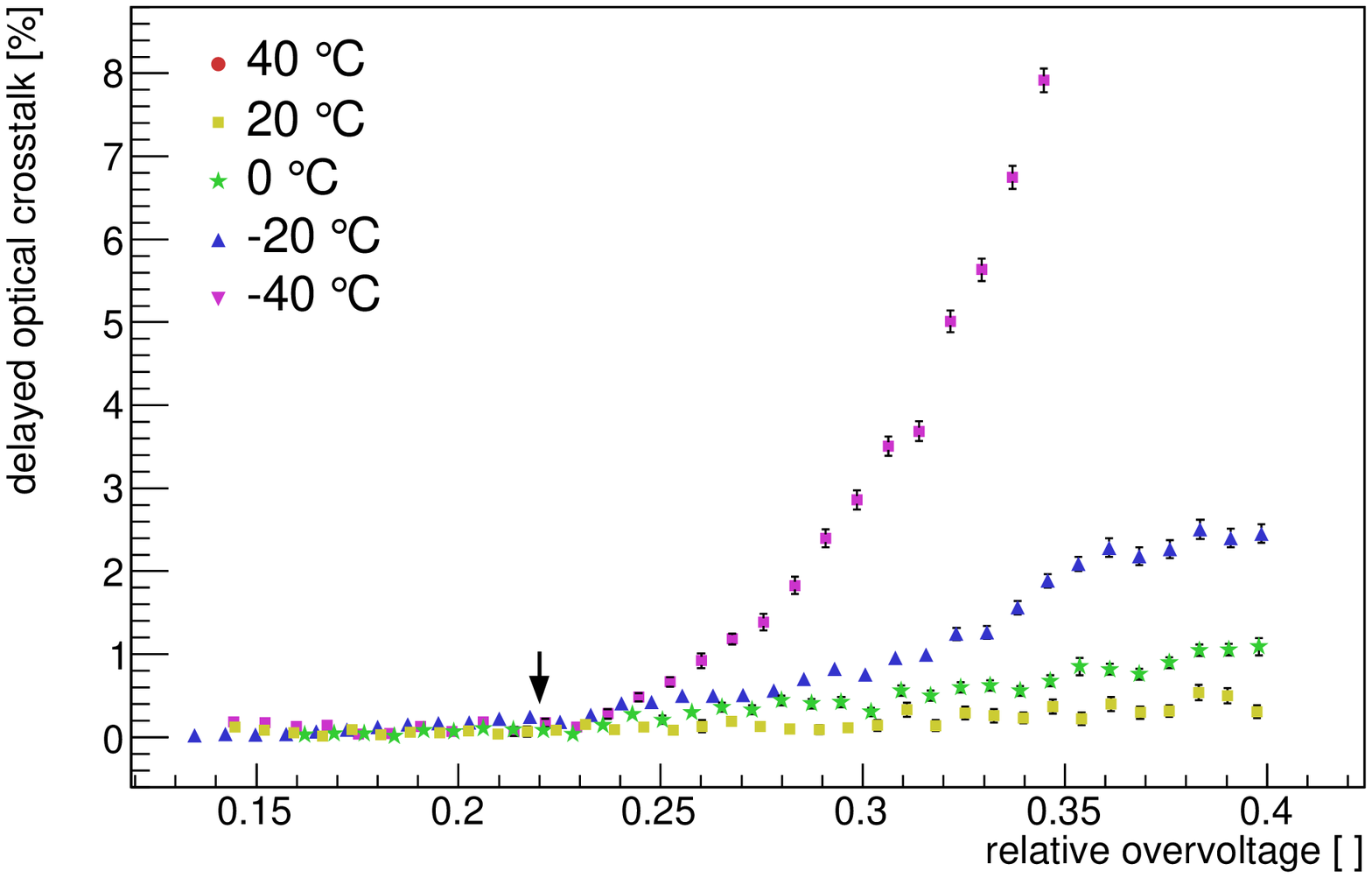}}\\
\subfloat[Hamamatsu S13360-3050CS]{\includegraphics[width=\columnwidth]{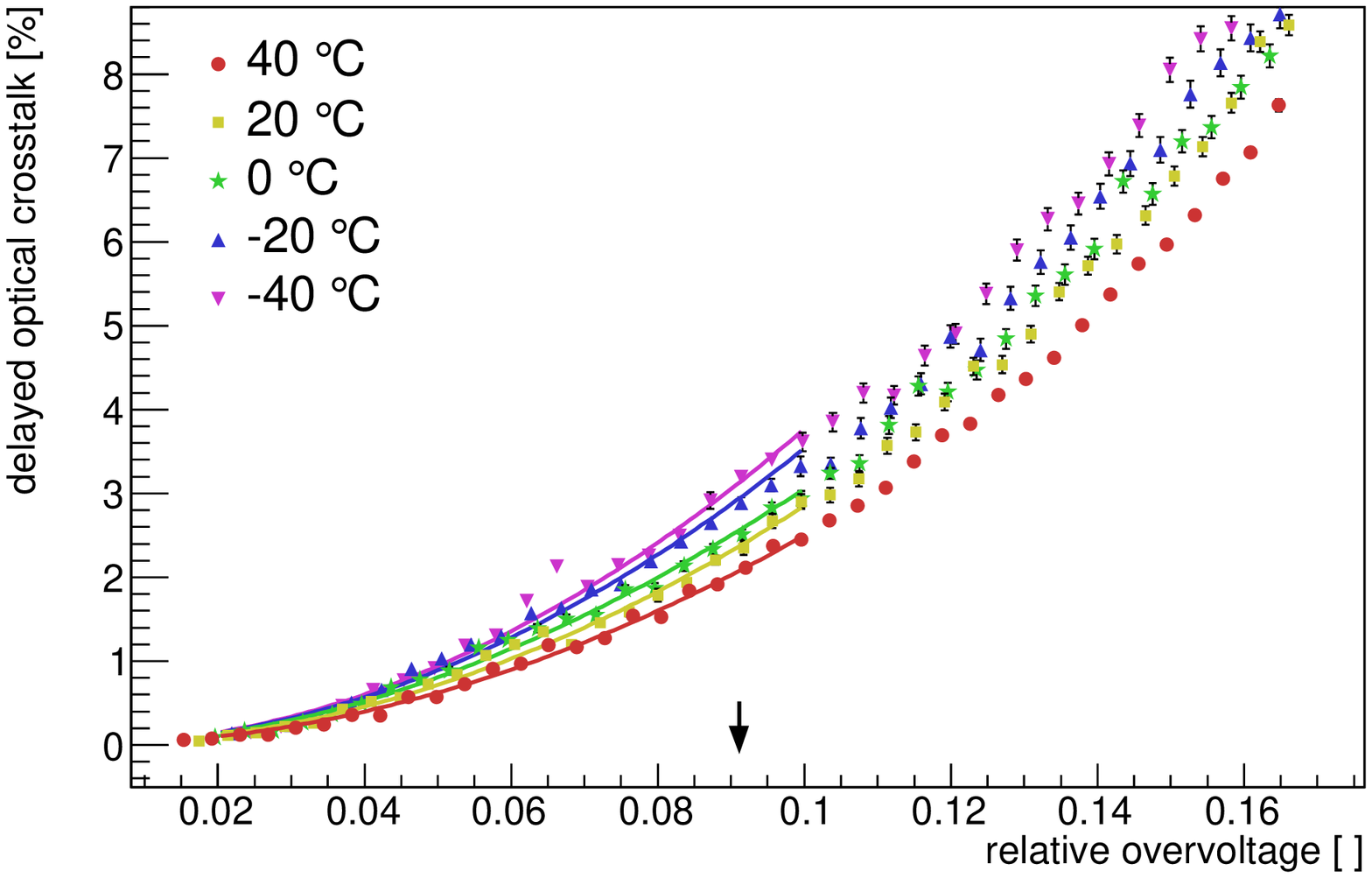}}\\
  \subfloat[SensL J-series 30035]{\includegraphics[width=\columnwidth]{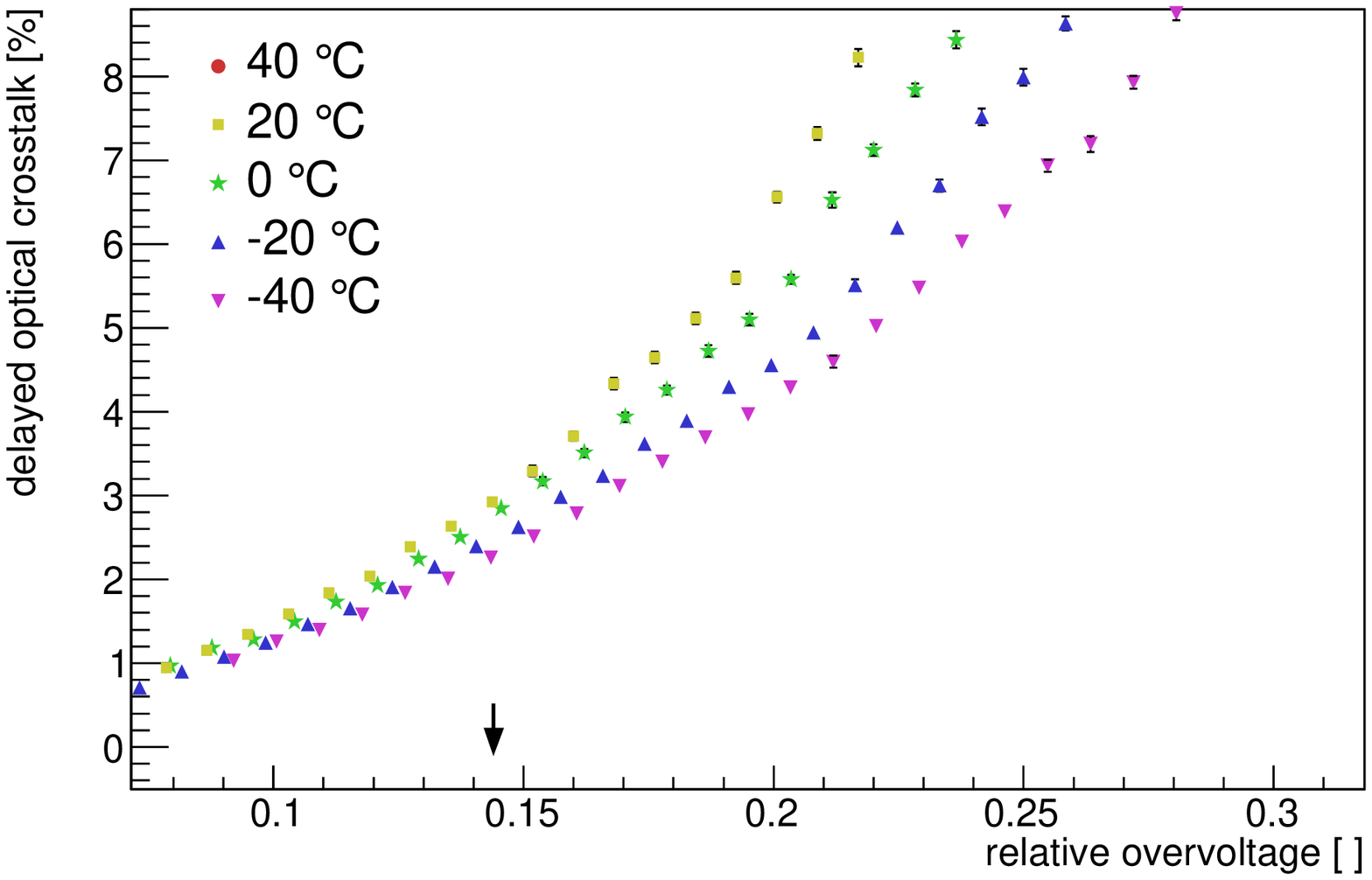}}
\caption{Delayed optical crosstalk. The arrow marks the nominal operating bias of each device.} 
  \label{DOC}
\end{figure}

\begin{figure}[!tb]
  \centering
  \subfloat[FBK NUV-HD]{\includegraphics[width=\columnwidth]{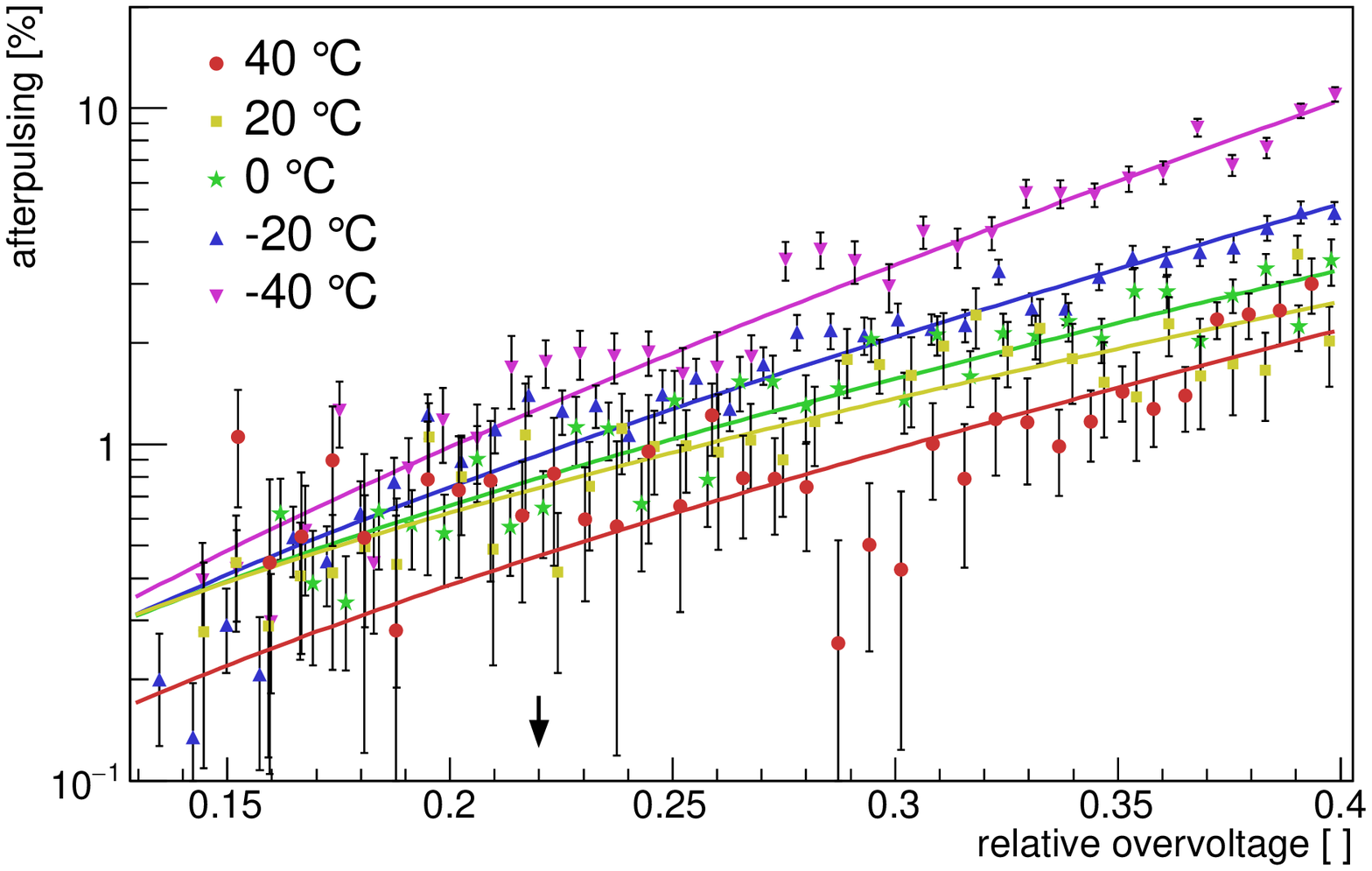}}\\
\subfloat[Hamamatsu S13360-3050CS]{\includegraphics[width=\columnwidth]{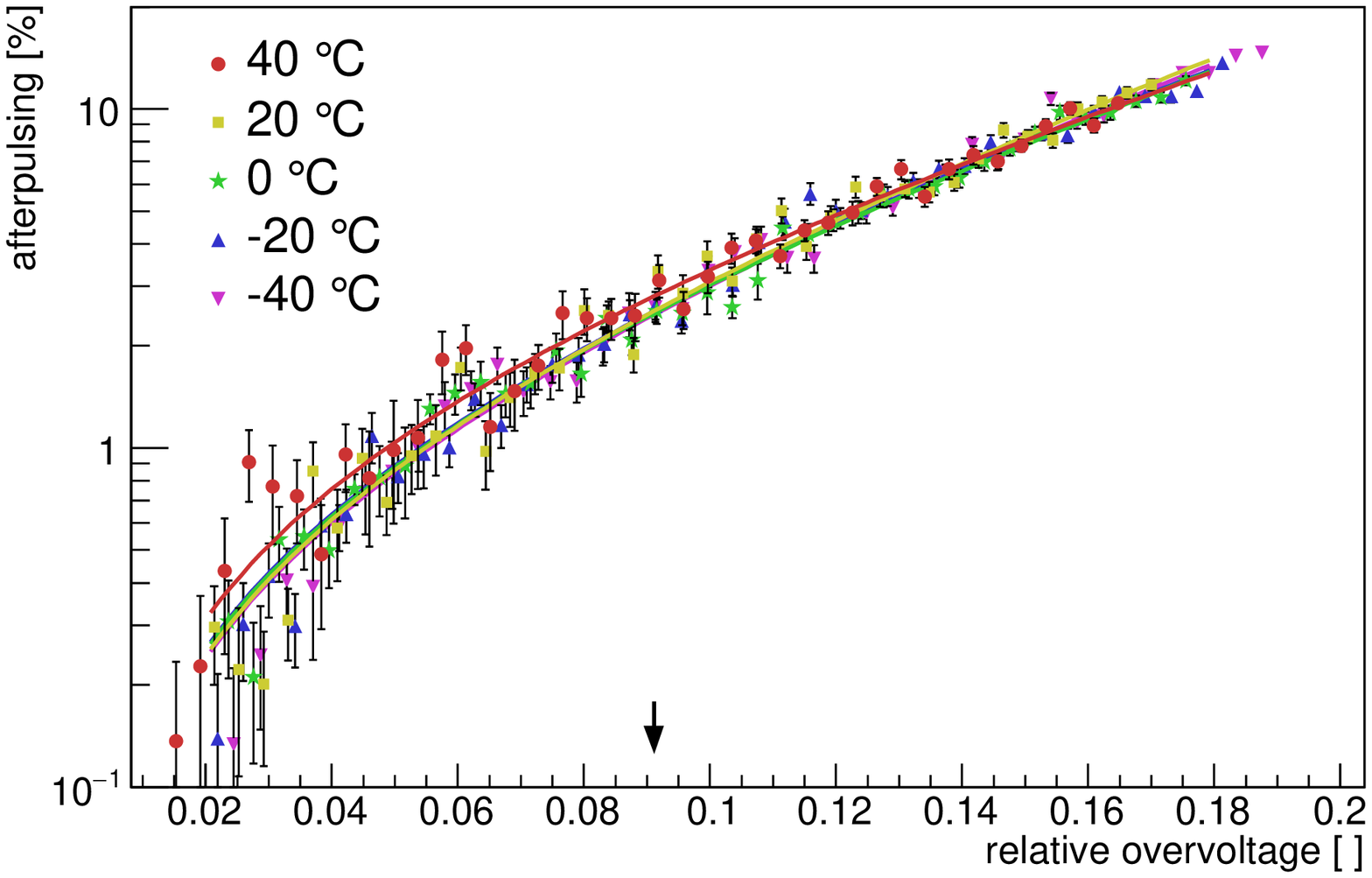}}\\
  \subfloat[SensL J-series 30035]{\includegraphics[width=\columnwidth]{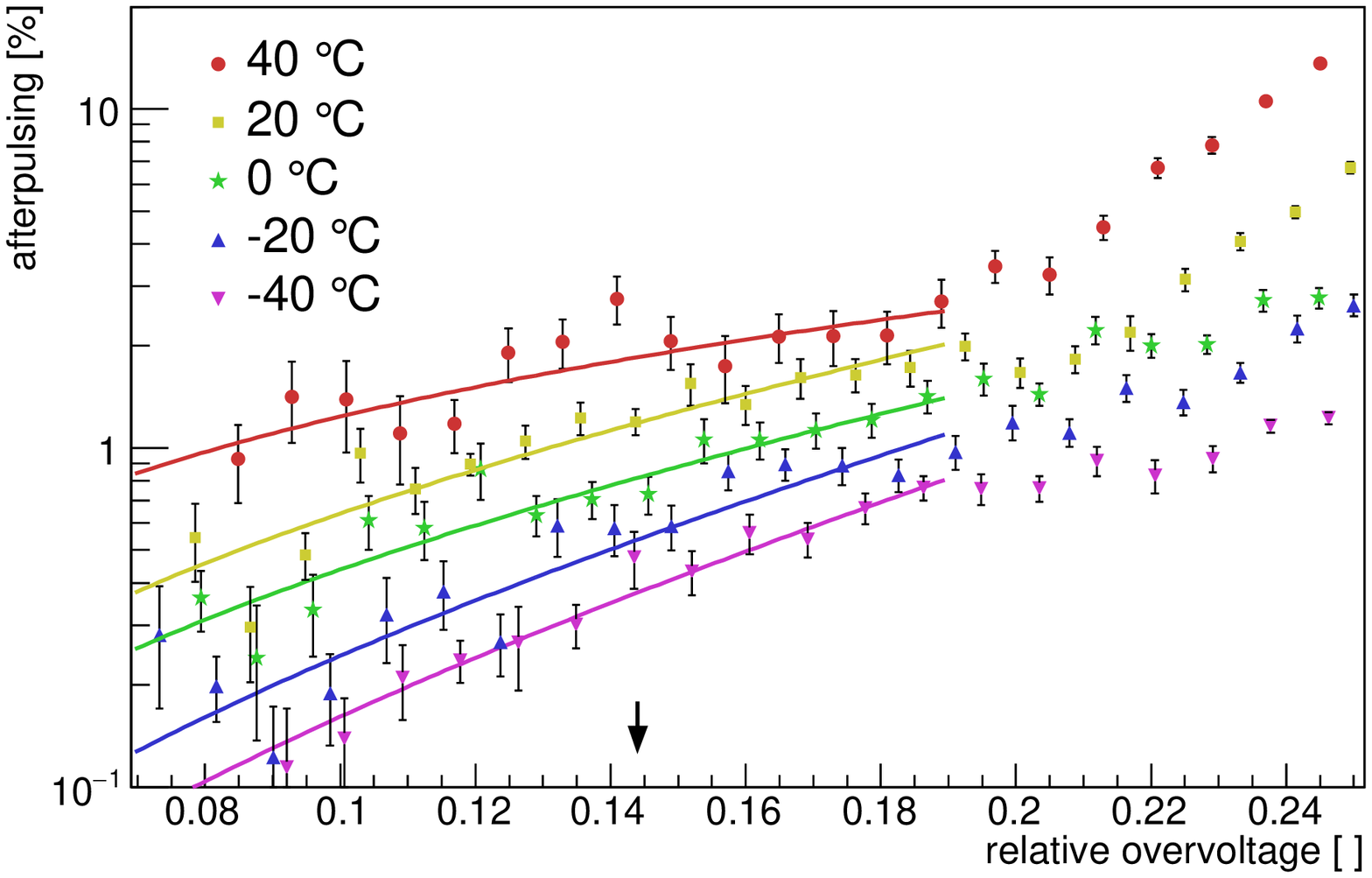}}
\caption{Afterpulsing. The arrow marks the nominal operating bias of each device.} 
  \label{AP}
\end{figure}

Figures \ref{DOC} and \ref{AP} show the delayed OC and afterpulsing probabilities, respectively. At their respective operating voltages all devices have a probability for delayed OC of about
2\%. The afterpulsing probability is less than 2\% for the Hamamatsu SiPM and less than 1\% for the FBK and SensL SiPM. Again we note that the delayed  OC has to be understood as a lower
limit due to the inefficiencies of extracting pulses with time differences that are less than 10\,ns. The afterpulsing probabilities on the other hand are likely overestimated by about 20\%
because of the hard cut that is applied in the residuals to divide the two components. The best separation between the two components is achieved in the measurement of the Hamamatsu device
and is thus the least affected by an overspill of OC events.

From the point of view of judging the performance of the three SiPMs in an application, the afterpulsing and delayed OC probabilities at the operating voltages are sufficiently low that it
is
in fact not necessary to perform a more detailed analysis of, for example, the afterpulsing trapping time constants. 

The overvoltage dependence of the delayed OC can be expected to be described in the same way as the direct OC, \emph{i.e.}, with Equation \ref{OCf}. Fits to the Hamamatsu data are shown in
the Figure \ref{DOC}. However, due to the inefficiency in our pulse-extraction algorithm, we could not extract meaningful parameters from the fit, which is also reflected by a poor probability of the
fit.

The afterpulsing vs.\ overvoltage data are fit with the function
\begin{equation}
AP(U_{\mbox{\footnotesize rel}}) = A\cdot e^{\left(U_{\mbox{\footnotesize rel}}/\delta\right)} \cdot \left[1-e^{\left(-U_{\mbox{\footnotesize rel}}/\alpha\right)}\right]\,,
\end{equation}
where $A$ is a normalization, and the second term describes the bias dependence of the afterpulsing probability. The last term has to be understood as an effective breakdown probability
because it averages over all possible times when afterpulses can happen during the recovery of a cell. Because individual trapping times are exponentially distributed, the majority of the
trapped charges are released shortly after the breakdown of a cell has stopped. This means that the breakdown probability is small at the time when most afterpulse are released and $\alpha$,
therefore, expected to be large. 

The afterpulsing as function of bias does not show a dependence on temperature for the Hamamatsu SiPM. We note that trapping time constants decrease exponentially with increasing
temperature. It is thus expected that afterpulsing decreases with increasing temperature because more trapped carriers are released before the cell recovers to a meaningful breakdown
probability. The expected temperature behavior is observed in the FBK device but not in the SensL device. We cannot rule out that the observed behaviour is due to a contamination of
afterpulses with delayed optical crosstalk events.

For the FBK, Hamamatsu, and SensL SiPMs, the fit values averaged over all temperatures for $\alpha$ are 80, 80, and 100, respectively. For $\delta$ they are 0.2, 0.09, and 0.15,
respectively.
The uncertainties are fairly large and hide any temperature dependencies.

\subsection{Cell Recovery Times}

The last quantity measured is the cell recovery time. Cell recovery times can be measured by flashing an SiPM with two fast consecutive pulses and recording how the second SiPM signal
amplitude
changes as a function of the time difference between the two pulses. The recovery time can also be measured by analyzing the amplitude vs.\ time characteristics of afterpulses, which is
expected to be described with
\begin{equation}
A(t) = A_0 \left[1-e^{t/\tau}\right]\,, 
\end{equation}
where $\tau$ is the time constant of the recovery time. We measured the recovery time using the latter method. The black dots in Fig.\  \ref{ampltimes} are  afterpulses selected to be fit with the above function, which is shown as the solid
black line in the figure.

\begin{figure}[!tb]
  \centering
  \subfloat[FBK NUV-HD]{\includegraphics[width=\columnwidth]{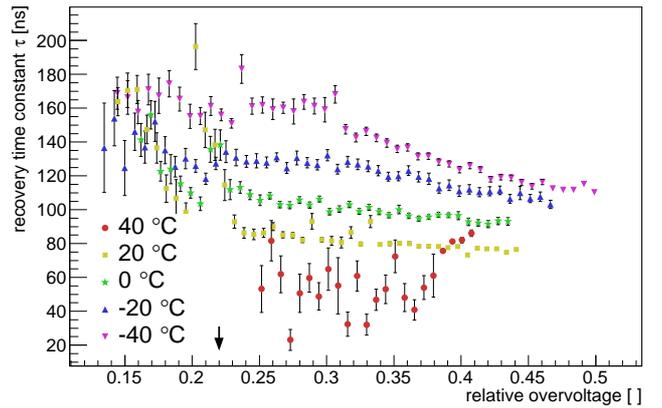}}\\
  \subfloat[Hamamatsu S13360-3050CS]{\includegraphics[width=\columnwidth]{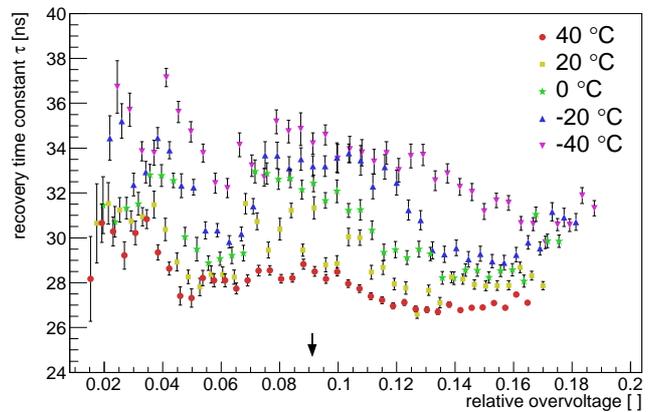}}\\
  \subfloat[SensL J-series 30035]{\includegraphics[width=\columnwidth]{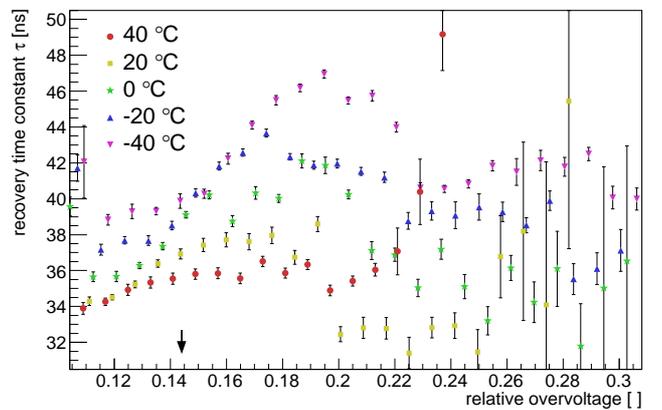}}
\caption{Recovery times. The arrow marks the nominal operating bias of each device.} 
  \label{rectimes}
\end{figure}

The measured recovery time constants are shown in Fig.\  \ref{rectimes} for all devices. At the operating voltages, the time constants are in good agreement with the product of the cell
capacitance and quench resistors. 

An expected trend that is observed for all devices is the decrease of the recovery time with increasing temperature, which is due to the decreasing value of the quench resistor. (s.\ Fig.\ 
\ref{QRs}).

\section{Discussion}
In this paper we presented the characterization of three recent, blue-sensitive SiPMs from FBK, Hamamatsu, and SensL. All three devices show superior performance in terms of their optical
and electrical characteristics with respect to past generations of SiPMs. 

The very good performance of the three devices motivated us to investigate how to best parameterize SiPM characteristics as a function of bias and temperature. We believe that standardizing
the parameterization of SiPMs will become increasingly important as the community of SiPM users is constantly growing, and not everyone has in-house capabilities to perform in-depth device
studies. Furthermore, the optimal operating point of an SiPM varies from application to application, which requires knowledge of SiPM parameters over a wide range of temperature and bias.
With a standardized SiPM parameterization at hand, the user can focus on the application and with the help of the model find the optimal SiPM and its operating point.

We have found parameterizations of the breakdown probability, optical crosstalk, dark rate, and afterpulsing as a function of temperature and bias that can be applied to all three tested
SiPMs. The parameterization also allows extraction of physical parameters like the location of the high-field region using $\alpha$ in the breakdown probability, or the optical crosstalk
efficiency factor $\gamma$.

The choice of comparing device characteristics at the bias where the PDE at 400\,nm reaches 90\% breakdown probability is driven by our ultimate desire to obtain SiPMs with the highest
optical efficiency and, at the same time, sufficiently low nuisance parameters.  If one has to select one of the three devices for an application, detailed end-to-end
simulations are needed that find the bias that results in the best compromise between PDE and nuisance parameters. Such a study is not within the scope of this paper. Instead we discuss how
well the tested devices match the requirements for Cherenkov telescopes when the SiPMs are operated at 90\% breakdown probability, and we point out the remaining shortcomings that
prevent the tested devices from being perfect photon detectors for Cherenkov telescopes when operated at that bias.

Reduced optical crosstalk, afterpulsing, and dark-count rates allow the operation of all three devices at much higher relative overvoltages, thus yielding breakdown probabilities of more
than 90\%
for blue photons. Not only does a 90\% breakdown probability provide a significant boost in PDE, but it also reduces the sensitivity of gain and PDE on temperature changes. Using that one
degree change in temperature shifts the breakdown voltage by 0.1\% for all three devices; the gain of an SiPM changes by 1\%/$^{\circ}$C if it is operated at 10\% overvoltage. If a device is
operated at 20\% overvoltage, the gain changes by only 0.5\%/$^{\circ}$C. The three tested devices operate in between these limits. 

The temperature dependence of the PDE is even smaller because the breakdown probability is in saturation. With our parameterization of the breakdown probability it can be calculated that
the relative PDE changes between 0.2\%/$^{\circ}$C and 0.3\%/$^{\circ}$C for the three tested devices if they are operated at 90\% breakdown probability. These values are on par with typical values for bialkali photomultiplier tubes
\cite{hamamatsuPMThandbook}. Measures to temperature-stabilize SiPMs in applications or to correct data offline is, therefore, not necessary anymore, or the requirements to
temperature-stabilize devices can be much more relaxed. 

The peak PDE of the three devices ranges between 40\% and 50\%, which, again, is a huge improvement compared to the PDEs of devices available just 10 years ago. Being able to operate at
90\% breakdown probability is
certainly one main reason for the high PDEs, but it is worth noting that the spectral response has shifted considerably into the blue/UV region. Considering that the maximum
achievable geometrical fill factor is probably around 80\%, the maximum possible PDE that can be expected for SiPMs is around 65\% assuming a 90\% breakdown probability and a 90\% quantum
efficiency. In fact, FBK recently presented results of SiPMs with a peak PDE of more than 60\% PDE  \cite{PCClaudio}.  Enhancing the blue efficiency of SiPMs further and shifting their peak
efficiency toward lower wavelengths is likely to be realized by thinning the passivation layer and the first implant, which will be technological challenges.

Optical crosstalk, dark rates, and afterpulsing are also much reduced in comparison to older devices. Dark rates are typically a few ten kHz/mm$^2$, whereas early devices typically had rates
of one MHz/mm$^2$. Optical crosstalk has been lowered by reducing cell capacitances, introducing trenches between cells, and optimizing the layout of structures. Each tested devices has
successfully implemented one or more of the aforementioned measures, and direct optical crosstalk ranges between 6\% and 20\% at 90\% breakdown probability.

Delayed optical crosstalk and afterpulsing are two more nuisance parameters that could be considerably improved, with typical values being $\sim2$\%.

Parameters that are well within the requirements are cell recovery time and
gain. A lower gain and a shorter cell recovery time in future devices is perfectly acceptable. A lower gain would
reduce power dissipated by the SiPM, which is a plus when SiPMs are used  in environments with intense photon backgrounds.

Given all of these improvements, only a short list of desirable changes remain:
\begin{itemize}
\item The sensitivity should be highest between 250\,nm and 550\,nm if possible with a flat response. Above 550\,nm the sensitivity should cut off sharply. Such a spectral response would
maximize the detection of Cherenkov light and at the same time efficiently reject ambient light coming from the night sky, which dominates at long wavelengths. Of the three tested devices,
the FBK device comes closest to the ideal response, but improvements would still be desirable to further suppress the response at long wavelengths.
\item Direct optical crosstalk is one of the main factors limiting the lowest achievable trigger threshold. The majority of trigger concepts used in Cherenkov telescopes employ an n-fold
coincidence of neighboring camera pixels. In the coincidence, each pixel has to have a signal above a certain threshold. How low that threshold can be set depends ideally only on the maximum
acceptable trigger rate due to statistical up-fluctuations in the ambient light. For most operating or planned Cherenkov telescopes, a direct optical crosstalk of 3\% would double that
trigger rate which would be acceptable. It is of course desirable to minimize optical crosstalk as much as possible. With 6\% optical crosstalk, the Hamamatsu device is not far from an
optimal value.
\item Afterpulsing and delayed optical crosstalk add to the effective dark-count rate and contaminate the extracted Cherenkov signal by introducing a positive bias. With about 2\%
afterpulsing and delayed optical crosstalk, respectively, all three devices have acceptable values that can be dealt with at the stage of signal extraction. However, keeping both effects
below 1\% would simplify the data analysis and reduce systematic uncertainties in the energy scale of Cherenkov telescopes.
\item The cost of SiPMs is still a dominant contribution to the total per channel costs (readout electronics and photosensor). Considerable efforts have been made in the past to reduce the
cost of the readout electronics, and it is not unreasonable to assume that with new concepts costs of \$5 per readout channel can be realized in the future. SiPMs would have to cost about
\$$0.1/\mbox{mm}^{2}$ to contribute equally to the per channel costs. 
\end{itemize}  
All these items are major technological challenges, but it is not evident that fundamental physical limitations preclude one from surmounting them. Therefore, we are confident that new and improved devices will become available in the future.

\section*{Acknowledgment}
We are grateful to FBK, Hamamatsu, and SensL for providing us with samples of their latest developments. We thank J.\ Biteau for useful discussions and input that improved the extraction of the breakdown voltage from the IV curves. in our procedure to fit the IV curves This research was in part supported by the National Science Foundation under grant
no.\ PHYS-1505228. 

\section*{References}

\bibliography{references}

\end{document}